\documentclass[aps,prd,superscriptaddress,nofootinbib,amsfonts,amssymb,amsmath,notitlepage,onecolumn]{revtex4-1}

\usepackage[dvipdfmx]{graphicx}
\usepackage{amsmath}
\usepackage{amssymb} 
\usepackage{afterpage,dblfloatfix}
\usepackage{color}
\usepackage{xcolor}
 \usepackage{float}
 \usepackage{dblfloatfix}
\usepackage[colorlinks=true, allcolors=blue!75!black]{hyperref}
\usepackage{makecell}
\usepackage{multirow}
\usepackage{slashed}
\usepackage{wrapfig}
\usepackage{placeins}
\usepackage{bm}
\usepackage[nameinlink]{cleveref}
\usepackage[caption=false]{subfig}

\usepackage{autobreak}

\newcommand{\bvec}[1]{\mbox{\boldmath $#1$}}

\newcommand{\df}{\text{d}}

\newcommand{\pmat}[1]{\begin{pmatrix}#1\end{pmatrix}}

\newcommand{\Star}[1]{#1\ensuremath{^*}\kern-\scriptspace}

\graphicspath{{./figs/}}
\newbox{\ORCIDicon}
\sbox{\ORCIDicon}{\large
                  \includegraphics[width=0.8em]{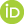}}

\allowdisplaybreaks[3]

\begin{document}

\title{
Columbia plot based on symmetry-improved CJT formalism in linear sigma model
}

\author{Yuepeng \surname{Guan}\,\href{https://orcid.org/0009-0007-8571-0931}{\usebox{\ORCIDicon}}}
\thanks{{\tt guanyp22@mails.jlu.edu.cn}} 
\affiliation{Center for Theoretical Physics and College of Physics, Jilin University, Changchun, 130012, China}

\author{Mamiya Kawaguchi\,\href{https://orcid.org/}{\usebox{\ORCIDicon}}}
\thanks{{\tt mamiya@aust.edu.cn}} 
\affiliation{ 
Center for Fundamental Physics, School of Mechanics and Physics,
Anhui University of Science and Technology, Huainan, Anhui 232001, People’s Republic of China
} 

\author{Shinya Matsuzaki\,\href{https://orcid.org/}{\usebox{\ORCIDicon}}}
\thanks{{\tt synya@jlu.edu.cn}}
\affiliation{Center for Theoretical Physics and College of Physics, Jilin University, Changchun, 130012, China}%

\author{Akio Tomiya\,\href{https://orcid.org/}{\usebox{\ORCIDicon}}}
\thanks{{\tt akio@yukawa.kyoto-u.ac.jp}} 
\affiliation{Department of Information and Mathematical Sciences, Tokyo Woman’s Christian  University, Tokyo 167-8585, Japan} 
\affiliation{RIKEN Center for Computational Science, Kobe 650-0047, Japan}

\begin{abstract}

We study the Columbia plot for the chiral phase transition in the framework of a three-flavor linear sigma model based on the Cornwall-Jackiw-Tomboulis (CJT) formalism.
The conventional CJT approach with the Hartree truncation suffers from artificial chiral breaking, leading to the violation of the Nambu-Goldstone theorem and the (anomalous) chiral Ward-Takahashi identities.
We apply the symmetry-improved CJT formalism to resolve this issue. 
We observe a first-order phase transition and a tricritical point in the light-quark mass regime, which is relatively independent of the size of the sigma meson, in contrast to the conventional CJT approach. 
The tricritical point, found on the $m_s$ axis, is at $m_s^{\rm tri}/m_s^{\rm phys.} = 0.175$ with $m_s^{\rm phys.}$ being the physical strange quark mass in real-life QCD. 
The critical pion mass in the three-flavor symmetric limit, on the second-order boundary, is measured at $m_\pi \sim 52.4$ MeV, with the critical temperature $T_c \sim 51.7$ MeV. 

\end{abstract}
\maketitle

\section{Introduction}

Quantum Chromodynamics (QCD) exhibits a rich phase structure at high temperature associated with chiral symmetry restoration and deconfinement.
A deep understanding of the chiral phase transition 
is of importance to explore the origin of mass and matter. 
The study along this line has attracted a growing interest in not only the current terrestrial experiments, but also astrophysical and cosmological observations; heavy-ion collisions~\cite{STAR:2010vob,Mohanty:2011nm,Luo:2017faz,Nonaka:2019fad}, neutron star interiors~\cite{Buballa:2003qv,Fukushima:2013rx,Oertel:2016bki}, 
and the thermal history of the early universe~\cite{Planck:2018vyg,Zheng:2024tib}.

In the general QCD parameter space, the chiral phase transition can be characterized by the lightest quark mass $m_u=m_d\equiv m_l$ and the strange quark mass $m_s$. The parameter space is called the Columbia plot~\cite{Pisarski:1983ms,Brown:1990ev}. 
The physical point at which QCD of the Standard Model is located, thus, is merely a single point in the continuous quark mass parameter space. 
Going off the physical point therefore brings a broader and deeper view of how the chiral phase transition, hence the origin of mass and matter, is realized 
in real-life QCD.

The phase structure of the Columbia plot 
for the small $m_l$ regime is still an open question. 
 It is typically classified into three ``scenarios"~\cite{Bernhardt:2023hpr,Pasztor:2024dpv,Klinger:2025xxb}: 
\begin{itemize}
    \item[i)]
    the \textit{tricritical point} at $m_s=m_s^{\rm tri}$ on the $m_s$-axis with $m_u = m_d = 0$ is present, at which point the first order regime is closed and the phase transition becomes second order, which ever persists toward the two-flavor chiral limit, $m_s \to \infty$, reflecting the universality class of the $O(4)$ theory with no $U(1)$ axial symemtry restoration~\cite{Brown:1990ev,Resch:2017vjs,deForcrand:2017cgb,Dini:2021hug,Bernhardt:2023hpr}; 
    
    \item[ii)] 
    the tricritical point is absent and $U(1)$ axial symmetry is restored; hence, that is not in the $O(4)$ universality class, so that the first-order regime can be observed till the two-flavor chiral limit~\cite{deForcrand:2017cgb,Dini:2021hug}; 
    \item[iii)] 
    no first-order regime emerges and the second-order regime shows up only on the $m_s$-axis~\cite{Cuteri:2021ikv,Fejos:2022mso,Bernhardt:2023hpr}. 
\end{itemize}
Upon the current status of the lattice simulations, for $2+1$ flavor QCD with small $m_l$ and large $m_s$, either scenarios i) or ii) tend to be favored~\cite{Dick:2015twa,Iwasaki:1996ya,DElia:2004uwa,DElia:2005nmv,Kogut:2006gt,Bonati:2014kpa,Philipsen:2016hkv,Cuteri:2017gci,Ding:2018auz}. 
The functional renormalization group (fRG) approach to QCD has also realized scenario i) for the two-flavor~\cite{Braun:2009gm} and $2+1$-flavor cases~\cite{Braun:2020ada}.  
In the three-flavor symmetric case with $m_l=m_s$, 
Scenario iii) has been supported~\cite{Cuteri:2021ikv,Dini:2021hug,Fejos:2022mso,Bazavov:2017xul,Dini:2021hug,Kuramashi:2020meg}. 
The first-order signal has not yet been observed even for a small pion mass, down to $m_\pi \lesssim 50$~MeV with highly improved staggered fermions~\cite{Bazavov:2017xul,Dini:2021hug} and 
$m_\pi \lesssim 110$~MeV with improved Wilson fermions~\cite{Kuramashi:2020meg}. 
A recent analysis using the Dyson-Schwinger equation has also supported a second-order phase transition for three massless quark flavors~\cite{Bernhardt:2023hpr}, namely, implying no first-order regime in the small-quark mass regime.

Alternatively to lattice QCD studies, 
low-energy effective field theories (LEFTs) can also provide a complementary view of the investigation on the chiral phase structure on the Columbia plot~\cite{Lenaghan:2000kr,Kovacs:2006ym,Lenaghan:2000ey,Fukushima:2008wg,Schaefer:2008hk,Mitter:2013fxa,Grahl:2013pba,Eser:2015pka,Resch:2017vjs,Giacosa:2024orp,Fejos:2024bgl,Ahmed:2024rbj,Fejos:2023lvw,Pisarski:2024esv,Tiwari:2024jfd,Tiwari:2025mnr}. 
These theories qualitatively capture the relevant infrared dynamics through effective degrees of freedom (light mesons and quarks), keeping the global symmetries such as the chiral symmetry and its violation as present in QCD.
The linear sigma model (LSM) is one candidate description as such, which serves as a particularly valuable framework for studying chiral phase transitions.

In employing the LSM, one notices one remark made in the literature~\cite{Resch:2017vjs}, showing that fluctuations like mesonic loop corrections substantially modify the result in the mean-field approximation projected onto the Columbia plot. 
To properly incorporate the potentially nonperturbative mesonic loop corrections,  
one can use the Cornwall-Jackiw-Tomboulis (CJT) method in two-particle (2PI) irreducible effective action formulations. 
Plugging this formalism into an LSM, one will solve the gap equation for 
two-point functions of meson fields, which involve resummed quantum-dressed propagators. 
Accordingly, the ground state in the CJT formalism will be determined at the stationary point for not only the chiral-order parameter but also the meson propagators. 
This formalism would thus provide an alternative approach to beyond-perturbation study the Columbia plot based on a different systematics compared to other nonperturbative methods, such as the fRG and Dyson-Schwinger equations, and also lattice simulations.

However, it has also been pointed out that the CJT method inevitably suffers from the difficulty that the Nambu-Goldstone (NG) theorem or the threshold property are violated~\cite{Lenaghan:2000ey,Lenaghan:2000kr,Pilaftsis:2013xna}.  
This is because a non-systematic truncation of the 2PI digarams, e.g., often referred to as the Hartree approximation, 
is required to apply in practically evaluating the CJT effective action at the two-loop level; hence non-chiral invariant set of the two-loop corrections generates artificial chiral breaking, which feeds back to the violent deviation from the low-energy theorem 
and the NG theorem in the chiral limit as well. 
This issue is therefore crucial at around the critical temperature of the chiral phase transition. In fact, the violation of the NG theorem is reflected in generating the massive NG boson at finite temperature even in the case without the explicit symmetry breaking (e.g., the chiral limit for the QCD case). 
In the literature~\cite{Pilaftsis:2013xna}, it has been clarified that the loopwise truncation deviates the Ward-Takahashi identities (WTIs) of the one-particle irreducible (1PI) from those of 2PI actions, with the former protecting the NG theorem at any order of truncation.

Thus, the so-called {\it symmetry-improved (SI) CJT} formalism has been proposed with the enforcement of the 1PI WTIs to preserve the original global symmetry ($O(N)$ symmetry as a reference model)~\cite{Pilaftsis:2013xna}. 
An application of this original SICJT formalism to $O(4)$ LSM has been performed in ~\cite{Mao:2013gva}, where the two-flavor chiral phase transition has been 
addressed in the chiral limit with the exact NG boson even at finite temperature.

In this paper, we apply the SICJT formalism to a three-flavor LSM 
and explore the chiral phase structure in the small-quark mass regime of the Columbia plot. We observe a first-order regime and a tricritical point consistently with  Scenario i) above. 
This is a definitive conclusion fairly insensitive to the size of the sigma meson mass, in contrast to the conventional CJT approach in~\cite{Lenaghan:2000ey,Lenaghan:2000kr}.

This paper is organized as follows.
In Sec.~\ref{sec:LinearSigmaModelAndCJTFormalism}, we introduce a three-flavor 
LSM and summarize what is necessary to address the chiral phase transition in the CJT formalism at zero and finite temperature. 
This section is essentially just a review of the literature~\cite{Lenaghan:2000ey, Lenaghan:2000kr,Kawaguchi:2020qvg}. 
In Sec.~\ref{sec:WTISsndSI}, applying the argument in~\cite{Pilaftsis:2013xna} to the present LSM case, we discuss the violation of the NG theorem by demonstrating the discrepancy between the WTIs associated with the 1PI and the 2PI formalisms.
Then, we introduce the SICJT formulation established in~\cite{Pilaftsis:2013xna}.
In Sec.~\ref{sec:Results}, we show our results on the chiral phase transitions and projection onto the Columbia plot, and discuss the related physical consequences, in comparison with the results based on the conventional CJT approach. 
We also present the critical exponents around several critical points in the Columbia plot evaluated from the SICJT formalism.
The conclusion of this paper is provided in Sec.~\ref{sec:conclusion}. 
In Appendices~\ref{app:VacuumPhenomenologyOfLinearSigmaModel}, \ref{app:CJTFormalismAtFinitTemperature}, \ref{app:DerivationsOfWTIS}, and \ref{app:PhaseTransitionsAtSomeTypicalPoints}, we also present supplementary materials for more details relevant to the present work: 
tree-level formulae in the LSM (\ref{app:VacuumPhenomenologyOfLinearSigmaModel}); 
the CJT formalism (\ref{app:CJTFormalismAtFinitTemperature}); 
the WTIs with and without explicit symmetry breaking (\ref{app:DerivationsOfWTIS});    
benchmarked temperature dependences for the chiral phase transitions at certain  critical points in the Columbia plot  (\ref{app:PhaseTransitionsAtSomeTypicalPoints}).

\section{An LSM and CJT formalism}
\label{sec:LinearSigmaModelAndCJTFormalism}

We begin by introducing a phenomenologically successful three-flavor LSM proposed in~\cite{Kuroda:2019jzm} and analyzed at finite temperature in~\cite{Kawaguchi:2020qvg} within the conventional CJT formalism~\cite{Cornwall:1974vz}. 
This LSM can, in particular, reproduce what is called the inverse mass hierarchy between 
$a_0(950)$ and $\kappa(700)=K_0^{*}(700)$ spectra, due to the introduction of 
the explicit-chiral breaking contribution related to the $U(1)$-axial 
anomaly-induced term (see Eq.~\eqref{k-term})~\cite{Kuroda:2019jzm}. 
Some of the detailed formulae can be found in Appendices~\ref{app:VacuumPhenomenologyOfLinearSigmaModel} and \ref{app:CJTFormalismAtFinitTemperature}.

The central building block of the model is the LSM field $\Phi$, which transforms 
bifundamentally under the full $U(3)_L \times U(3)_R$ symmetry as 
\begin{align}
    \Phi \rightarrow g_L \cdot \Phi \cdot g_R^\dagger\,, 
    \label{eq:chiralTransformation}
\end{align}
with $g_{L,R} \in U(3)_{L,R}$.  
Throughout the present paper, we work in Euclidean spacetime. 
The model Lagrangian is given as  
\begin{align}
    \mathcal{L}_{\rm LSM} = \operatorname{tr} \Big[ \partial_\mu \Phi^\dagger \partial_\mu \Phi \Big] + V\big(\Phi,\Phi^\dagger\big),
    \label{eq:lagrangianLSM}
\end{align}
The mesonic field $\Phi$ is parametrized as $\Phi = \phi_a T^a = ( \sigma_a + i \pi_a ) T^a$ for $a = 0, \cdots,8$, in which $\sigma_a$ and $\pi_a$ denote the scalar and pseudoscalar fields, respectively. 
$T^a = \lambda^a/2$ represents generators of the $U(3)$ group, in which $\lambda^{a=1,\cdots,8}$ are the Gell-Mann matrices, and $\lambda^0 = \sqrt{2/3} \, \textbf{1}_{3 \times 3}$.
The mesonic potential $V\big(\Phi,\Phi^\dagger\big)$ in Eq.~\eqref{eq:lagrangianLSM} is given by
\begin{align}
    V\big(\Phi,\Phi^\dagger\big) = V_0 + V_{\rm anom} + V_{\rm SB} + V_{\rm SB-anom}.
    \label{eq:LSMTreePotential}
\end{align}
The first term $V_0$ in Eq.~\eqref{eq:LSMTreePotential} corresponds to the $U(3)_L \times U(3)_R$ invariant part. 
This potential term is, up to mass dimension four, given as 
\begin{align}
    V_0 = \mu^2 \operatorname{tr} \big[ \Phi^\dagger\Phi \big] + \lambda_1 \operatorname{tr} \big[ \big( \Phi^\dagger\Phi \big)^2 \big] + \lambda_2 \big( \operatorname{tr} \big[ \Phi^\dagger\Phi \big] \big)^2
\,. \label{V0}
\end{align}

The second term $V_{\rm anom}$ in Eq.~\eqref{eq:LSMTreePotential} arises from the QCD instanton contribution associated with the $U(1)$ axial anomaly. 
It takes the determinant form, the so-called Kobayashi-Maskawa-'t Hooft (KMT) term~\cite{Kobayashi:1970ji,Kobayashi:1971qz,tHooft:1976rip,tHooft:1976snw},   
\begin{align}
    V_{\rm anom} = - B \big( \det\big[ \Phi \big] + \det\big[ \Phi^\dagger \big] \big)
\,. \label{Vanom}
\end{align}

The third term $V_{\rm SB}$ in Eq.~\eqref{eq:LSMTreePotential} is the leading-order explicit chiral-breaking term 
tied with the low-energy theorem, 
\begin{align}
    V_{\rm SB} = -c \operatorname{tr} \big[ \mathcal{M} \Phi^\dagger + \mathcal{M}^\dagger \Phi \big]
\,. 
\end{align}
Here ${\cal M}$ is the spurion field, which transforms in the same way as $\Phi$ in Eq.~\eqref{eq:chiralTransformation}, 
\begin{align}
    \mathcal{M} \rightarrow g_L \cdot \mathcal{M} \cdot g_R^\dagger\,, 
\label{trans-calM}
\end{align}
and takes the vacuum expectation value (VEV) $\mathcal{M} \equiv \operatorname{diag} \{ m_u, m_d, m_s \}$ to incorporate the proper explicit-chiral breaking into mesons arising from the current quark masses present in QCD. 

The fourth term $V_{\rm SB-anom}$ in Eq.~\eqref{eq:LSMTreePotential} is characteristic to the present LSM, 
\begin{align}
    V_{\rm SB-anom} = -k c \Big[ \epsilon_{abc} \epsilon^{def} \mathcal{M}^a_d \Phi^b_e \Phi^c_f + {\rm h.c.} \Big]
\,.  
\label{k-term}
\end{align}
This is the axial-anomaly induced flavor-breaking term, which controls the inverse mass hierarchy for $a_0$ and $K_0^*$ masses~\cite{Kuroda:2019jzm}.

The dynamical chiral symmetry breaking is 
monitored by the chiral order parameter, i.e., 
the VEV, or the thermal average of $\Phi$, which is 
parameterized as
\begin{align}
    \langle \Phi \rangle \equiv \bar \Phi = \bar \sigma_a T^a,
\end{align}
without loss of generality.
The isospin symmetry is adopted as $m_u = m_d \equiv m_l \neq m_s$, such that the VEVs of the $\sigma$-fields read 
\begin{align}
    \bar \sigma_a T^a = \bar \sigma_0 T^0 + \bar \sigma_8 T^8 = \operatorname{diag} \{ \bar \Phi_1, \bar \Phi_1, \bar \Phi_3 \},
\end{align}
together with the following relations: 
\begin{align}
    \bar \Phi_1 &= \frac{1}{\sqrt{6}} \bar \sigma_0 + \frac{1}{2\sqrt{3}} \bar \sigma_8, \nonumber\\
    \bar \Phi_8 &= \frac{1}{\sqrt{6}} \bar \sigma_0 - \frac{1}{\sqrt{3}} \bar \sigma_8.
\label{Phi-VEVs}
\end{align}

At tree-level, the stationary condition reads
\begin{align}
    \frac{\delta S_{\rm LSM}}{\delta \Phi_1} \Biggl|_{\Phi = \bar\Phi} = 0, \qquad \frac{\delta S_{\rm LSM}}{\delta \Phi_3} \Biggl|_{\Phi = \bar\Phi} = 0\,,
\end{align}
where $S_{\rm LSM} = \int_x \mathcal L_{\rm LSM}$ is the Euclidean action, with the shorthand notation $\int_x\equiv \int \df \tau \int \df^3\bvec{x}$. 
The meson mass matrices are given by 
\begin{align}
    \Big[ m_S^2(\bar\sigma) \Big]^{ab} &= \frac{\partial^2 V_{\rm LSM}}{\partial \sigma_a \partial \sigma_b} \Biggl|_{\sigma_a = \bar\sigma_a}, \nonumber\\
    \Big[ m_P^2(\bar\sigma) \Big]^{ab} &= \frac{\partial^2 V_{\rm LSM}}{\partial \pi_a \partial \pi_b} \Biggl|_{\sigma_a = \bar\sigma_a}.
\end{align}
The explicit expressions and derivations for those formulae can be found in Appendix~\ref{app:VacuumPhenomenologyOfLinearSigmaModel}.
For the scalar mesons, $\Big[ m_S^2(\bar\sigma) \Big]^{11}$ are 
$\Big[ m_S^2(\bar\sigma) \Big]^{44}$ respectively identified as the squared masses for $a_0(980)$, $m^2_{a_0}$, and 
$\kappa(700)=K^*_0(700)$, $m^2_{\kappa}$.
The $f_0(500)$ and $f_0(980)$ masses arise as the mass eigenvalues from the isospin-singlet scalar mass matrix through a mixing angle $\theta^0_S$ defined with an orthogonal transformation matrix in Eq.~\eqref{eq:orthogonalTransScalarVacuum}.
Similarly to the scalar mesons, $\Big[ m_P^2(\bar\sigma) \Big]^{11} = m^2_{\pi}$ and $\Big[ m_P^2(\bar\sigma) \Big]^{44} = m^2_{K}$, respectively. 
The $\eta^\prime$ and $\eta$ masses are read as the mass eigenvalues for the isospin-singlet pseudoscalar-mass matrix with a mixing angle $\theta^0_P$, 
as in Eq.~\eqref{eq:orthogonalTransPseudoScalarVacuum}.

To evaluate the quantum and thermal corrections arising in the present LSM 
at finite temperature $T$, 
we apply the CJT formalism~\cite{Cornwall:1974vz, Roder:2003uz}.  
We start from the general loop expansion of the 2PI effective potential for the LSM with $N_f$ flavors, which goes like  
\begin{align}
    &V_{\rm CJT}[\sigma, S, P] \nonumber\\
    =& V(\sigma) + \frac{1}{2} \int_k \operatorname{tr} \Big[ \log S^{-1}(k) + \log P^{-1}(k) \Big] \nonumber\\
    &+ \frac{1}{2} \int_k \operatorname{tr} \Big[ \overline{S}^{-1}(k;\sigma)S(k) + \overline{P}^{-1}(k;\sigma)P(k) - 2 \cdot \textbf{1}_{\rm adj} \Big] \nonumber\\
    &+ V_2[\sigma,S, P],
    \label{eq:GeneralCJTPotnetialMainText}
\end{align}
where $\sigma$ is a general background field; 
the trace is taken in the flavor space; 
$V(\sigma)$ is the tree-level potential defined in Eq.~\eqref{eq:LSMTreePotential}, 
and $V_2[\sigma, S, P]$ is the sum of all 2PI  diagrams, which are constructed by the full-meson propagators $\mathcal S$ and $\mathcal P$; 
$\textbf{1}_{\rm adj}$ is the unit matrix in the adjoint representation of $SU(3)$ having the size of $N_f^2 \times N_f^2$; the full scalar (pseudoscalar) propagators are denoted as $S$ ($P$), which are treated as independent variables in the present framework. 
In Eq.~\eqref{eq:GeneralCJTPotnetialMainText}, we have abbreviated the loop momentum integral as 
$\int_k f(k) \equiv T \sum_n\int \df^3 \bvec{k}/(2\pi)^3 f(\omega_n,\bvec{k})$ 
with $\omega_n$ being the bosonic Matsubara frequency. 
In Eq.~\eqref{eq:GeneralCJTPotnetialMainText}, $\overline{S}^{-1}(k;\sigma)$ and $\overline{P}^{-1}(k;\sigma)$ are the tree-level mesonic inverse propagators induced from the tree-level action, which reads 
\begin{align}
    \Big[ \overline{S}^{-1}(k;\sigma) \Big] ^{ab}&= k^2 \delta^{ab} + \Big[ m_S^2(\sigma) \Big]^{ab}, \nonumber\\
    \Big[ \overline{P}^{-1}(k;\sigma) \Big]^{ab} &= k^2 \delta^{ab} + \Big[ m_P^2(\sigma) \Big]^{ab}.
\end{align}

\begin{figure}[t!]
    \centering
    \includegraphics[width=0.4\linewidth]{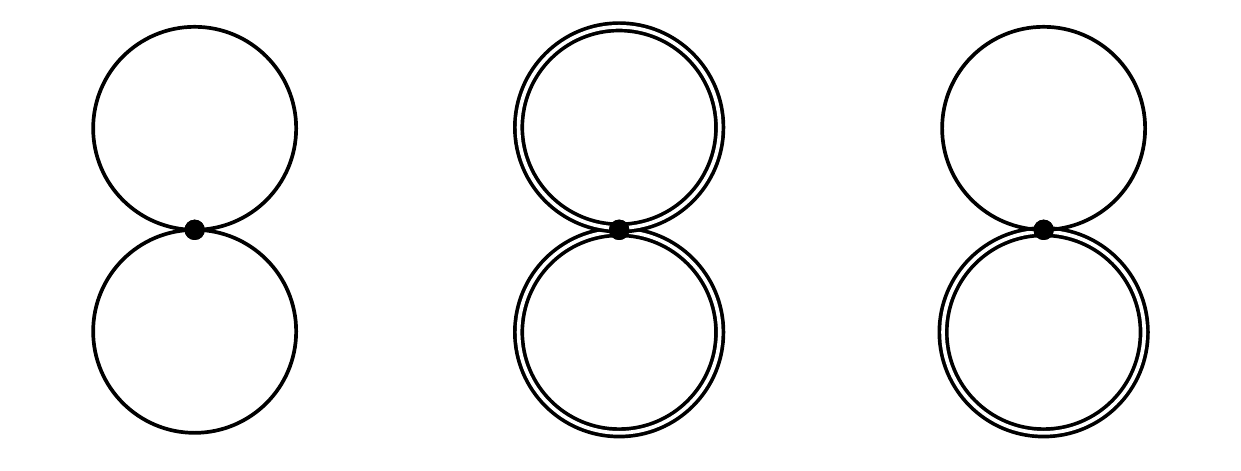}
    \caption{
    Feynman diagrams in the Hartree approximation contributing to $V_2$ in Eq.~\eqref{eq:V2UnderHartreeMainText}. 
    The solid lines represent the scalar meson propagators $S_{ab}$, and the doubled solid lines the pseudoscalar meson propagators $P_{ab}$. 
    The black blob stands for the bare vertices, which are denoted in the text as $\mathcal{F}^{abcd}$ and $\mathcal{H}^{abcd}$, respectively. 
    }
    \label{fig:HartreeApprox}
\end{figure}

In the present work, we adopt the Hartree approximation, by which the sum of the 2PI diagram terms $V_2$ in Eq.~\eqref{eq:GeneralCJTPotnetialMainText} is truncated at the two-loop level as follows: 
\begin{align}
    &V_2[S, P] \nonumber\\
    &= \mathcal{F}^{abcd} \biggl[ \int_k S_{ab}(k) \int_q S_{cd}(q) + \int_k P_{ab}(k) \int_q P_{cd}(q) \biggl] \nonumber\\
    &+ 2 \mathcal{H}^{abcd} \int_k S_{ab}(k) \int_q P_{cd}(q),
    \label{eq:V2UnderHartreeMainText}
\end{align}
where $\mathcal{F}^{abcd}$ and $\mathcal{H}^{abcd}$ are associated with the four-meson couplings $\lambda_1$ and $\lambda_2$ in Eq.~\eqref{V0}, and the explicit expressions are given in Eq.~\eqref{eq:definitionOfHandF}. 
The corresponding Feynman diagrams are illustrated in Fig.~\ref{fig:HartreeApprox}.
In the current setup, the daisy-type diagrams are resumed systematically by solving the gap equation of the meson propagators (see also Eq.~\eqref{eq:gapEquationsInCJTMainText}).

The nonperturbative VEVs (or the thermal averages) of $\sigma_a$ and the full scalar and pseudoscalar meson propagators $S_{ab}$ and $P_{ab}$ are determined by the stationary condition for $V_{\rm CJT}$ at $\sigma = \bar{\sigma}, S={\cal S}$, and $P={\cal P}$ as  
%
\begin{align}
    \frac{\partial V_{\rm CJT}[\sigma, S, P]}{\partial \sigma_a} \Biggl|_{\sigma = \bar \sigma, S = {\mathcal{S}}, P = {\mathcal{P}}} &= 0, \nonumber\\
    \frac{\delta  V_{\rm CJT}[\sigma, S, P]}{\delta S_{ab}}\Biggl|_{\sigma = \bar \sigma, S = {\mathcal{S}}, P = {\mathcal{P}}} &= 0, \nonumber\\
    \frac{\delta  V_{\rm CJT}[\sigma, S, P]}{\delta P_{ab}}\Biggl|_{\sigma = \bar \sigma, S = {\mathcal{S}}, P = {\mathcal{P}}} &= 0.
    \label{eq:quantumEOMsinCJTMainText}
\end{align}
See also Eq.~\eqref{eq:generalGapEquations}. 
In the Hartree approximation, as evident from Fig.~\ref{fig:HartreeApprox}, the loop corrections are all contact such that no external momentum dependence is renormalized from the tree-level one 
for meson-two point functions. 
Therefore, we parameterize the nontrivial solution for (inverse) meson propagators, 
${\cal S}_{ab}$ and ${\cal P}_{ab}$, 
only with momentum-independent dynamical mass terms as follows: 
\begin{align}
    \Big[ {\mathcal S}^{-1}(k) \Big]^{ab} &= k^2 \delta^{ab} + \Big[ M_S^2 \Big]^{ab}, \nonumber\\
    \Big[ {\mathcal P}^{-1}(k) \Big]^{ab} &= k^2 \delta^{ab} + \Big[ M_P^2 \Big]^{ab}.
    \label{eq:parameterizationOfFullPropagatorsMainText}
\end{align}
Then, the stationary condition in Eq.~\eqref{eq:quantumEOMsinCJTMainText} 
is cast into the form,  
\begin{align}
    0 = \frac{\partial V(\bar\sigma)}{\partial \bar\sigma_0} &- 3 \mathcal G^{0bc} \biggl[ \int_k {\mathcal S}_{cb}(k) - \int_k {\mathcal P}_{cb}(k) \biggl]\nonumber\\
    &+ 4 \mathcal F^{0bcd} \bar\sigma_b \int_k {\mathcal S}_{dc}(k) + 4 \mathcal H^{0bcd} \bar\sigma_b \int_k {\mathcal P}_{dc}(k), \nonumber\\
    0 = \frac{\partial V(\bar\sigma)}{\partial \bar\sigma_8} &- 3 \mathcal G^{8bc} \biggl[ \int_k {\mathcal S}_{cb}(k) - \int_k {\mathcal P}_{cb}(k) \biggl]\nonumber\\
    &+ 4 \mathcal F^{8bcd} \bar\sigma_b \int_k {\mathcal S}_{dc}(k) + 4 \mathcal H^{8bcd} \bar\sigma_b \int_k {\mathcal P}_{dc}(k),
    \label{eq:stationaryCOnditionsInCJTMainText}
\end{align}
and  
\begin{align}
    \Big[ M_S^2 \Big]^{ab} =& \Big[ m_S^2(\bar\sigma) \Big]^{ab}\nonumber\\
    &+ 4 \mathcal{F}^{abcd} \int_k {\mathcal S}_{cd}(k) + 4 \mathcal{H}^{abcd} \int_k {\mathcal P}_{cd}(k), \nonumber\\
    \Big[ M_P^2 \Big]^{ab} =& \Big[ m_P^2(\bar\sigma) \Big]^{ab}\nonumber\\
    &+ 4 \mathcal{F}^{abcd} \int_k {\mathcal P}_{cd}(k) + 4 \mathcal{H}^{abcd} \int_k {\mathcal S}_{cd}(k).
    \label{eq:gapEquationsInCJTMainText}
\end{align}
The detailed expression of $\mathcal G^{abc}$, which has been introduced in Eq.~\eqref{eq:stationaryCOnditionsInCJTMainText} is given in Eq.~\eqref{eq:definitionOfG}, which is associated with the $U(1)$ axial anomaly, i.e. the KMT determinant term proportional to $B$ in Eq.~\eqref{Vanom}.

\section{(Anomalous) WTIs, NG theorem, and symmetry improvement} 
\label{sec:WTISsndSI}

In this section, following~\cite{Pilaftsis:2013xna}, we review the intrinsic issue of the 2PI formalism on the violation of the NG theorem and the (anomalous) WTIs.
For readability, we will pick up only the resultant key equations 
and the detailed derivations of the WTIs are referred to 
as in Appendix.~\ref{app:DerivationsOfWTIS}.

Consider an action $S[\phi]$ having a symmetry associated with the infinitesimal transformation of a set of general fields $\phi_a$ ($a=1,\cdots$) as $\phi_a \rightarrow \phi_a + i \epsilon \delta_\epsilon \phi_a$, where $\delta_\epsilon \phi_a$ is the corresponding generator in the field space. 
Momentarily, no explicit breaking is assumed for the original symmetry. 
Through the symmetry transformation of the generating functional (with the source $J^a$), we have the following WTI 
\begin{align}
    J^a\cdot \langle \delta_\epsilon \phi_a \rangle_J = 0.
    \label{eq:WTIatGeneratingFunctionalLevelMainText}
\end{align}
If the generator belongs to the linear symmetry representation in the corresponding Lie group, such as 
$\delta_\epsilon \phi_a = d_a^{\,\, b} \phi_b$ with $d$ being an arbitrary coefficient matrix, we then get the following WTI for the 1PI effective action $\Gamma$: 
\begin{align}
     \frac{\delta \Gamma}{\delta (\phi_{\rm cl})_a} \cdot \delta_\epsilon (\phi_{\rm cl})_a = 0,
     \label{eq:WTIfor1PIEAMainText}
\end{align}
where 
$(\phi_{\rm cl})_a \equiv \langle \phi_a \rangle_J$ is the VEV including the source field $J$.

In the case of the chiral-$SU(3)$ transformation given in Eq.\eqref{eq:chiralTransformation}, we find 
\begin{align}
    \delta^\alpha \sigma_a = d^{\alpha b}_{\quad a} \pi_b, \qquad \delta^\alpha \pi_a = - d^{\alpha b}_{\quad a} \sigma_b,
    \label{eq:transLawOfFieldsMainText}
\end{align}
%
Following Eq.~\eqref{eq:WTIfor1PIEAMainText}, in the case without explicit breaking, i.e., in the three-flavor chiral limit, the WTIs for these chiral rotations are then read off as  
\begin{align}
    \frac{\delta \Gamma[\sigma_{\rm cl},\pi_{\rm cl}]}{\delta (\sigma_{\rm cl})_a} \cdot d^{\alpha b}_{\quad a} (\pi_{\rm cl})_b + \frac{\delta \Gamma[\sigma_{\rm cl},\pi_{\rm cl}]}{\delta (\pi_{\rm cl})_a} \cdot d^{\alpha b}_{\quad a} (\sigma_{\rm cl})_b = 0.
    \label{eq:WTIsInChiralLimitMainText}
\end{align}
%
Passing through a couple of algebraic computations (see Appendix~\ref{app:DerivationsOfWTIS}), Eq.~\eqref{eq:WTIsInChiralLimitMainText} can finally be reduced to the following simple relations: 
\begin{align}
    M_\pi^2 \bar\Phi_1 = 0, \qquad  M_K^2 \Big( \bar \Phi_1 + \bar \Phi_3 \Big) = 0,
    \label{eq:NGtheoremIn!PI}
\end{align}
where we have also used Eq.~\eqref{Phi-VEVs}. 
The $M_{\pi}$ and $M_{K}$ denote the full nonperturbative pion and kaon masses arising from the full propagator $P_{ab}$. 
Equation (\ref{eq:NGtheoremIn!PI}) manifests itself that in the chiral limit, the spontaneously chiral broken phase with $\bar{\Phi}_1 \neq 0$ or $\bar{\Phi}_3 \neq 0$ yields $M_\pi = 0 $ and $M_K =0$, precisely reflecting the NG theorem. 
The chiral-limit WTI only for the two-flavor chiral rotation has been discussed in~\cite{Mao:2013gva}, which has been currently extended by including the $M_K$ term in Eq.~\eqref{eq:NGtheoremIn!PI}. 

The above formulations can be further generalized to the case with explicit chiral symmetry breaking supplied from the spurion field for the current quark masses, $\mathcal M$.
The action $S$, corresponding to the LSM Lagrangian \eqref{eq:lagrangianLSM}, is $SU(3)$ chiral invariant by including the (infinitesimal) transformation of ${\cal M}$ given in Eq.~\eqref{trans-calM} together with Eq.\eqref{eq:transLawOfFieldsMainText}. 
The WTIs \eqref{eq:WTIfor1PIEAMainText} are then generalized into the so-called 
anomalous chiral WTIs~\footnote{
Another type of anomalous chiral WTIs would be available in Refs.~\cite{GomezNicola:2016ssy,GomezNicola:2017bhm}}:  
\begin{align}
    \biggl\langle \frac{\delta S}{\delta \mathcal{M}} \biggl\rangle_J \cdot \delta_\epsilon \mathcal{M} + \biggl\langle \frac{\delta S}{\delta \mathcal{M}^\dagger} \biggl\rangle_J \cdot \delta_\epsilon \mathcal{M}^\dagger + J^a \cdot \langle \delta_\epsilon \phi_a \rangle_J = 0\,. 
    \label{eq:WTIatQuantumLevelMainText}
\end{align}
After somewhat lengthy calculations (see Appendix~\ref{app:DerivationsOfWTIS}),  
we see that Eq.~\eqref{eq:WTIatQuantumLevelMainText} gives a naturally extended formulae  
in the presence of the current quark masses for $M_\pi$ and $M_K$ from the chiral limit case in Eq.~\eqref{eq:NGtheoremIn!PI}:  
\begin{align}
    M_\pi^2 \bar\Phi_1 &= c m_l \Big( 1 + 2k \bar \Phi_3 \Big) \,, \nonumber\\
    M_K^2 \Big( \bar \Phi_1 + \bar \Phi_3 \Big) &= \big( c m_l + cm_s \big) \Big( 1 + 2 k \bar \Phi_1 \Big).
    \label{eq:GMORrelations1PIMainText}
\end{align} 

To see the more physical consequences of the constraint relations in Eq.~\eqref{eq:GMORrelations1PIMainText}, we shall read the light and strange quark condensates, defined as the response of the light and strange quark masses, from the LSM Lagrangian in Eq.~\eqref{eq:lagrangianLSM} as  
\begin{align}
    \langle \bar \ell \ell \rangle &= \biggl\langle \frac{\partial \mathcal L_{\rm LSM}}{\partial m_l} \biggl\rangle = -2 c \big( \bar\Phi_1 + 2 k \bar \Phi_1 \bar\Phi_3 \big), \nonumber\\
    \langle \bar s s \rangle &= \biggl\langle \frac{\partial \mathcal L_{\rm LSM}}{\partial m_s} \biggl\rangle = -2 c \big( \bar\Phi_3 + 2 k \bar \Phi_1 ^2 \big)\,.  
\label{condensates}
\end{align}
We also read the pion and kaon decay constants $f_\pi$ and $f_K$, which 
arise as the overlap amplitudes between  
the corresponding axial vector currents and pseudoscalars, as 
\begin{eqnarray}
f_\pi&=&2\bar\Phi_1,\nonumber\\
f_K&=&\bar\Phi_1+\bar\Phi_3.
\label{decay_constants}
\end{eqnarray} 
From these, we find that two constraint relations in Eq.~\eqref{eq:GMORrelations1PIMainText} take the 
\begin{eqnarray}
f_\pi^2 m_\pi^2&=&-2m_l\langle \bar ll \rangle,\nonumber\\
f_K^2 m_K^2&=&-\frac{m_l+m_s}{2}(\langle \bar ll \rangle+\langle \bar ss \rangle),
\label{GMOR}
\end{eqnarray} 
which are precisely the Gell-Mann-Oakes-Renner (GMOR) relations. 
Thus, the low-energy theorem is intact  
even in the presence of the intrinsically flavor-violating $k$-term~\cite{Kuroda:2019jzm,Kawaguchi:2020qvg}.


The \textit{threshold property} of the (pseudo) NG bosons 
($\bar{\Phi}_{1,3} \to 0$)
is thus described by the anomalous WTIs in 1PI formalism. 
The masses of the NG modes are constrained to zero in the chiral broken phase ($\bar{\Phi}_1\neq0$, $\bar{\Phi}_3\neq 0$), or proportional to the explicit breaking current masses. 
Since the relations \eqref{eq:NGtheoremIn!PI} or \eqref{eq:GMORrelations1PIMainText} hold before performing the loop expansion in the 1PI effective action, we expect that the threshold property holds even when we truncate the 1PI effective action loopwise.
Hereafter, we shall call both the anomalous chiral WTIs and the chiral-limit WTIs simply the WTIs, otherwise specialized.

\begin{figure}[t!]
    \centering
    \includegraphics[width=0.4\linewidth]{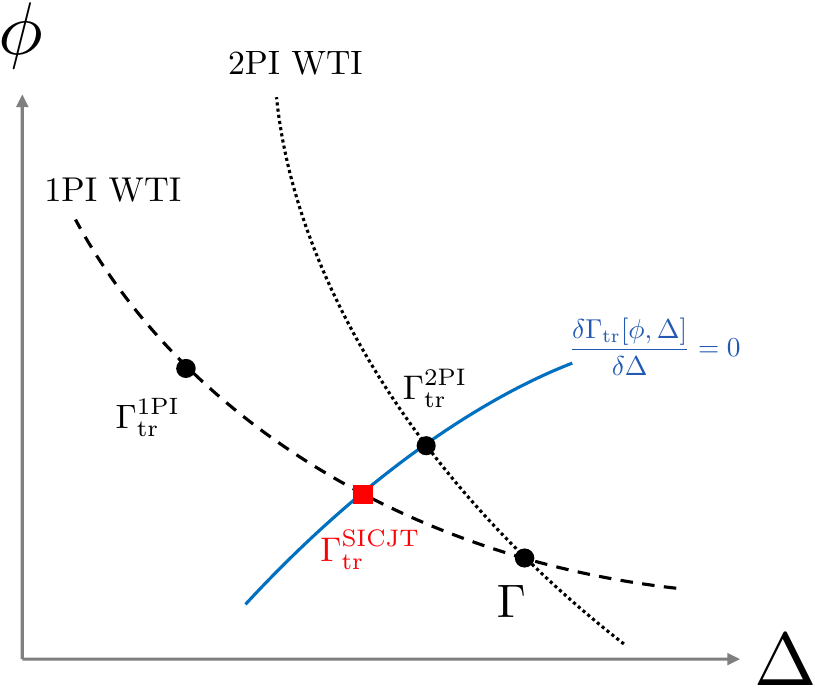}
    \caption{
    A schematic sketch of the SICJT formalism.
    The figure is drawn in the $(\phi,\Delta)$-plane where $\phi$ is the sourced field VEV and $\Delta$ denotes the sourced propagator, describing the variables of the effective actions.
    The black blobs (from right to left) represent the solution obtained from the full-quantum action $\Gamma$, the truncated 1PI effective action $\Gamma_{\rm tr}^{\rm 1PI}$, and the truncated 2PI effective action $\Gamma_{\rm tr}^{\rm 2PI}$.
    The black-dashed and -dotted curve respectively denote the constraint trajectories from the 1PI or the 2PI formalism, and the blue-solid line denotes the gap equation of the propagators $\Delta$ with arbitrary $\phi$.
    The red square mark points to the solution realized by the SICJT formalism.
    }
    \label{fig:ScenarioOfSICJT}
\end{figure}

Examining the WTIs derived in the 2PI formalism, however, we realize that the situation would be drastically changed. 
In the chiral limit, the WTIs for the 2PI effective action read
\begin{align}
    0 &= \frac{\delta \Gamma[\phi_{\rm cl}, \Delta]}{\delta (\phi_{\rm cl})_a} \cdot d_a^{\,\, b} ( \phi_{\rm cl} )_b \nonumber\\
    &\qquad + \frac{\delta \Gamma[\phi_{\rm cl}, \Delta]}{\delta \Delta_{ab}} \cdot \Big( d_b^{\,\, c} \Delta_{ac} + d_a^{\,\, c} \Delta_{cb} \Big),
    \label{eq:WTIat2PIMainText}
\end{align}
which leads to the second derivatives;  
\begin{align}
    0 =& \frac{\delta^2 \Gamma[\phi_{\rm cl}, \Delta]}{\delta (\phi_{\rm cl})_c \delta (\phi_{\rm cl})_a} \cdot d_a^{\,\, b} ( \phi_{\rm cl} )_b + \frac{\delta \Gamma[\phi_{\rm cl}, \Delta]}{\delta (\phi_{\rm cl})_a} \cdot d_a^{\,\, c} \nonumber\\
    &+ \frac{\delta^2 \Gamma[\phi_{\rm cl}, \Delta]}{\delta (\phi_{\rm cl})_c \delta \Delta_{ab}} \cdot \Big( d_b^{\,\, c} \Delta_{ac} + d_a^{\,\, c} \Delta_{cb} \Big).
    \label{eq:2PIWTIderivativeMainText}
\end{align}
Due to the third term appearing in the second line of Eq.~\eqref{eq:2PIWTIderivativeMainText}, this identity does not provide a direct constraint to prevent the NG boson from being massive.
If the 1PI and 2PI effective actions are evaluated exactly without any truncation, the solutions of 
$\phi_{\rm cl}^a$ and the propagators $\Delta_{ab}$ should be the same since they come from the same sourceless generating functional.
In this case, the WTIs in the 1PI formalism (e.g., Eq.~\eqref{eq:WTIatQuantumLevelMainText} without $\mathcal M$-term) and the one in the 2PI formalism \eqref{eq:WTIat2PIMainText} should be satisfied simultaneously at the solution to the stationary condition.
However, truncating both effective actions at the loop level ($\Gamma_{\rm tr}^{\rm 1PI}$ and $\Gamma_{\rm tr}^{\rm 2PI}$), 
we find that they will follow different WTIs.
Then, since the 2PI WTI does not directly constrain the NG boson mass, the threshold property or the NG theorem would be violated.
The schematic sketch of the above scenario is shown in Fig.~\ref{fig:ScenarioOfSICJT}.
This issue would be crucial for the phase transition of the system, since the massless nature is directly connected to the scale-invariance of the correlators.

To resolve this issue, we employ the SICJT formalism \cite{Pilaftsis:2013xna}, for which the strategy is to replace the stationary condition in Eq.\eqref{eq:stationaryCOnditionsInCJTMainText} by the generalized GMOR relation in Eq.\eqref{eq:GMORrelations1PIMainText} derived from the WTIs of the 1PI formalism, while keeping the second constraint with the gap equations for meson masses in Eq.\eqref{eq:gapEquationsInCJTMainText}.
The solution found in this procedure is denoted as $\Gamma_{\rm tr}^{\rm SICJT}$ in Fig.~\ref{fig:ScenarioOfSICJT}.
This completes the current formulation, and we shall apply this method to study the Columbia plot.

\section{Numerical analysis}
\label{sec:Results}

In this section, we show our numerical results on the $T$-dependences for the 
chiral order parameters, masses of scalar and pseudoscalar meson nonets, 
by solving the gap equations of the meson propagators in Eq.~\eqref{eq:gapEquationsInCJTMainText} coupled with the generalized GMOR relations in Eq.~\eqref{eq:GMORrelations1PIMainText} based on the SICJT formalism. 
The analysis is performed both at the physical point and in the chiral limit. 
We then discuss the Columbia plot and make a comparison with the results based on the conventional CJT formalism, 
which follows by solving the gap equations coupled to the stationary conditions in Eq.~\eqref{eq:stationaryCOnditionsInCJTMainText}, 
which significantly breaks the NG theorem and threshold property fixed as in the generalized GMOR relations in Eq.~\eqref{eq:GMORrelations1PIMainText}.  
The criticalities at the tricritical point and the second-order phase boundary, as well as the critical pion mass in the three-flavor symmetric limit, are also addressed in depth.

\subsection{Numerical implements}

Before presenting our results, we briefly summarize the numerical strategy applied in the current work.
To solve the coupled equations, we shall adapt the iteration method and decompose the problem into three layers to gain better numerical stability. 
Given a set of test values for the VEVs $\bar \Phi_{1,3}$ and full meson masses $M_{S,P}$, we fix the values of the mixing angles $\theta_{S,P}$ through Eq.~\eqref{eq:MixingElementOfGapEq}, by using the Newtonian iteration method with $\theta^0_{S,P}$ in Eqs.~\eqref{thetaS0} and \eqref{thetaP0} used as initial conditions. 
Then we solve the gap equations~\eqref{eq:gapEquationsInCJTMainText} using the fixed-point iteration method with the tree-level masses $m_{S,P}$, which can be found in Appendix~\ref{App-A-1}, used as the initial conditions. 
Finally, we solve either the stationary condition in Eq.~\eqref{eq:stationaryCOnditionsInCJTMainText} or the GMOR relations in Eq.~\eqref{eq:GMORrelations1PIMainText} by using the Newtonian method with tests of different initial guesses, to search for different phases at a given temperature.

\subsection{Parameter setup at physical point}
\label{sec:ParameterSetups}

We take two different sets of model parameters, characterized by a different $f_0(500)$ mass as an input, to examine the sensitivity of the size of the $f_0(500)$ mass to 
the order of phase transition or the phase structure of the Columbia plots, 
as was discussed earlier~\cite{Lenaghan:2000kr,Lenaghan:2000ey} based on the conventional CJT formalism.

\subsubsection{Parameter set I} 
\label{Para-I}

The first set of parameters is chosen following the literature~\cite{Kuroda:2019jzm, Kawaguchi:2020qvg} as 
\begin{gather}
    \mu^2 = 1.02 \times 10^4~{\rm MeV^2}, \quad \lambda_1 = 11.8, \quad \lambda_2 = 20.4, \nonumber\\
    c m_l^{\rm phys.} = 6.11\times10^5~{\rm MeV^3},\nonumber\\
    c m_s^{\rm phys.} = 198\times10^5~{\rm MeV^3}, \nonumber\\
    B = 3.85\times 10^3~{\rm MeV}, \quad k = 3.40\times 10^{-3}~{\rm MeV^{-1}},
\end{gather}
which provides the pion and kaon decay constants as $f_\pi = 92.1~{\rm MeV}$ and $f_K = 109.8~{\rm MeV}$ in Eq.~\eqref{decay_constants}, and the meson mass spectra
\begin{gather}
    m[f_0(500)] = 672.4~{\rm MeV}, \quad m[f_0(980)] = 990.4~{\rm MeV}, \nonumber\\
    m_{a_0} = 937.6~{\rm MeV}, \quad m_\kappa = 863.4~{\rm MeV}, \nonumber\\
    m_{\eta^\prime} = 958.2~{\rm MeV}, \quad m_\eta = 552.9~{\rm MeV}, \nonumber\\
    m_\pi = 137.9~{\rm MeV}, \quad m_K = 494.1~{\rm MeV}.
\end{gather}
The mixing angles at vacuum are
\begin{align}
    \theta_S^0 = 44.3^{\circ}, \qquad \theta_P^0 = 6.80^{\circ}.
\end{align}

\subsubsection{Parameter set II}
\label{para-II}

We re-fit the parameters following the $\chi^2$-fit procedure as performed in Ref.~\cite{Kuroda:2019jzm}, with 
replacing the $f_0(500)$ mass by  
$m[f_0(500)] = 800~{\rm MeV}$. 
Then we get the following best-fit values for the model parameters: 
%
\begin{gather}
    \mu^2 = -1.79 \times 10^5~{\rm MeV^2}, \quad \lambda_1 = 30.9, \quad \lambda_2 = 22.2, \nonumber\\
    c m_l^{\rm phys.} = 6.11\times10^5~{\rm MeV^3}, \nonumber\\
    c m_s^{\rm phys.} = 198\times10^5~{\rm MeV^3}, \nonumber\\
    B = 3.86 \times 10^3~{\rm MeV}, \quad k = 1.51 \times 10^{-3}~{\rm MeV^{-1}},
\end{gather}
which provides the pion and kaon decay constants as $f_\pi = 92.1~{\rm MeV}$ and $f_K = 109.8~{\rm MeV}$, and the meson mass spectra
\begin{gather}
    m[f_0(500)] = 797.1~{\rm MeV}, \quad m[f_0(980)] = 1151~{\rm MeV}, \nonumber\\
    m_{a_0} = 943.1~{\rm MeV}, \quad m_\kappa = 966.5~{\rm MeV}, \nonumber\\
    m_{\eta^\prime} = 897.0~{\rm MeV}, \quad m_\eta = 513.9~{\rm MeV}, \nonumber\\
    m_\pi = 125.8~{\rm MeV}, \quad m_K = 460.1~{\rm MeV}.
\end{gather}
The mixing angles in the vacuum are
\begin{align}
    \theta_S^0 = 45.0^{\circ}, \qquad \theta_P^0 = 0.291^{\circ}.
\end{align}
Notice that the physical observables obviously deviate from the inputs, which indicates that the current parameter space may not fully cover the input observables.

\subsection{Chiral phase transition at the physical point}

\begin{figure}[t]
    \centering
    \includegraphics[width=0.32\linewidth]{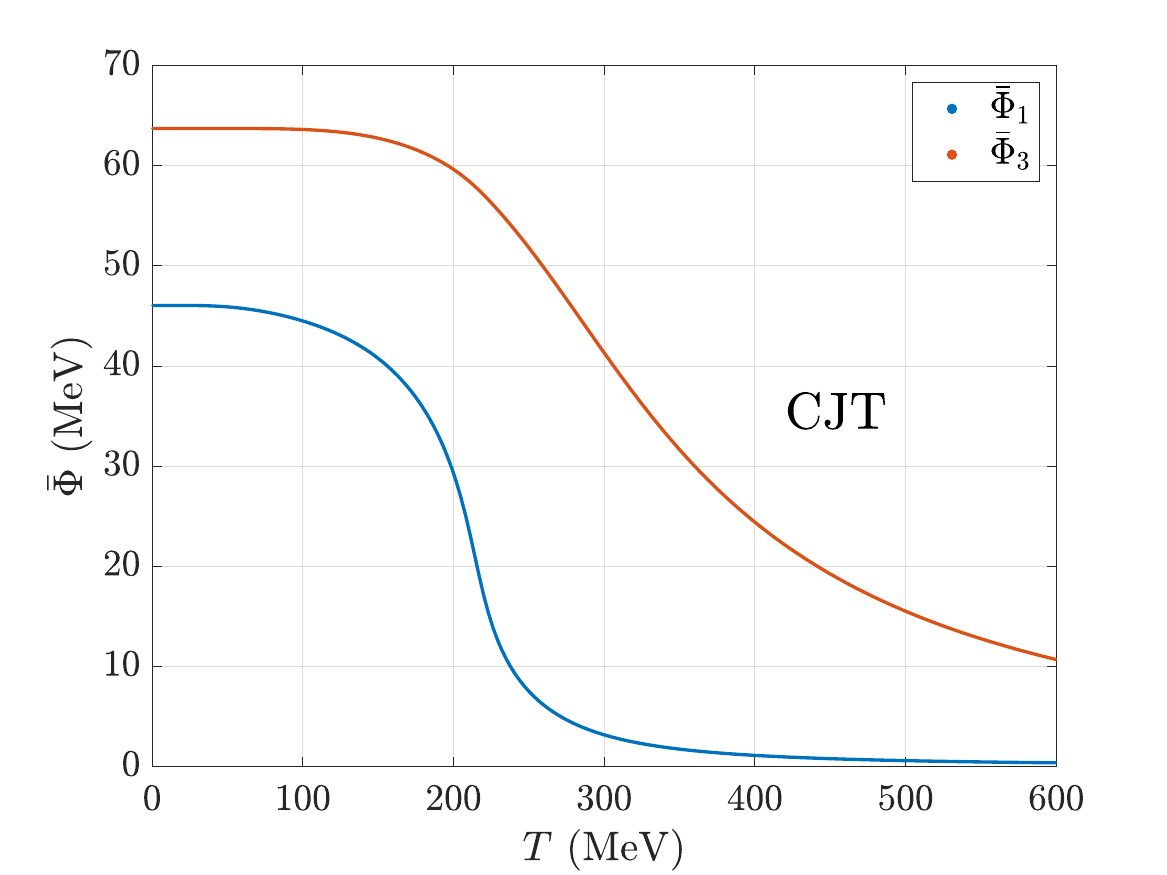}
    \includegraphics[width=0.32\linewidth]{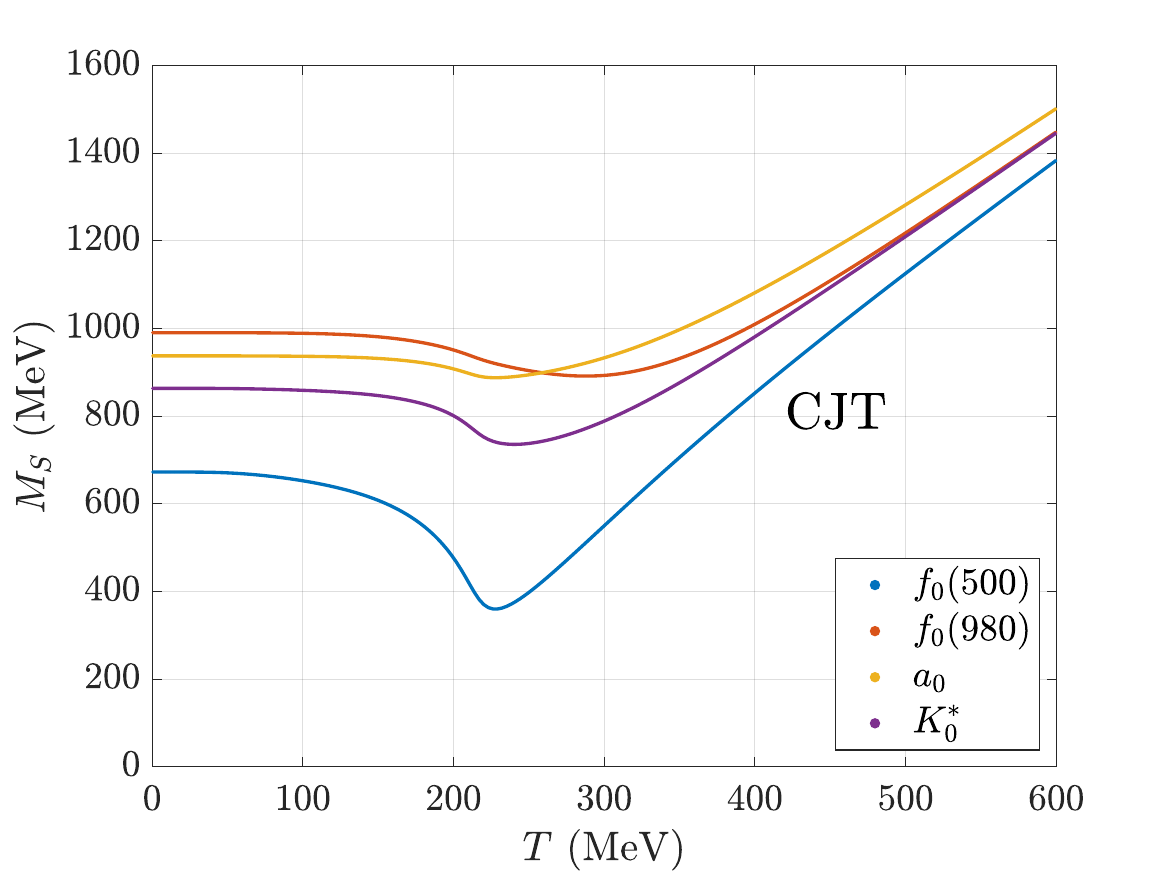}
    \includegraphics[width=0.32\linewidth]{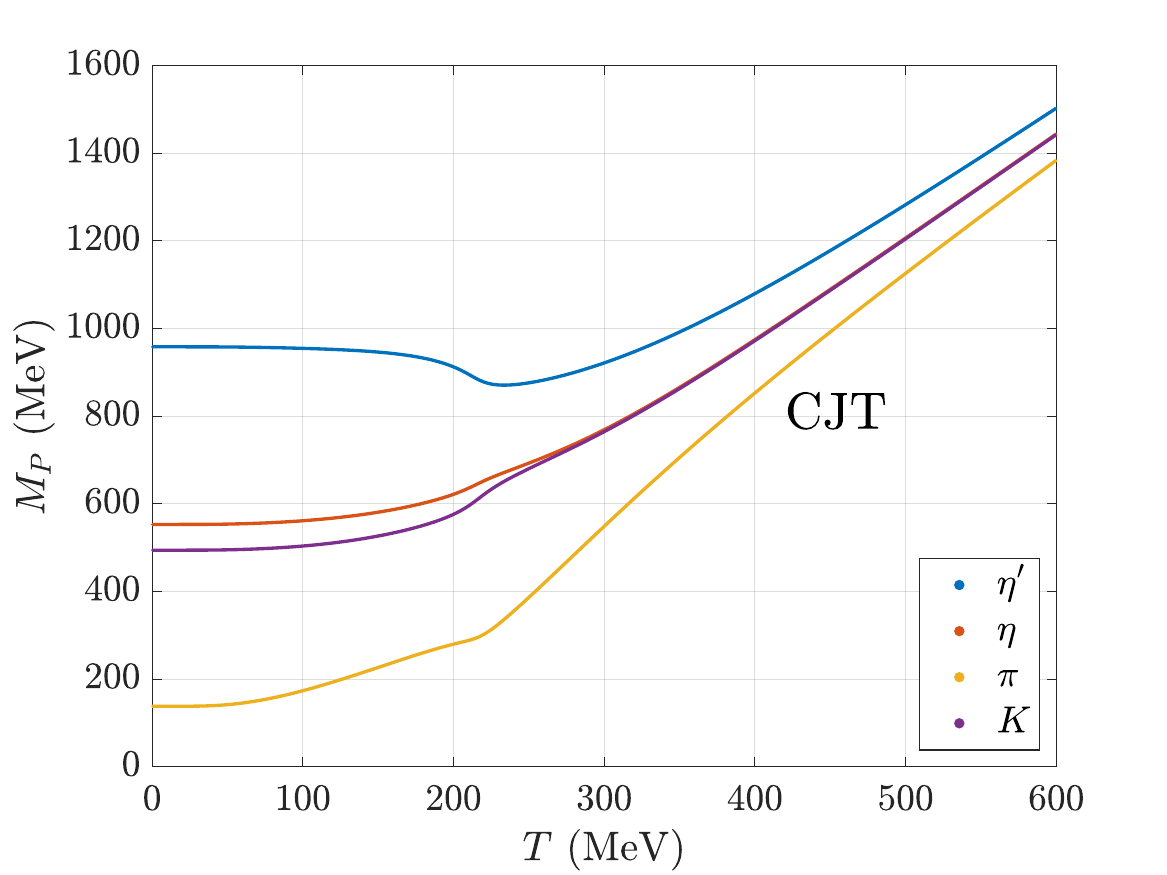}
    \\
    \includegraphics[width=0.32\linewidth]{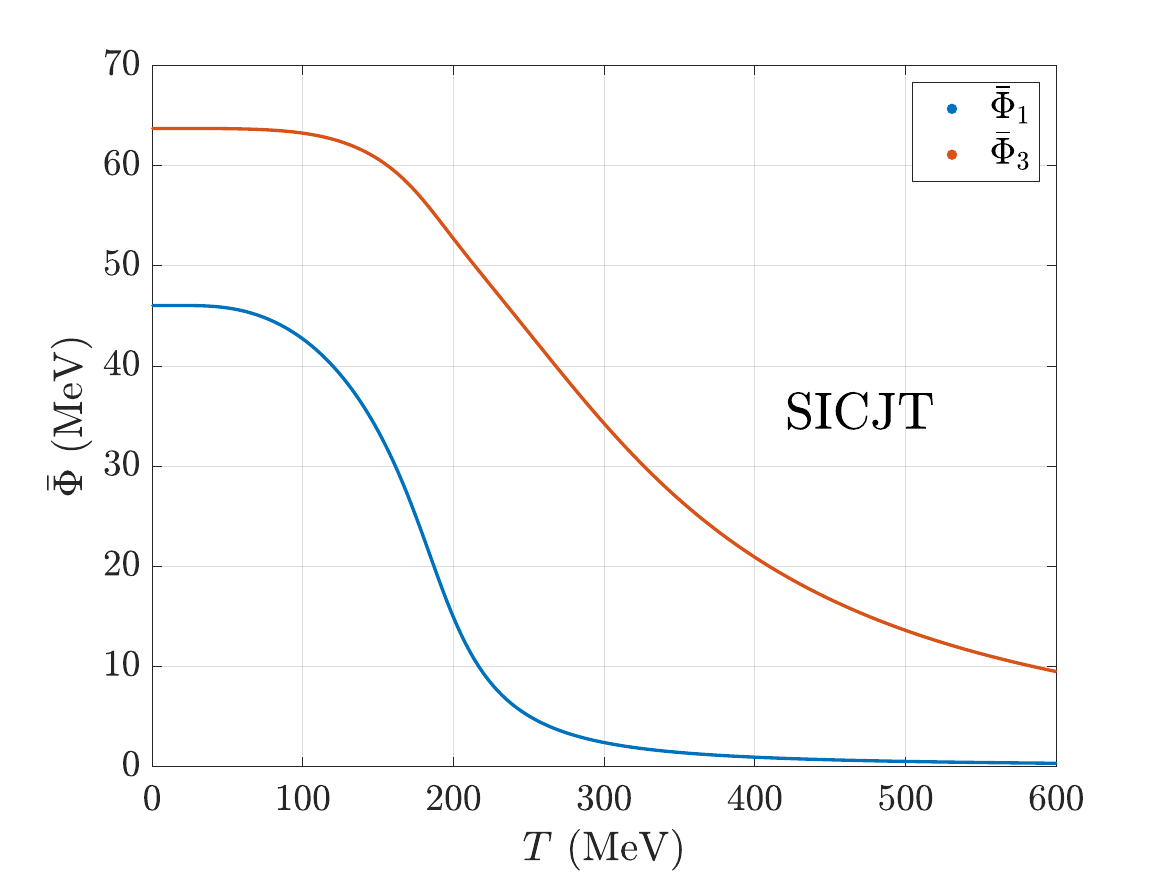}
    \includegraphics[width=0.32\linewidth]{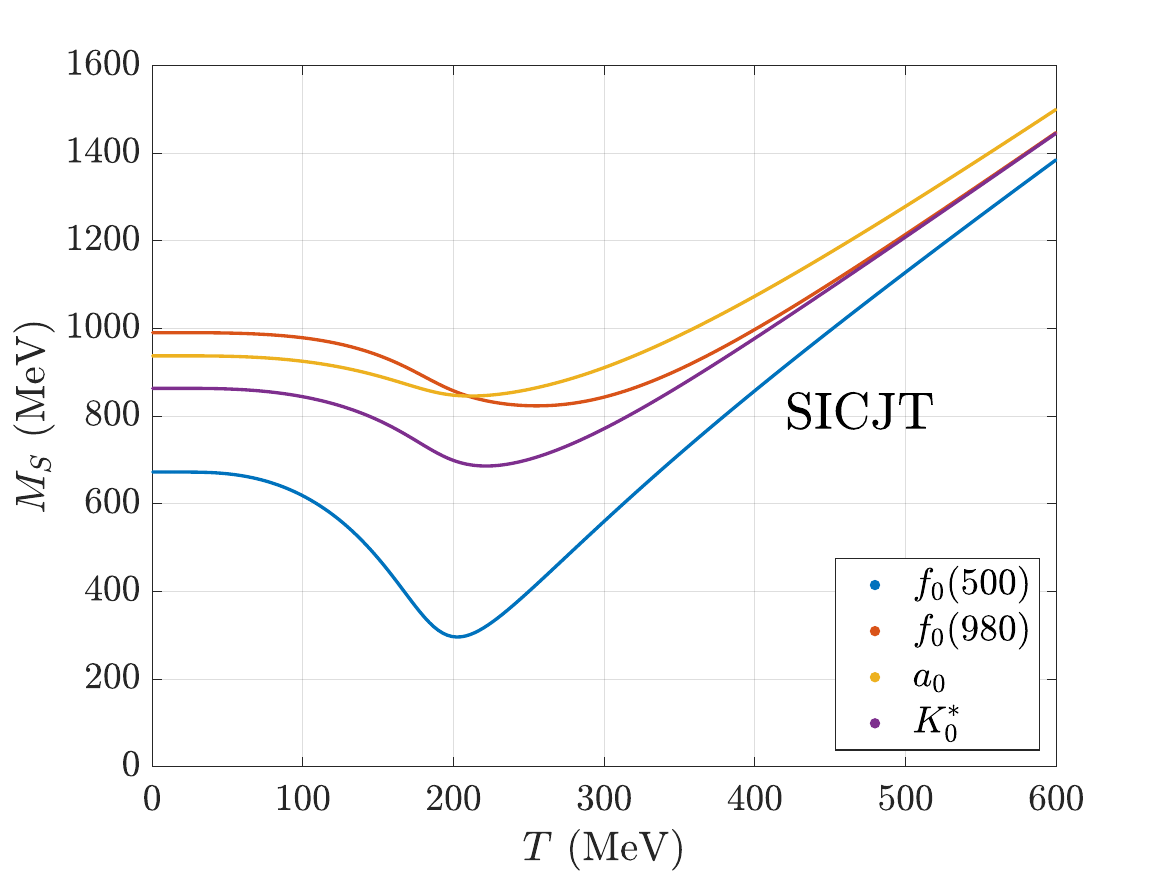}
    \includegraphics[width=0.32\linewidth]{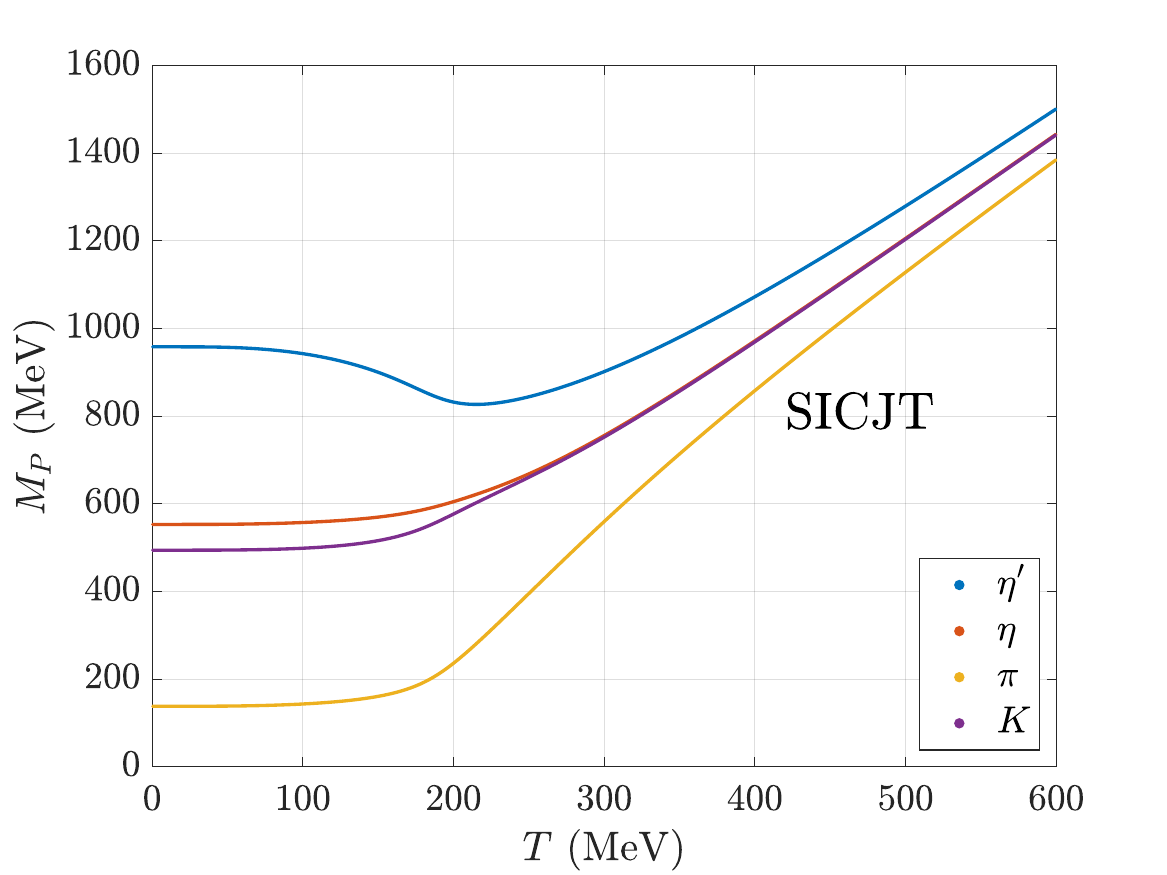}
    \caption{
    Plots of 
    $T$-dependences on $\bar{\Phi}_{1,3}$ (left panel) and masses of scalar (middle) and pseudoscalar mesons (right), 
    at the physical point, in the cases of the conventional CJT (upper panels) and SICJT (lower panels) formalisms applied to the present LSM with the parameter set I in Sec.~\ref{Para-I}. 
    }
    \label{fig:PhysicalPointCJTvsSICJT}
\end{figure}


In Fig.~\ref{fig:PhysicalPointCJTvsSICJT}
we show the $T$-dependences on $\bar{\Phi}_{1,3}$, and the scalar and pseudoscalar meson masses $M_S$ and $M_P$ at the physical point $(m_l, m_s) = (m_l^{\rm phys.}, m_s^{\rm phys.})$ with the parameter set I provided in Sec.~\ref{Para-I}. In the figure, the upper panel refers to the conventional CJT method, 
while the lower panel corresponds to the result from the SICJT method. 
Comparing both methods, we find the following qualitative characteristics:
\begin{itemize}
    \item The pseudocritical temperature of the light quark condensate, $T_{\rm pc}$, which is defined by the temperature with the maximum thermal susceptibility $\chi_T = - \partial \langle \bar\ell \ell \rangle_T / \partial T$, is around $\sim 185 \ {\rm MeV}$ in the SICJT formalism, which is reduced compared to the one estimated from the conventional CJT formalism, $T_{\rm pc} \sim 215 \ {\rm MeV}$. 
    Those values of $T_{\rm pc}$ are anyhow higher than the ones obtained from other nonperturbative approaches, such as fRG (e.g., Ref.~\cite{Fu:2019hdw}), Dyson-Schwinger equations (e.g., Ref.~\cite{Gao:2020qsj}) or lattice QCD (e.g., Ref.~\cite{Aoki:2006we}).
    This higher $T_{\rm pc}$ can be understood by the lack of dynamical quark degrees of freedom. 
    The extended studies including quarks would be performed in the future; 

    \item The pion mass $M_\pi$ shows relatively smoother $T$-dependence for $T \lesssim T_{\rm pc}$ in the SICJT formalism, compared to the conventional CJT case. before the chiral restoration happens.
    This is due to the realization of the proper threshold property and the manifest NG theorem: the introduction of loop corrections in the conventional CJT method breaks the low-energy theorem, i.e., the GMOR relations, such that (pseudo) NG boson mass spectra can develop artificially and unphysically with finite $T$. 
\end{itemize}
Both cases show the degeneracy of the meson mass spectra at $T > T_{\rm pc}$, which precisely reflects the effective restoration of the chiral symmetry due to the reduction of the size of the light quark condensate. 
The parameter set II in Sec.~\ref{para-II} yields essentially the same results. 

\subsection{Chiral phase transition in chiral limit}

\begin{figure}[t]
    \centering
    \includegraphics[width=0.35\linewidth]{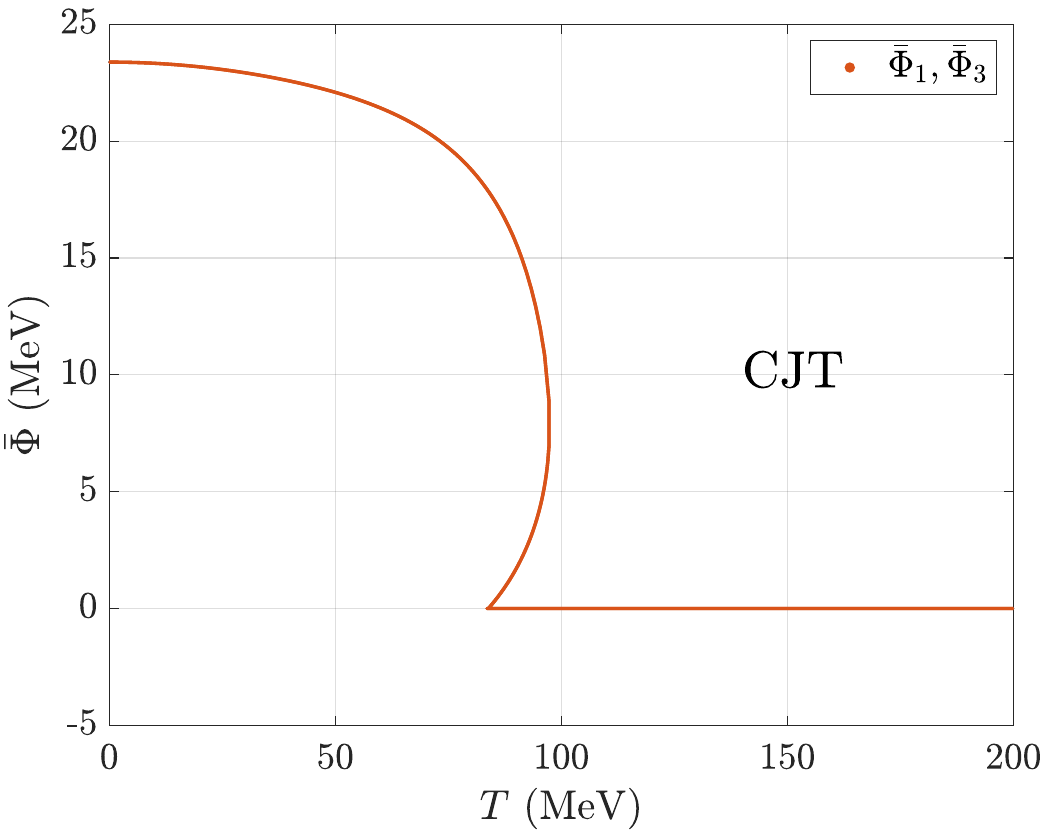}
    \qquad
    \includegraphics[width=0.35\linewidth]{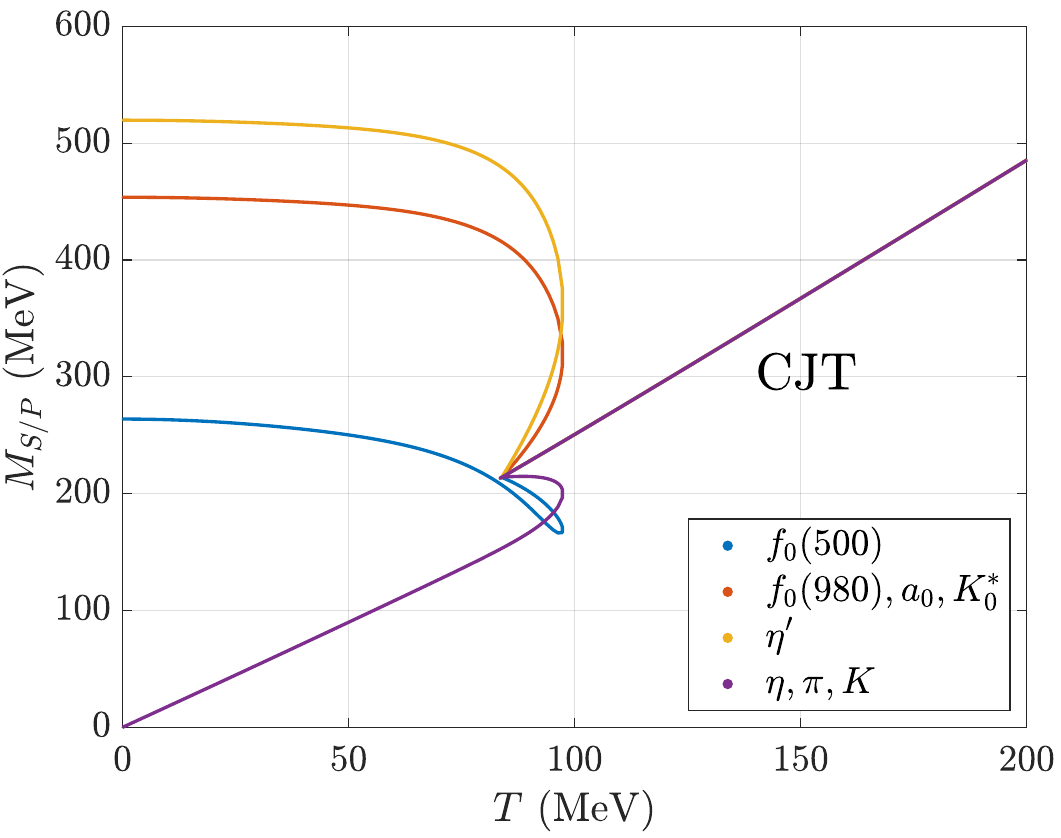}
    \\
    \includegraphics[width=0.35\linewidth]{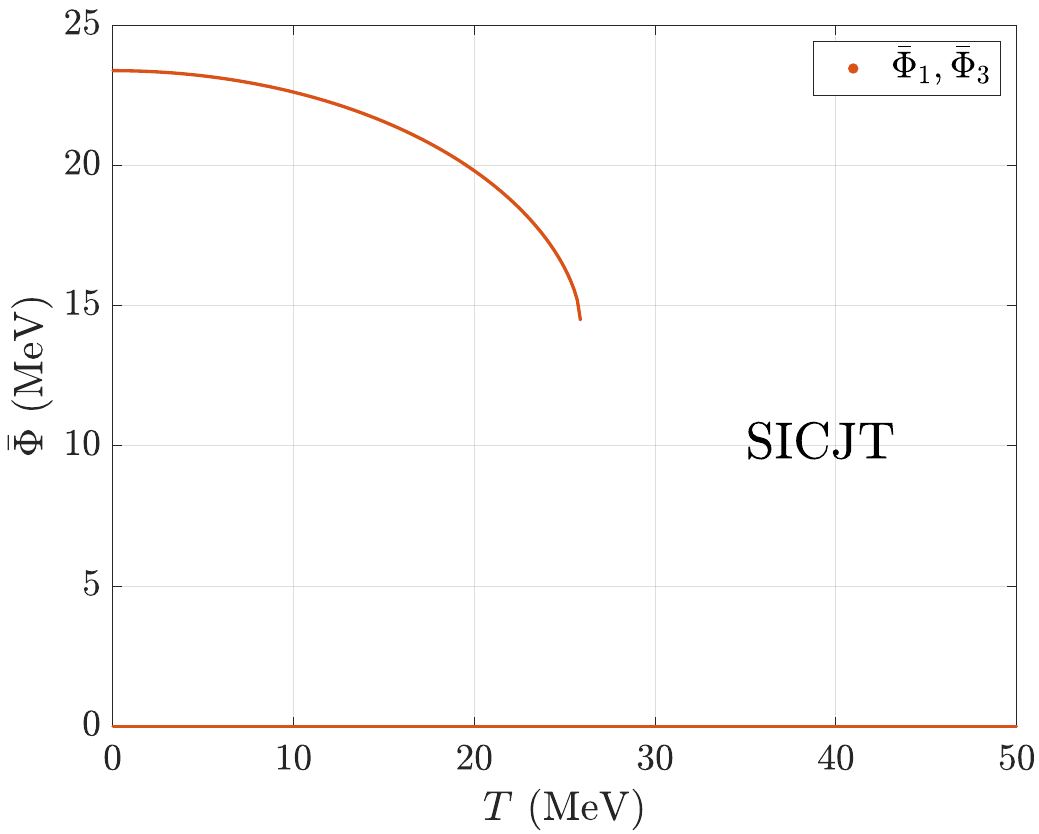}
    \qquad
    \includegraphics[width=0.35\linewidth]{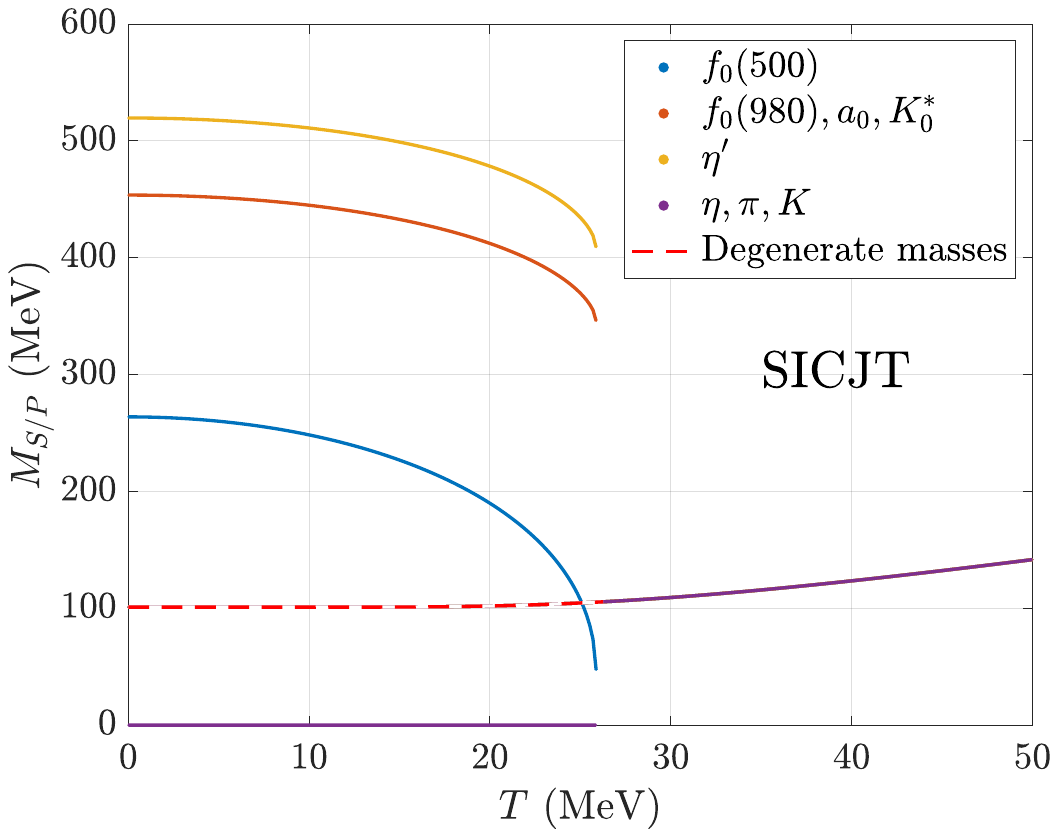}
    \caption{
    The same as Fig.~\ref{fig:PhysicalPointCJTvsSICJT}, for the chiral limit case, with the plots of the scalar and pseudoscalar meson masses combined into a single one, which has been summarized in the right panel. 
    The red-dashed line in the bottom-right panel denotes the degenerate masses of all mesons in the chirally symmetric phase below the critical temperature.}
    \label{fig:ChiralLimitCJTvsSICJT}
\end{figure}


We also explore the chiral phase transitions in the chiral limit at $(m_l, m_s) = (0,0)$ for both the conventional CJT and SICJT formalisms.  
See Fig.~\ref{fig:ChiralLimitCJTvsSICJT}, where the plots have been created by taking the parameter set I in Sec.~\ref{Para-I}, but with $m_l = m_s = 0$.

In the conventional CJT formalism, we have examined all stationary points, including the local minimum as well as the local maximum. For the latter point, those maximum states connect the vanishing point of the chiral broken phase and the emerging point of the chiral symmetric phase.
In the SICJT case, the tachyonic modes with negative curvatue masses (corresponding to the local maximum states) show up, which can be reflected as the smoother dropping blue curve in the bottom-right panel of Fig.~\ref{fig:ChiralLimitCJTvsSICJT}: around the phase transition temperature $T_c \sim 25 \ {\rm MeV}$, the $f_0(500)$ mass shows the trend to be vanishing. 
This brings further numerical complexity in evaluating the meson loops, but makes milder effects on the scope of the current work; thus, we drop the local maximum state in evaluating the chiral order parameter in the SICJT case.
For the mass spectra in both cases, all meson masses are merged into one, denoted by the purple line after the chiral restoration happens, and linked with the unified mass spectra in the false vacuum, which is marked by the red-dashed line in the right-bottom panel of Fig.~\ref{fig:ChiralLimitCJTvsSICJT}.

We find the following characteristics for the phase transition in the chiral limit. 
%
\begin{itemize}
    \item From the left panels of both CJT and SICJT cases, we find that the chiral order parameter discontinuously changes with temperature around some critical values, and the local maximum state is clearly seen in the conventional CJT case. 
    Those behaviours of the chiral order parameters indicate the orders of the phase transitions in both cases to be of the first-order. 
    The phase transition takes place at the critical temperature $T_c \sim 25 \ {\rm MeV}$ in the case of the SICJT formalism (the bottom-left panel of Fig.~\ref{fig:ChiralLimitCJTvsSICJT}). 
    This $T_c$ is much smaller than the one measured based on the conventional CJT formalism, which yields $T_c \sim 97 \ {\rm MeV}$ (see the top-left panel). 
    This reduction can be understood by the proper reflection of the NG theorem in the chiral limit even at finite $T$: thermal loop corrections arising from lighter mesons make the critical temperature lower. 
    The presence of the first-order phase transition in both two formalisms is essentially due to the non-restoration of the $U(1)$ axial anomaly, which is currently encoded in the KMT determinant (the $B$ term in Eq.~\eqref{Vanom}) providing a negatively large cubic potential along the chiral-order parameter direction $\bar{\Phi}_1$ (See also the parameter set I and II in Secs.~\ref{Para-I} and \ref{para-II}).

    \item The masses of the pion $M_\pi$ and the kaon $M_K$ are fixed to vanish in the SICJT formalism below the exact chiral restoration (see the bottom-right panel of Fig.~\ref{fig:ChiralLimitCJTvsSICJT}), while they start to develop in the CJT case at finite $T$, even before the chiral restoration happens (see the top-right panel). 
    The former feature is closely related to the realization of the proper threshold property and the NG theorem, which is violated in the CJT formalism. More remarkably, at $T=T_c$ in the chiral limit for the SICJT formalism, the $f_0(500)$ mass goes to zero, reflecting the chiral partner of the massless pion. 
\end{itemize}

Recently, based on highly improved staggered quarks on lattices with temporal extent 
$N_\tau = 8$~\cite{Dini:2021hug} -- which is an extension of the study with $N_\tau=6$~\cite{Bazavov:2017xul} -- the three-flavor lattice QCD study has reported that no evidence of the first-order signal is observed for $m_\pi \gtrsim  80$ MeV. 
The critical temperature in the three-flavor chiral limit has also been estimated via the finite-size scaling analysis, to give $T_c= 98^{+3}_{-6}$ MeV, based on the data and statistical averaging over different fit ansatz with $m_\pi$ in a range of $80\,{\rm MeV} \lesssim m_\pi \lesssim 140$ MeV.
The present model with the SICJT formalism thus tends to give a lower $T_c$ than the lattice result, which, however, may only be of limited relevance because of different model setups and systematics.

\subsection{Columbia plots}


\begin{figure}[t]
    \centering
    \includegraphics[width=0.4\linewidth]{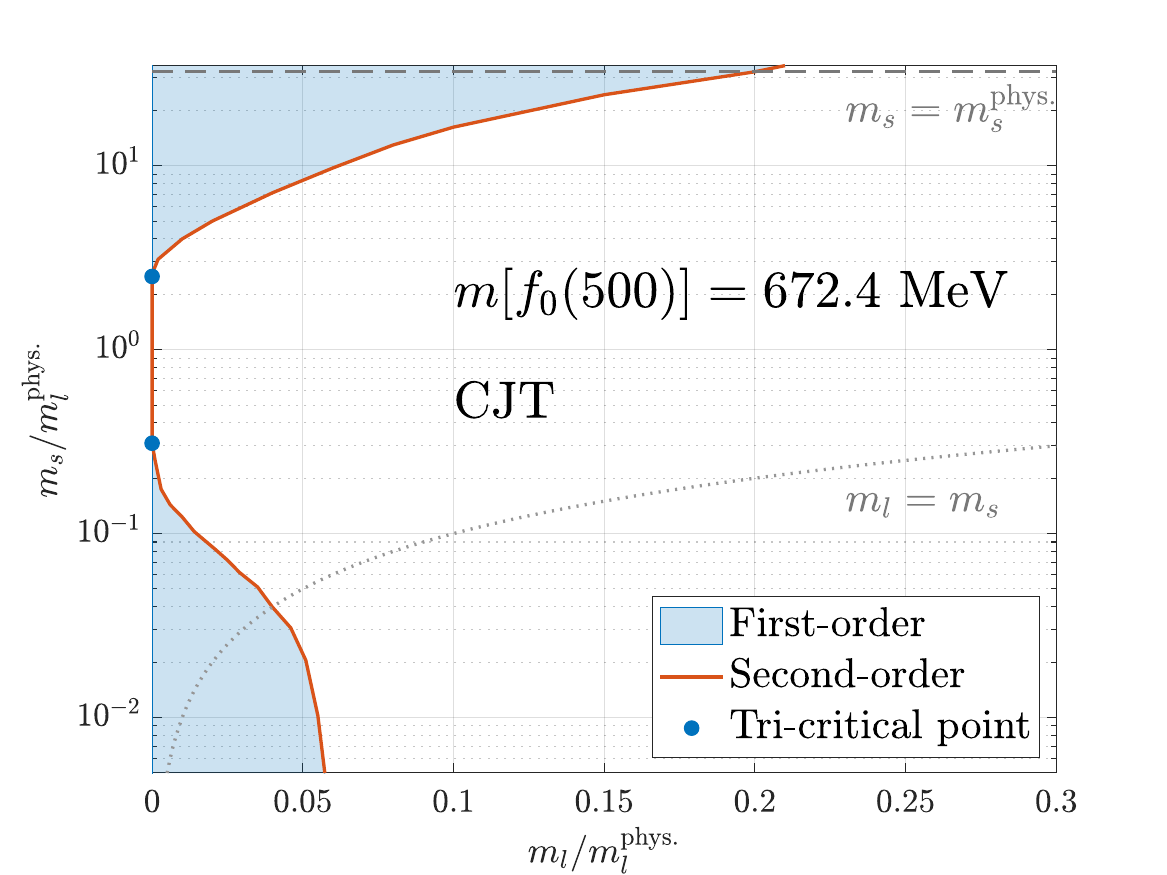}
    \includegraphics[width=0.4\linewidth]{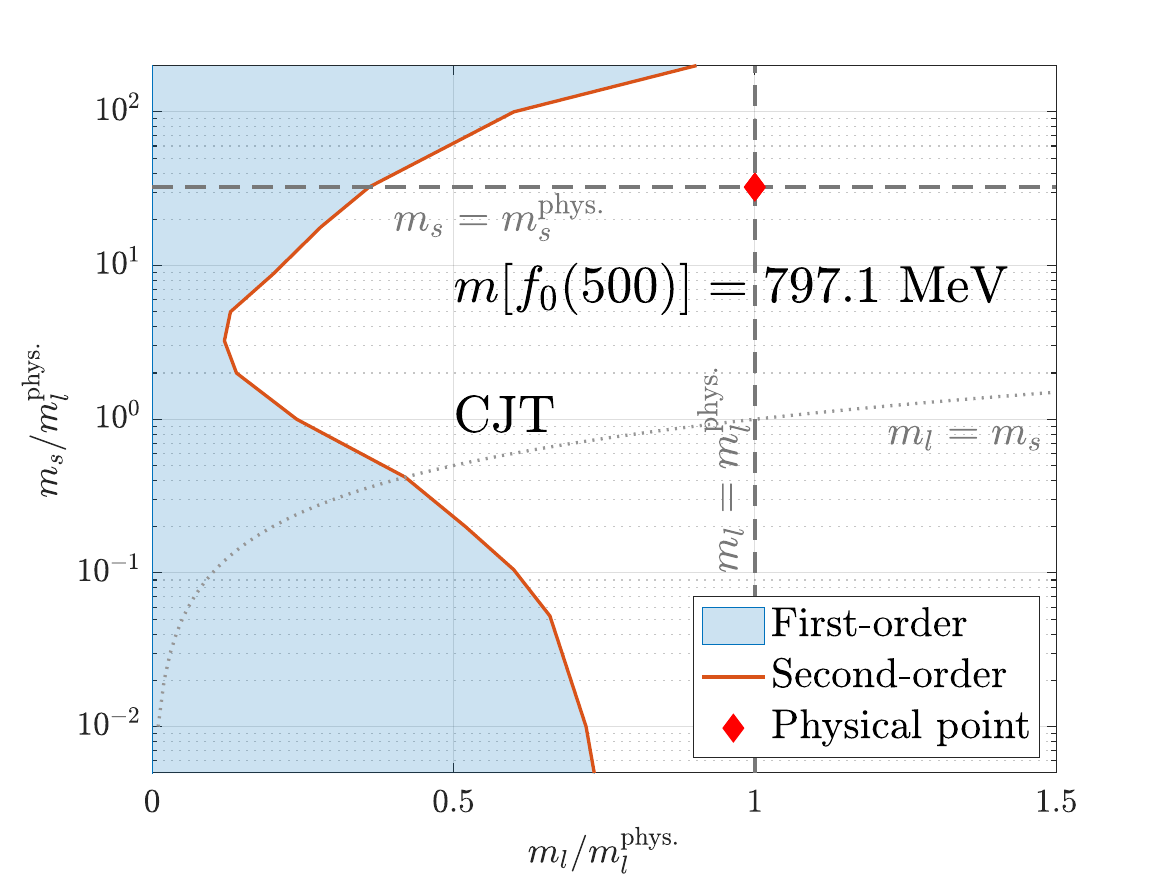}
    \\
    \includegraphics[width=0.4\linewidth]{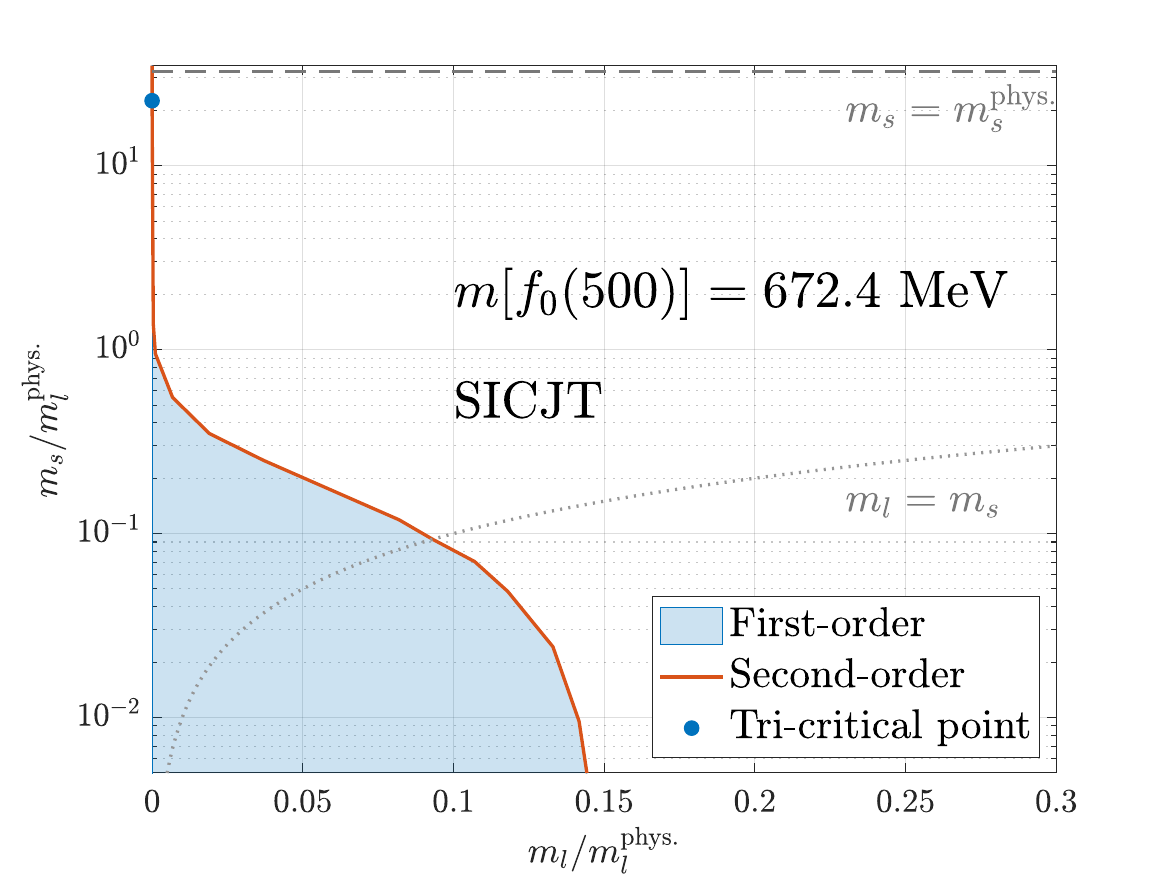}
    \includegraphics[width=0.4\linewidth]{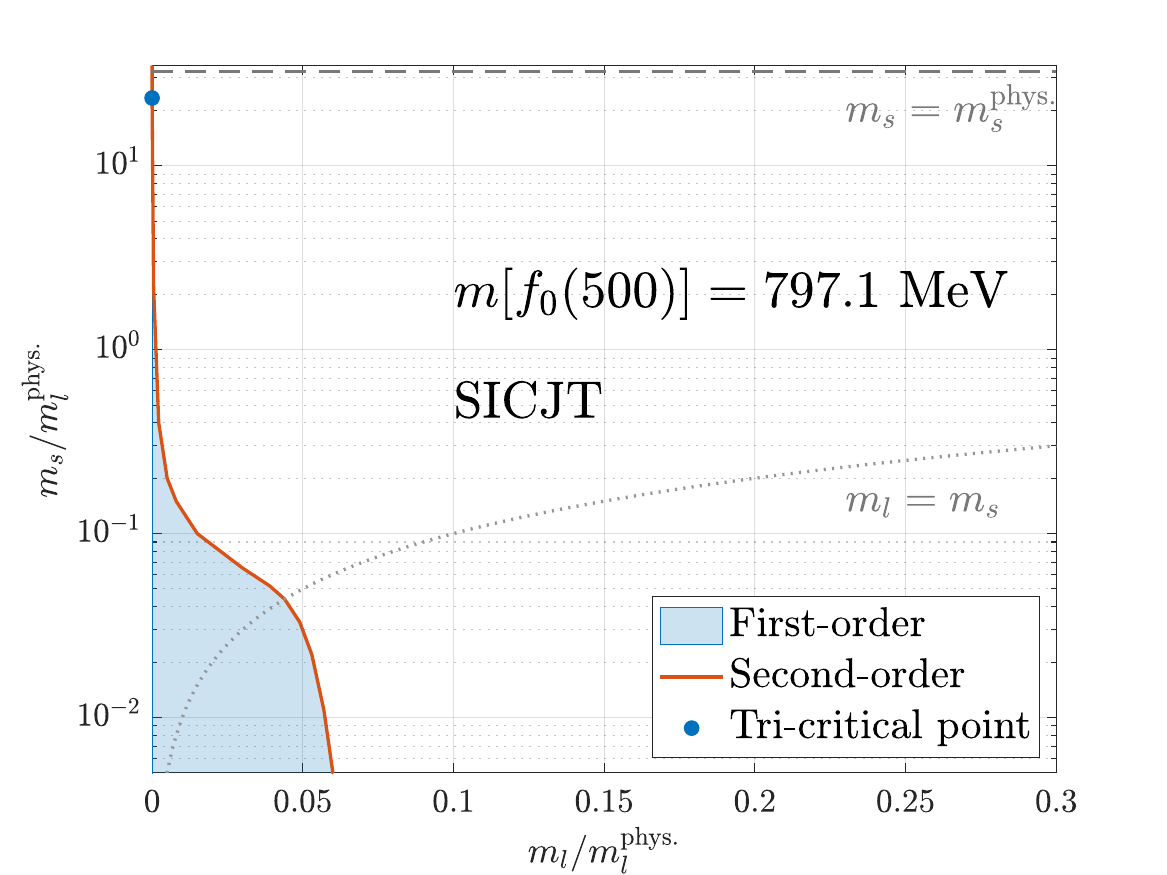}
    \caption{
    The conventional CJT result projected onto the Columbia plots with the parameter sets I in Sec.~\ref{Para-I} (top-left panels) and II in Sec.~\ref{para-II} (top-right panel), in comparison with the SICJT results (bottom-left and -right panels) on the same parameter setup. 
    The orange-solid curves denote the second-order phase boundaries, the blue-shaded areas stand for the first-order domains, and blank areas represent the crossover regions.
    The three-flavor symmetric limit (the case with $m_l=m_s$) has been drawn by the dotted curves, and the red-diamond marker corresponds to the physical point where $(m_c, m_s) = (m_l^{\rm phys.},m_s^{\rm phys.})$. 
    The $m_l$ and $m_s$ axes have been normalized by $m_l^{\rm phys.}$. 
    }
    \label{fig:ColumbiaPlotsSigma}
\end{figure}

In Fig.~\ref{fig:ColumbiaPlotsSigma}, we make the Columbia plots based on the SICJT formalism (bottom panels) in comparison with the conventional CJT case (top panels). 
The parameter sets I and II in Secs.~\ref{Para-I} and~\ref{para-II} have been reflected, respectively, in the left and right panels~\footnote{The present analysis has been limited by the large size of $m_s$ 
up to $2 m_s^{\rm phys.}$, above which a part of the $k$-term proportional to $m_s$ (Eq.~\eqref{k-term}) will cause an instability in the system, because of the artificial non-decoupling 
effect of the strange quark in the $k$-term. This issue needs to be fixed in a separate publication.}.

In the literature~\cite{Lenaghan:2000kr,Lenaghan:2000ey}, 
the Columbia plot has already been discussed based on a three-flavor LSM with the conventional CJT formalism, which is similar to the present model except for the $k$-term (Eq.~\eqref{k-term}), where the input observables to fix the model parameters are somewhat, but not substantially different from those used in Secs.~\ref{Para-I} and \ref{para-II}. 
In the top panels of Fig.~\ref{fig:ColumbiaPlotsSigma}, we have reproduced the resulting phase structure fully consistent with the one observed in the references, including the overall characteristics: 
the first-order regime is present, and two tricritical points emerge at around $m_s = 
{\cal O}(0.1 m_s^{\rm phys.})$ and $\gtrsim {\cal O}(m_s^{\rm phys.})$ for lighter $f_0(500)$, 
while those two points get merged to be gone as $m_{f_0(500)}$ is heavier and heavier, still keeping the presence of the first-order regime.

We find that the SICJT formalism removes the artificial first-order regime and gives the {\it symmetry-improved} Columbia plot (bottom panels in Fig.~\ref{fig:ColumbiaPlotsSigma}): 
no significant $m_{f_0(500)}$ dependence is seen, and the first-order regime as well as the single tricritical point are stably present. 
This also implies no existence of the extra first-order regime which has been observed in the conventional CJT formalism. 
Those features have been measured due to the proper realization of the low-energy theorem at finite $T$, which protects the pion to be light enough and safely suppresses the other heavier meson contributions in the thermal bath, including those from $f_{0}(500)$, up to $T \lesssim T_{({\rm p})c}$.  
It has been claimed in Refs.~\cite{Lenaghan:2000kr,Lenaghan:2000ey} that the existence of the extra first-order regime in the conventional CJT formalism is tied with the Hartree-level approximation. 
Our present result supports another possibility that working on the Hartree-level approximation is not the main reason for generating the extra first-order regime; rather, the violation of the NG theorem may be the reason.

We also still see from Fig.~\ref{fig:ColumbiaPlotsSigma} that 
in the SICJT case, the first-order regime gets shrunken when $m_{f_0(500)}$ gets larger.
This is simply due to the generation of more stringent convex negative curvature 
(by the negative $\mu^2$ term in Eq.~\eqref{V0} with the sign choices given in Secs.\ref{Para-I} and \ref{para-II}). 
On more physics grounds, the generation of heavier $f_0(500)$ tends to require dynamical chiral symmetry breaking essentially triggered by the negative $\mu^2$ term, while in the lighter $f_0(500)$ case, the $U(1)$ axial anomaly essentially triggers the dynamical chiral symmetry breaking even with a positive $\mu^2$ term.

When taking a particular look at the three-flavor symmetric limit where  
$m_l = m_s$, we find the critical quark mass on the second-order boundary (red curves in 
Fig.~\ref{fig:ColumbiaPlotsSigma}) to be $m_l^c = m_s^c = 0.0931 \, m_l^{\rm phys.}$, 
(e.g., for the parameter set I case, in Sec.~\ref{Para-I}, with $m[f_0(500)] = 672.4 \ {\rm MeV}$ (bottom-left panel)). 
This quark mass value is translated into the critical pion mass at the vacuum as $m_\pi^{c} = 52.4 \ {\rm MeV}$. 
The chiral restoration is observed at $T_c \sim 51.6 \ {\rm MeV}$.



\subsection{Critical exponents}

In this section, we present our result on the criticality of the light quark condensate in Eq.~\eqref{condensates} evaluated from the SICJT formalism. 
The scaling behavior of the light quark condensate $\langle \bar \ell \ell \rangle$ with respect to the temperature $T$ and the current quark mass $m_l$ in the vicinity of the criticality is characterized by the critical exponents $\beta$ and $\delta$. 
Suppose the critical point of the second-order phase transition is located at $(T,m_l) = (T^c, m_l^c)$ with the critical current mass for the strange quark, $m_s^c$, we define the following dimensionless variables as
\begin{align}
    t \equiv \frac{T - T^c}{T^c}, \qquad \tilde m_l \equiv \frac{m_l - m_l^c}{m_l^{\rm phys.}},
\end{align}
with the following scaling function
\begin{align}
    \langle \bar \ell \ell \rangle(t,\tilde m_l) - \langle \bar \ell \ell \rangle(0,0) = f_l(t,\tilde m_l),
\end{align}
where $f_l(t,\tilde m_l)$ is a dimensionful function and satisfies $f_l(0,0) = 0$.
The critical exponents of the thermal and the external source response are then given by
\begin{align}
    f_l(t \to 0,0) \propto (-t)^\beta, \qquad f_l(0,\tilde m_l \to 0) \propto \tilde m_l^{1/\delta}.
    \label{eq:scalingAnsatz}
\end{align}
We summarize the method and related details of the evaluation of the critical exponents in the Appendix~\ref{app:EvaluationOfTheCriticalExponents}.

Based on the symmetry argument, we expect the universality classes on the second-order line in the Columbia plot as follows.
For the critical points located at the boundary dividing the first-order and the crossover areas (except for the tricritical point), the chiral symmetry is explicitly broken with the presence of three massive quarks.
In this case, the chiral symmetry is reduced to remain as the discretized one, i.e., the $Z_2$ symmetry.
We shall expect the $Z_2$ universality class on this boundary, described by the three-dimensional Ising model~\cite{Campostrini:2002cf}. 
As to the critical points lying on the massless light flavor axis (the $y$-axis in the Columbia plot above the tricritical point), the criticalities are expected to be described by the $O(4)$ universality classes~\cite{Rajagopal:1992qz}, since the massless two-flavor theory possesses the $SU(2)_L \times SU(2)_R \sim O(4)$ symmetry.
Furthermore, if one assumes that the $U(1)_A$ symmetry is restored around the critical temperature of the chiral phase transition, the $U(2) \times U(2)$ universality class would also be possible to be found~\cite{Pelissetto:2013hqa,Butti:2003nu}.
Regarding the tricritical point, which connects the line of the first-order and second-order regions along the $y$-axis in the Columbia plot, the Ginzburg-Landau description of the free energy around this point is read as the following tricritical $\sigma^6$ theory:  
\begin{align}
    \mathcal F(\sigma) = a_2(T,m_s) \sigma^2 + a_4(T,m_s) \sigma^4 + a_6 \sigma^6 + \cdots,
\end{align}
where $\sigma \sim \langle \bar \ell \ell \rangle$ denotes the light quark condensate respectively.
The tricritical point is located where $a_4(T,m_s^{\rm tri})= 0$, leading to the interplay between the first- and second-order phase transitions.
The upper critical dimension of this tricritical $\sigma^6$ theory is $d_c = 3$, which is the same as the spatial dimension concerned in the present work, indicating that fluctuations become irrelevant and merely referring to mean-field exponents is validated. 
Thus, we shall expect the mean-field universality class at the tricritical point in the Columbia plot.

In Table~\ref{tab:CriticalExponents}, we list the critical exponents evaluated at several critical points, as well as the aforementioned universality classes.
For the critical exponents at the point where $m_s^c=0$, the results fit well with the $Z_2$ universality class.
For the flavor-symmetric point ($m_l^{c} = m_s^{c} = 0.0931 \, m_l^{\rm phys.}$) and the tricritical point ($m_l^{\rm tri} = 0$, $m_s^{\rm tri} = 0.696 \, m_s^{\rm phys.}$), the values of $\beta$ are relatively close to the ones expected from the $Z_2$ and $O(4)$ universality classes, while the $\delta$ starts to deviate from those benchmark values.
For the values obtained at the critical point on the $y$-axis above the tricritical point ($m_l^c = 0$, $m_s^c = m_s^{\rm phys.}$), our result does not meet any universality class mentioned above.


Given the result of the criticality in the present work, we find that the current framework of the analysis based on the SICJT formulation may not quantitatively capture the critical exponents. 
We speculate that the observed deviation from the expected universality class may originate from three sources. 
First, the truncation scheme adopted in this work corresponds to the Hartree–Fock level, which incorporates only momentum-independent corrections to the meson propagators. 
Because mesonic fluctuations play a central role near criticality, it would be valuable to explore higher-order effects, such as wave-function (field-renormalization) corrections to the mesons. 
Second, the inclusion of the axial-anomaly–induced flavor-breaking term in Eq.~(\ref{k-term}) introduces an additional source of the chiral symmetry breaking. 
In this setup, the current quark masses no longer couple linearly to the vacuum expectation value of the mesonic field, thereby modifying the system’s response to external sources. 
Finally, the evaluation of the gap equations for meson masses in the concave region is also crucial near critical points. 
The treatment of this region influences the convergence of the solutions and can alter the resulting universality class.
Thus, the present setup and analysis can be used for qualitative studies, such as the analysis of the transition temperature or the meson-mass spectra properties; however, it needs further improvement in several aspects to be capable of accurately evaluating the QCD criticality.

\begin{table}[t]
\begin{tabular}{l|cc|cc}
\hline \hline
Model & \hspace{10pt} $\beta$ \hspace{10pt}  & \hspace{10pt}  $\delta$  \hspace{10pt}  & \hspace{10pt}  $m_\pi \, ({\rm MeV})$ \hspace{10pt}  & \hspace{10pt}  $T_c \, ({\rm MeV})$ \\
\hline
SICJT: ($m_l^c = 0.1474 \, m_l^{\rm phys.}$, $m_s^c = 0$) & 0.3235 & 4.7907  & 65.4 & 51.6 \\
SICJT: ($m_l^c = m_s^c = 0.0931 \, m_l^{\rm phys.}$) & 0.3222 & 3.2128 & 52.1 & 51.6 \\
SICJT: ($m_l^{\rm tri} = 0$, $m_s^{\rm tri} = 0.696 \, m_s^{\rm phys.}$) & 0.2842 & 7.4597 & 0.0 & 113.5 \\
SICJT: ($m_l^c = 0$, $m_s^c = m_s^{\rm phys.}$) & 0.6532 & 3.7805 & 0.0 & 125.4 \\
\hline 
$O(4)$ universality class~\cite{Rajagopal:1992qz} & 0.38 & 4.82 & - & - \\
$Z_2$ universality class~\cite{Campostrini:2002cf}  & 0.326 & 4.78 & - & - \\
$U(2) \times U(2)$ universality class~\cite{Pelissetto:2013hqa,Butti:2003nu}  & 0.388 & 4.789 & - & - \\
3D $XY$ universality class~\cite{Hasenbusch:1999cc} & 0.3490 & 4.7798 & - & - \\
Mean-field universality class & 0.25 & 5 & - & - \\
\hline \hline 
\end{tabular}
\caption{
The list of the critical exponents 
at several critical points estimated from the present LSM based on 
the SICJT formalism with the parameter set I in Sec.~\ref{Para-I}. 
The resulting numbers are compared to those from typical statistical models, referred to as the critical points when specified: $O(4)$, $Z_2$, and 3D XY-universality classes. 
Also have been shown the corresponding pion mass $m_\pi$ in the vacuum and the critical temperatures $T_c$ observed in the present SICJT formalism (See also the bottom-left panel in Fig.~\ref{fig:ColumbiaPlotsSigma}). 
}
\label{tab:CriticalExponents}
\end{table}

\section{Conclusion}
\label{sec:conclusion}

In this work, we have investigated the structure of the Columbia plot within a three-flavor LSM monitoring the underlying QCD, which exhibits the global chiral $U(3)_L \times U(3)_R$ symmetry, explicitly broken by current quark masses and by the $U(1)$ axial anomaly. 
We have included both leading-order anomaly terms and anomaly-induced flavor-breaking interactions, motivated by the phenomenological success of the model shown in the literature~\cite{Kuroda:2019jzm}. 
Our analysis was carried out within the CJT framework for the 2PI effective action, considering both the conventional and the symmetry-improved formulations, the latter being dubbed the SICJT~\cite{Pilaftsis:2013xna}.

A primary focus of the present work is to address the known issues in truncated 2PI approaches, particularly the violation of the NG theorem and the threshold property in the chiral broken phase. 
These violations become significant near the criticality of the chiral phase transition. 
By enforcing the WTIs for the 1PI effective action through the SICJT formalism, we have refined the conventional CJT approach on the three-flavor LSM 
and analyzed the consequences for the chiral phase structure and the projection onto the Columbia plot.

Our analysis has clarified that an enhanced first-order region, which has been detected in the conventional CJT method with the standard Hartree-level approximation, is merely an artifact due to the lack of the manifest NG and low-energy theorems at finite temperature. 
This result suggests that a consistent incorporation of mesonic fluctuations that preserves the chiral symmetry is crucial for properly capturing the chiral phase structure, even within relatively simple truncation schemes.

Still, our current approach should potentially be applicable as a tool for studying other qualitative behaviours of critical phenomena, such as the search for the critical endpoint in QCD-like theories and their LEFTs.
Beyond mean-field and large-$N$ approximations, the SICJT approach would sufficiently incorporate mesonic fluctuation contributions through the self-consistent resummation of meson two-point functions by respecting the necessary symmetry constraints. 
Thus, our present work would be a definite step forward in improving the reliability of exploring meson mass spectra near the chiral critical temperature.

The symmetry-improved 2PI formalism would also offer a useful framework for exploring the QCD phase diagram in terms of LEFTs, particularly for applications at some critical point where the proper treatment of soft modes and collective excitations is essential. 
Other possible and simple-minded outlooks may also be involved: extending the current framework by including vector and axialvector mesons; incorporation of the coupling to Polyakov loops; finite chemical potential; the convolution with other nonperturbative methods such as the fRG; 
the studies of the electroweak phase transition and/or dark chiral phase transitions including physics beyond the Standard Model. In that case, in a way consistent with the (anomalous) WTIs, the SICJT formalism would help perform a proper resummation of 2PI diagram series, improving the radius of convergence like the conventional daisy resummation, and the validity of the conventional perturbative formulations.

\section*{Acknowledgments} 
We thank Zheng-Ze Li for fruitful discussions and contributions 
in the early stage of the present work. 
This work was supported in part by the National Science Foundation of China (NSFC) under Grant No.11747308, 11975108, 12047569, 
and the Seeds Funding of Jilin University (S.M.). 
The work by M.K. is supported by RFIS-NSFC under Grant No. W2433019.
The work of A.T. was partially supported by JSPS  KAKENHI Grant Numbers 20K14479, 22K03539, 22H05112, and 22H05111, and MEXT as ``Program for Promoting Researches on the Supercomputer Fugaku'' (Simulation for basic science: approaching the new quantum era; Grant Number JPMXP1020230411, and 
Search for physics beyond the standard model using large-scale lattice QCD simulation and development of AI technology toward next-generation lattice QCD; Grant Number JPMXP1020230409).

\onecolumngrid
\appendix
\section{The three-flavor LSM at tree-level}
\label{app:VacuumPhenomenologyOfLinearSigmaModel}

In this appendix, we summarize the tree-level formulae in the three-flavor LSM employed in the present paper. 

\subsection{The vacuum constraint by the stationary condition}
\label{App-A-1}

At tree-level, the potential \eqref{eq:LSMTreePotential} along the vacuum direction specialied only by the  $(\bar\Phi_1,\bar\Phi_3)$-background fields reads 
\begin{align}
    V(\bar\Phi) =& \mu^2 \Big( 2 \bar \Phi_1^2 + \bar\Phi_3^2 \Big) + \lambda_1 \Big( 2 \bar \Phi_1^4 + \bar\Phi_3^4 \Big) + \lambda_2 \Big( 2 \bar \Phi_1^2 + \bar\Phi_3^2 \Big)^2 \nonumber\\
    &- 2B \bar \Phi_1^2 \bar \Phi_3 - 2\Big( 2 cm_l \bar\Phi_1 + cm_s \bar\Phi_3 \Big) - 4k \Big( cm_s \Phi_1^2 + 2 cm_l \bar\Phi_1 \bar\Phi_3 \Big). 
\end{align}
%
The slopes of the classical action $S_{\rm LSM} = \int d^4 V(\bar{\Phi})$ 
are then given as 
\begin{align}
    \frac{\delta S_{\rm LSM}}{\delta \Phi_1} \Biggl|_{\Phi = \bar\Phi} &= 4 \Big( \mu^2 \bar\Phi_1 + 2 \lambda_1 \bar\Phi_1^3 + 4 \lambda_2 \bar\Phi_1^3 + 2 \lambda_2 \bar\Phi_1 \bar\Phi_3^2 - B \bar\Phi_1 \bar\Phi_3 - cm_l - 2 k cm_l \bar\Phi_3 - 2kcm_s \bar\Phi_1 \Big), \nonumber\\
    \frac{\delta S_{\rm LSM}}{\delta \Phi_3} \Biggl|_{\Phi = \bar\Phi} &= 2 \Big( \mu^2 \bar\Phi_3 + 2 \lambda_1 \bar\Phi_3^3 + 4 \lambda_2 \bar\Phi_1^2 \bar\Phi_3 - B \bar\Phi_1^2 - cm_s - 4 k cm_l \bar\Phi_1 \Big).
    \label{eq:SlopPhibasis}
\end{align}

Alternatively, using Eq.~\eqref{Phi-VEVs}, 
the tree-level potential $V(\bar{\Phi})$
in terms of 
the $(\bar\sigma_0,\bar\sigma_8)$-background fields 
reads  
\begin{align}
    V(\bar\sigma) =& \frac{1}{2} \mu^2 \Big( \bar\sigma_0^2 + \bar\sigma_8^2 \Big) + \lambda_1 \biggl[ \frac{1}{12} \bar\sigma_0^4 + \frac{1}{2} \bar\sigma_0^2 \bar\sigma_8^2 - \frac{1}{3\sqrt{2}} \bar\sigma_0 \bar\sigma_8^3 + \frac{1}{8} \bar\sigma_8^4 \biggl] + \frac{\lambda_2}{4} \Big( \bar\sigma_0^2 + \bar\sigma_8^2 \Big)^2 \nonumber\\
    &- B \biggl[ \frac{1}{3\sqrt{6}}\bar\sigma_0^3 - \frac{1}{2\sqrt{6}}\bar\sigma_0 \bar\sigma_8^2 - \frac{1}{6\sqrt{3}}\bar\sigma_8^3 \biggl] - \sqrt{\frac{2}{3}} (2cm_l + cm_s)\bar\sigma_0 - \frac{2}{\sqrt{3}} (cm_l - cm_s)\bar\sigma_8 \nonumber\\
    &- k \biggl[ \frac{cm_l}{3} \Big( 4 \bar\sigma_0^2 - 2\sqrt{2} \bar\sigma_0\bar\sigma_8 - 4\bar\sigma_8^2 \Big) + \frac{cm_s}{3} \Big( 2\bar\sigma_0^2 + 2\sqrt{2} \bar\sigma_0 \bar\sigma_8 + \bar\sigma_8^2 \Big) \biggl]. 
    \label{eq:treePotentialAroundMeanField}
\end{align}
The corresponding slopes of the classical action are then given as 
\begin{align}
    \frac{\delta S_{\rm LSM}}{\delta \sigma_0} \Biggl|_{\sigma_a = \bar\sigma_a} =& \mu^2 \bar\sigma_0 + \lambda_1 \biggl[ \frac{1}{3} \bar\sigma_0^3 + \bar\sigma_0 \bar\sigma_8^2 - \frac{1}{3\sqrt{2}}\bar\sigma_8^3  \biggl] + \lambda_2 \Big( \bar\sigma_0^2 + \bar\sigma_8^2 \Big) \bar\sigma_0 - B \biggl[ \frac{1}{\sqrt{6}}\bar\sigma_0^2 - \frac{1}{2\sqrt{6}} \bar\sigma_8^2 \biggl] \nonumber\\
    &- \sqrt{\frac{2}{3}} (2cm_l + cm_s) - k \biggl[ \frac{cm_l}{3} \Big( 8 \bar\sigma_0 - 2\sqrt{2} \bar\sigma_8 \Big) + \frac{cm_s}{3} \Big( 4\bar\sigma_0 + 2\sqrt{2} \bar\sigma_8 \Big) \biggl] \,,\nonumber\\
    \frac{\delta S_{\rm LSM}}{\delta \sigma_8} \Biggl|_{\sigma_a = \bar\sigma_a} =& \mu^2 \bar\sigma_8 + \lambda_1 \biggl[ \bar\sigma_0^2 \bar\sigma_8 - \frac{1}{\sqrt{2}} \bar\sigma_0 \bar\sigma_8^2 + \frac{1}{2} \bar\sigma_8^3 \biggl] + \lambda_2 \Big( \bar\sigma_0^2 + \bar\sigma_8^2 \Big) \bar\sigma_8 + B \biggl[ \frac{1}{\sqrt{6}}\bar\sigma_0 \bar\sigma_8 + \frac{1}{2\sqrt{3}}\bar\sigma_8^2 \biggl] \nonumber\\
    &- \frac{2}{\sqrt{3}} (cm_l - cm_s) - k \biggl[ \frac{cm_l}{3} \Big( - 2\sqrt{2} \bar\sigma_0- 8\bar\sigma_8 \Big) + \frac{cm_s}{3} \Big( 2\sqrt{2} \bar\sigma_0 + 2 \bar\sigma_8 \Big) \biggl].
    \label{slope-2}
\end{align}
Note that those two sets of slopes in Eqs.~\eqref{eq:SlopPhibasis} and \eqref{slope-2} are related to each other through the linear transformation arising from the chain rule as 
\begin{align}
    \pmat{\frac{\delta S_{\rm LSM}}{\delta \Phi_1}  \\ \frac{\delta S_{\rm LSM}}{\delta \Phi_3}}
    =
    \pmat{2\sqrt{\frac{2}{3}} & \frac{2}{\sqrt{3}} \\ \sqrt{\frac{2}{3}} & -\frac{2}{\sqrt{3}} }
    \pmat{\frac{\delta S_{\rm LSM}}{\delta \sigma_0} \\ \frac{\delta S_{\rm LSM}}{\delta \sigma_8}}.
    \label{eq:linearTransOfSlops}
\end{align}

At the classical level, the equilibrium criteria, or the stationary conditions, read
\begin{align}
    \frac{\delta S_{\rm LSM}}{\delta \Phi_1} \Biggl|_{\Phi = \bar\Phi} = 0, \qquad \frac{\delta S_{\rm LSM}}{\delta \Phi_3} \Biggl|_{\Phi = \bar\Phi} = 0.
\end{align}
which is equivalent to
\begin{align}
    \frac{\delta S_{\rm LSM}}{\delta \sigma_0} \Biggl|_{\sigma_a = \bar\sigma_a} = 0, \qquad \frac{\delta S_{\rm LSM}}{\delta \sigma_8} \Biggl|_{\sigma_a = \bar\sigma_a} = 0
\end{align}
through Eq.~\eqref{eq:linearTransOfSlops}.

If we adopt the $SU(3)_V$ flavor symmetry, i.e., $m_l = m_s \equiv m$, we find that the slope of the $\sigma_8$-direction in Eq.~\eqref{eq:linearTransOfSlops} becomes
\begin{align}
    \frac{\delta S_{\rm LSM}}{\delta \sigma_8} \Biggl|_{\sigma_a = \bar\sigma_a} =& \bar\sigma_8 \Biggl\{ \mu^2 + \lambda_1 \biggl[ \bar\sigma_0^2 - \frac{1}{\sqrt{2}} \bar\sigma_0 \bar\sigma_8 + \frac{1}{2} \bar\sigma_8^2 \biggl] + \lambda_2 \Big( \bar\sigma_0^2 + \bar\sigma_8^2 \Big) + B \biggl[ \frac{1}{\sqrt{6}}\bar\sigma_0 + \frac{1}{2\sqrt{3}}\bar\sigma_8 \biggl] + 2 k cm  \Biggl\}. 
\end{align}
The stationary condition along this $\sigma_8$ direction thus includes the trivial solution $\bar\sigma_8 = 0$, i.e., $\bar\Phi_1 = \bar\Phi_3$, which precisely reflects the $SU(3)_V$ symmetry. 
This flavor symmetric solution can also be realized when we re-arrange Eq.~\eqref{eq:SlopPhibasis} as follows:  
\begin{align}
    \frac{1}{4} \frac{\delta S_{\rm LSM}}{\delta \Phi_1} \Biggl|_{\Phi = \bar\Phi} - \frac{1}{2}\frac{\delta S_{\rm LSM}}{\delta \Phi_3} \Biggl|_{\Phi = \bar\Phi} = &\Big( \bar\Phi_1 - \bar\Phi_3 \Big) \biggl[ \mu^2 + 2 \lambda_1 \Big( \bar\Phi_1^2 + \bar\Phi_1 \bar\Phi_3 + \bar\Phi_3^2 \Big)  \nonumber\\
    & + \lambda_2 \Big( 4\bar\Phi_1^2 + 2 \bar\Phi_3^2 \Big) + B \bar\Phi_1 + 2kcm_l \biggl] - (cm_l - cm_s) \Big( 1 - 2 k \bar\Phi_1 \Big)\,. 
\end{align}
We thus find the trivial solution $\bar\Phi_1 - \bar\Phi_3 = 0$ when $m_l = m_s$.

\subsection{Curvature masses}

The LSM field $\Phi$ includes the scalar and pseudoscalar meson fields as $\Phi = \phi_a T^a = ( \sigma_a + i \pi_a ) T^a$ for $a = 0, \cdots,8$.   
The Hessian matrix with respect to those meson fields 
is found by taking the second functional derivative for $S_{\rm LSM} = \int_x {\cal L}_{\rm LSM}$ with ${\cal L}_{\rm LSM}$ in Eq.~\eqref{eq:lagrangianLSM}.    
It is in four-dimensional momentum space given as 
\begin{align}
    \Big[ \bar S^{-1}(k, k^\prime; \bar\sigma) \Big]^{ab} &= \frac{\delta^2 S_{\rm LSM}}{\delta \sigma_a(-k^\prime) \delta \sigma_b(k)} = (2\pi)^4 \delta^{(4)}(k - k^\prime)\biggl( k^2 \delta^{ab} + \Big[ m_S^2(\bar\sigma) \Big]^{ab} \biggl) \nonumber\\
    &\equiv (2\pi)^4 \delta^{(4)}(k - k^\prime) \Big[ \bar S^{-1}(k; \bar\sigma) \Big]^{ab}, \nonumber\\
    \Big[ \bar P^{-1}(k, k^\prime; \bar\sigma) \Big]^{ab} &= \frac{\delta^2 S_{\rm LSM}}{\delta \pi_a(-k^\prime) \delta \pi_b(k)} = (2\pi)^4 \delta^{(4)}(k - k^\prime)\biggl( k^2 \delta^{ab} + \Big[ m_P^2(\bar\sigma) \Big]^{ab} \biggl) \nonumber\\
    &\equiv (2\pi)^4 \delta^{(4)}(k - k^\prime) \Big[ \bar P^{-1}(k; \bar\sigma) \Big]^{ab},
    \label{eq:TreePropagators}
\end{align}
where we have defined the tree-level curvature masses as 
\begin{align}
    \Big[ m_S^2(\bar\sigma) \Big]^{ab} = \frac{\partial^2 V_{\rm LSM}}{\partial \sigma_a \partial \sigma_b} \Biggl|_{\sigma_a = \bar\sigma_a} , \qquad \Big[ m_P^2(\bar\sigma) \Big]^{ab} = \frac{\partial^2 V_{\rm LSM}}{\partial \pi_a \partial \pi_b} \Biggl|_{\sigma_a = \bar\sigma_a}.
\label{tree-curv-m}
\end{align}
Notice that the mixing matrix elements such as $ \frac{\delta^2 S_{\rm LSM}}{\delta \pi_a(-k^\prime) \delta \sigma_b(k)}$ vanish because of the parity invariance. 
From Eq.~\eqref{tree-curv-m}, the matrix elements for the square of the nonet scalar meson masses read 
\begin{align}
    \Big[ m_S^2(\bar\sigma) \Big]^{00} &= \mu^2 + \lambda_1 \Big( \bar\sigma_0^2 + \bar\sigma_8^2 \Big) + \lambda_2 \Big( 3 \bar\sigma_0^2 + \bar\sigma_8^2 \Big) - B \sqrt{\frac{2}{3}} \bar\sigma_0 - \frac{4}{3} k (2 c m_l + c m_s), \nonumber\\
    \Big[ m_S^2(\bar\sigma) \Big]^{88} &= \mu^2 + \lambda_1 \biggl[ \bar\sigma_0^2 - \sqrt{2} \bar\sigma_0 \bar\sigma_8 + \frac{3}{2} \bar\sigma_8^2 \biggl] + \lambda_2 \Big( \bar\sigma_0^2 + 3 \bar\sigma_8^2 \Big) + \frac{B}{\sqrt{3}} \biggl[ \frac{1}{\sqrt{2}} \bar\sigma_0 + \bar\sigma_8 \biggl] + \frac{2}{3} k (4 c m_l - c m_s), \nonumber\\
    \Big[ m_S^2(\bar\sigma) \Big]^{08} &= \Big[ m_S^2(\bar\sigma) \Big]^{80} = \lambda_1 \biggl[ 2 \bar\sigma_0 \bar\sigma_8 - \frac{1}{\sqrt{2}} \bar\sigma_8^2 \biggl] + 2 \lambda_2 \bar\sigma_0 \bar\sigma_8 + \frac{B}{\sqrt{6}} \bar\sigma_8 + \frac{2 \sqrt{2}}{3} k (c m_l - c m_s), \nonumber\\
    \Big[ m_S^2(\bar\sigma) \Big]^{11} &= \Big[ m_S^2(\bar\sigma) \Big]^{22} = \Big[ m_S^2(\bar\sigma) \Big]^{33} \nonumber\\
    &= \mu^2 + \lambda_1 \biggl[ \bar\sigma_0^2 + \sqrt{2} \bar\sigma_0 \bar\sigma_8 + \frac{1}{2} \bar\sigma_8^2 \biggl] + \lambda_2 \Big( \bar\sigma_0^2 + \bar\sigma_8^2 \Big) + \frac{B}{\sqrt{3}} \biggl[ \frac{1}{\sqrt{2}} \bar\sigma_0 - \bar\sigma_8 \biggl] + 2 k c m_s, \nonumber\\
    \Big[ m_S^2(\bar\sigma) \Big]^{44} &=  \Big[ m_S^2(\bar\sigma) \Big]^{55} =  \Big[ m_S^2(\bar\sigma) \Big]^{66} =  \Big[ m_S^2(\bar\sigma) \Big]^{77} \nonumber\\
    &= \mu^2 + \lambda_1 \biggl[ \bar\sigma_0^2 - \frac{1}{\sqrt{2}} \bar\sigma_0 \bar\sigma_8 + \frac{1}{2} \bar\sigma_8^2 \biggl] + \lambda_2 \Big( \bar\sigma_0^2 + \bar\sigma_8^2 \Big) + \frac{B}{\sqrt{3}} \biggl[ \frac{1}{\sqrt{2}} \bar\sigma_0 + \frac{1}{2} \bar\sigma_8 \biggl] + 2 k c m_l, \nonumber\\
    {\rm (Others)} &= 0.
    \label{eq:ScalarCurvatureMasses}
\end{align}
The matrix elements $\Big[ m_S^2(\bar\sigma) \Big]^{11}$ and $\Big[ m_S^2(\bar\sigma) \Big]^{44}$ are identified as the square of the $a_0(980)$ mass, $m^2_{a_0}$, and the $\kappa(700)=K^*_0(700)$ mass, $m^2_{\kappa}$, respectively. 
The mixed $(\sigma_0, \sigma_8)$ sector can be diagnolized through an orthogonal transformation
\begin{align}
    \tilde \sigma_i &= \big( O_S^{-1} \big)_{i}^{\,\, a} \sigma_a, \nonumber\\
    \Big[ \tilde m_S^2(\bar\sigma) \Big]_{(i)} \delta^{ij} &= \big( O_S^{-1} \big)^{i}_{\,\, a} \Big[ m_S^2(\bar\sigma) \Big]^{ab} \big( O_S \big)_{b}^{\,\, j},
    \label{eq:OrthogonalTransForScalarsTree}
\end{align}
where $\Big[ \tilde m_S^2(\bar\sigma) \Big]_{(i)}$ denotes the diagonalized scalar mass matrix, and the orthogonal transformation matrix $O_S$ is given as 
\begin{align}
    O_S = \pmat{ \cos{\theta_S^0} & -\sin{\theta_S^0} \\ \sin{\theta_S^0} & \cos{\theta_S^0} }, 
    \label{eq:orthogonalTransScalarVacuum}
\end{align}
with the mixing angle, 
\begin{align}
    \theta_S^0 = \frac{1}{2} \arctan \left[ \frac{2 \Big[ m_S^2(\bar\sigma) \Big]^{08}}{\Big[ m_S^2(\bar\sigma) \Big]^{00} - \Big[ m_S^2(\bar\sigma) \Big]^{88}} \right]
\,.
\label{thetaS0}
\end{align}
Note that in the three-flavor symmetric limit where $\bar\sigma_8 = 0$ and $m_1=m_s=m$, 
the off diagonal element $\Big[ m_S^2(\bar\sigma) \Big]^{08}$ vanishes. 
The $f_0(500)$ and the $f_0(980)$ masses are identified through the lowest and highest eigenvalues, respectively, as 
\begin{align}
    m^2[f_0(500)] = \Big[ \tilde m_S^2(\bar\sigma) \Big]_{(0)} &= \Big[ m_S^2(\bar\sigma) \Big]^{00} \cos^2 \theta_S^0 + \Big[ m_S^2(\bar\sigma) \Big]^{88} \sin^2 \theta_S^0 + 2 \Big[ m_S^2(\bar\sigma) \Big]^{08} \cos \theta_S^0 \sin \theta_S^0, \nonumber\\
    m^2[f_0(980)] = \Big[ \tilde m_S^2(\bar\sigma) \Big]_{(8)} &= \Big[ m_S^2(\bar\sigma) \Big]^{00} \sin^2 \theta_S^0 + \Big[ m_S^2(\bar\sigma) \Big]^{88} \cos^2 \theta_S^0 - 2 \Big[ m_S^2(\bar\sigma) \Big]^{08} \cos \theta_S^0 \sin \theta_S^0. 
    \label{eq:TreeLevelDiagonalizationScalar}
\end{align}

Similarly, from Eq.~\eqref{tree-curv-m}, the matrix elements for the square of the nonet pseudoscalar meson masses are read off as   
\begin{align}
     \Big[ m_P^2(\bar\sigma) \Big]^{00} &= \mu^2 + \frac{\lambda_1}{3} \Big( \bar\sigma_0^2 + \bar\sigma_8^2 \Big) + \lambda_2 \Big( \bar\sigma_0^2 + \bar\sigma_8^2 \Big) + B \sqrt{\frac{2}{3}} \bar\sigma_0 + \frac{4}{3} k (2 c m_l + c m_s), \nonumber\\
    \Big[ m_P^2(\bar\sigma) \Big]^{88} &= \mu^2 + \lambda_1 \biggl[ \frac{1}{3}\bar\sigma_0^2 - \frac{\sqrt{2}}{3} \bar\sigma_0 \bar\sigma_8 + \frac{1}{2} \bar\sigma_8^2 \biggl] + \lambda_2 \Big( \bar\sigma_0^2 + \bar\sigma_8^2 \Big) - \frac{B}{\sqrt{3}} \biggl[ \frac{1}{\sqrt{2}} \bar\sigma_0 + \bar\sigma_8 \biggl] - \frac{2}{3} k (4 c m_l - c m_s), \nonumber\\
    \Big[ m_P^2(\bar\sigma) \Big]^{08} &= \Big[ m_P^2(\bar\sigma) \Big]^{80} = \frac{\lambda_1}{3} \biggl[ 2 \bar\sigma_0 \bar\sigma_8 - \frac{1}{\sqrt{2}} \bar\sigma_8^2 \biggl] - \frac{B}{\sqrt{6}} \bar\sigma_8 - \frac{2 \sqrt{2}}{3} k (c m_l - c m_s), \nonumber\\
    \Big[ m_P^2(\bar\sigma) \Big]^{11} &= \Big[ m_P^2(\bar\sigma) \Big]^{22} = \Big[ m_P^2(\bar\sigma) \Big]^{33} \nonumber\\
    &= \mu^2 + \frac{\lambda_1}{3} \biggl[ \bar\sigma_0^2 + \sqrt{2} \bar\sigma_0 \bar\sigma_8 + \frac{1}{2} \bar\sigma_8^2 \biggl] + \lambda_2 \Big( \bar\sigma_0^2 + \bar\sigma_8^2 \Big) - \frac{B}{\sqrt{3}} \biggl[ \frac{1}{\sqrt{2}} \bar\sigma_0 - \bar\sigma_8 \biggl] - 2 k c m_s, \nonumber\\
    \Big[ m_P^2(\bar\sigma) \Big]^{44} &=  \Big[ m_P^2(\bar\sigma) \Big]^{55} =  \Big[ m_P^2(\bar\sigma) \Big]^{66} =  \Big[ m_P^2(\bar\sigma) \Big]^{77} \nonumber\\
    &= \mu^2 + \frac{\lambda_1}{3} \biggl[ \bar\sigma_0^2 - \frac{1}{\sqrt{2}} \bar\sigma_0 \bar\sigma_8 + \frac{7}{2} \bar\sigma_8^2 \biggl] + \lambda_2 \Big( \bar\sigma_0^2 + \bar\sigma_8^2 \Big) - \frac{B}{\sqrt{3}} \biggl[ \frac{1}{\sqrt{2}} \bar\sigma_0 + \frac{1}{2} \bar\sigma_8 \biggl] - 2 k c m_l, \nonumber\\
    {\rm (Others)} &= 0.
    \label{eq:PseudoScalarCurvatureMasses}
\end{align}
The matrix elements $\Big[ m_P^2(\bar\sigma) \Big]^{11}$ and $\Big[ m_P^2(\bar\sigma) \Big]^{44}$ are identified as the squares of the pion mass, $m^2_\pi$, and the kaon mass $m^2_K$, respectively. 
The mixed $(\pi_0, \pi_8)$ sector can be diagonalized through an orthogonal transformation, similarly to the $(\sigma_0, \sigma_8)$ sector in Eq.~\eqref{eq:OrthogonalTransForScalarsTree} as 
\begin{align}
    \tilde \pi_i &= \big( O_P^{-1} \big)_{i}^{\,\, a} \pi_a, \nonumber\\
    \Big[ \tilde m_P^2(\bar\sigma) \Big]_{(i)} \delta^{ij} &= \big( O_P^{-1} \big)^{i}_{\,\, a} \Big[ m_P^2(\bar\sigma) \Big]^{ab} \big( O_P \big)_{b}^{\,\, j},
    \label{eq:OrthogonalTransForPseudoScalarsTree}
\end{align}
where $\Big[ \tilde m_P^2(\bar\sigma) \Big]_{(i)}$ denotes the diagonalized pseudoscalar meson mass matrix, and the transformation matrix $O_P$ reads
\begin{align}
    O_S = \pmat{ \cos{\theta_P^0} & -\sin{\theta_P^0} \\ \sin{\theta_P^0} & \cos{\theta_P^0} }.
    \label{eq:orthogonalTransPseudoScalarVacuum}
\end{align}
with the mixing angle,  
\begin{align}
    \theta_P^0 = \frac{1}{2} \arctan \left[ \frac{2 \Big[ m_P^2(\bar\sigma) \Big]^{08}}{\Big[ m_P^2(\bar\sigma) \Big]^{00} - \Big[ m_P^2(\bar\sigma) \Big]^{88}} \right]
\,. 
\label{thetaP0}
\end{align}
The off diagonal element $\Big[ m_P^2(\bar\sigma) \Big]^{08}$ vanishes in the three-flavor symmetric limit where $\bar\sigma_8 = 0$. 
Then, the $\eta^\prime$ and the $\eta$ masses are identified through the eigenvalues
\begin{align}
    m^2_{\eta^\prime} = \Big[ \tilde m_P^2(\bar\sigma) \Big]_{(0)} &= \Big[ m_P^2(\bar\sigma) \Big]^{00} \cos^2 \theta_P^0 + \Big[ m_P^2(\bar\sigma) \Big]^{88} \sin^2 \theta_P^0 + 2 \Big[ m_P^2(\bar\sigma) \Big]^{08} \cos \theta_P^0 \sin \theta_P^0, \nonumber\\
    m^2_{\eta} = \Big[ \tilde m_P^2(\bar\sigma) \Big]_{(8)} &= \Big[ m_P^2(\bar\sigma) \Big]^{00} \sin^2 \theta_P^0 + \Big[ m_P^2(\bar\sigma) \Big]^{88} \cos^2 \theta_P^0 - 2 \Big[ m_P^2(\bar\sigma) \Big]^{08} \cos \theta_P^0 \sin \theta_P^0\,. 
    \label{eq:TreeLevelDiagonalizationPsuedoScalar}
\end{align}
%

\section{CJT formalism at finite temperature}
\label{app:CJTFormalismAtFinitTemperature}

In this appendix, we summarize the formulations in the conventional CJT formalism 
applied to the present LSM at finite temperature.

\subsection{CJT formalism and Hartree approximation}

We begin by recasting the CJT effective potential given by Eq.~\eqref{eq:GeneralCJTPotnetialMainText}, i.e., 
\begin{align}
    V_{\rm CJT}[\sigma, S, P] =& V(\sigma) + \frac{1}{2} \int_k \operatorname{tr} \Big[ \log S^{-1}(k) + \log P^{-1}(k) \Big] \nonumber\\
    &+ \frac{1}{2} \int_k \operatorname{tr} \Big[ \bar{S}^{-1}(k;\sigma)S(k) + \bar{P}^{-1}(k;\sigma)P(k) - 2 \Big] + V_2[\sigma,S, P]\,. 
    \label{eq:GeneralCJTPotnetial}
\end{align}
%
Here, the shorthand notation of the momentum integral is defined as
\begin{align}
    \int_k \, f(k) = T \sum_{n = -\infty}^\infty \int \frac{\df^3 \bvec{k}}{(2\pi)^3} \, f(\omega_n, \bvec{k}),
\end{align}
%
with $\omega_n = 2 n \pi T$ being the bosonic Matsubara frequency.
The tree-level mesonic propagators $\bar{S}(k;\bar\sigma)$ and $\bar{P}(k;\bar\sigma)$ are defined as 
\begin{align}
    \Big[ \bar{S}^{-1}(k;\sigma) \Big] ^{ab}&= k^2 \delta^{ab} + \Big[ m_S^2(\sigma) \Big]^{ab}, \nonumber\\
    \Big[ \bar{P}^{-1}(k;\sigma) \Big]^{ab} &= k^2 \delta^{ab} + \Big[ m_P^2(\sigma) \Big]^{ab},
\end{align}
where $ m_S^2(\bar\sigma) $ and $ m_P^2(\bar\sigma) $ are the mesonic curvature masses which are read off from Eqs.~\eqref{eq:ScalarCurvatureMasses} and \eqref{eq:PseudoScalarCurvatureMasses}.

The VEV of $\sigma$ and the propagators ${\mathcal{S}}$ and $\bar{\mathcal{P}}$ are determined 
the stationary conditions and the gap equations
\begin{align}
    \frac{\partial V_{\rm CJT}[\sigma, S, P]}{\partial \sigma_a} \Biggl|_{\sigma = \bar \sigma, S = {\mathcal{S}}, P = {\mathcal{P}}} &= 0, \nonumber\\
    \frac{\delta  V_{\rm CJT}[\sigma, S, P]}{\delta S_{ab}}\Biggl|_{\sigma = \bar \sigma, S = {\mathcal{S}}, P = {\mathcal{P}}} &= 0, \nonumber\\
    \frac{\delta  V_{\rm CJT}[\sigma, S, P]}{\delta P_{ab}}\Biggl|_{\sigma = \bar \sigma, S = {\mathcal{S}}, P = {\mathcal{P}}} &= 0. 
    \label{eq:quantumEOMsinCJT}
\end{align}
The latter two gap equations can be expressed as 
\begin{align}
        \Big[ {\mathcal{S}}^{-1}(k) \Big]^{ab} &= \Big[ \bar{S}^{-1}(k;\sigma) \Big]^{ab} + \Sigma^{ab}, \nonumber\\
        \Big[ {\mathcal{P}}^{-1}(k) \Big]^{ab} &= \Big[ \bar{P}^{-1}(k;\sigma) \Big]^{ab} + \Pi^{ab},
        \label{eq:generalGapEquations}
\end{align}
where
\begin{align}
    \Sigma^{ab} &= 2 \frac{\delta V_2[\sigma,S, P]}{\delta S_{ba}} \Bigg|_{\sigma = \bar \sigma, S = {\mathcal{S}}, P = {\mathcal{P}}}, \nonumber\\
    \Pi^{ab} &= 2 \frac{\delta V_2[\sigma,S, P]}{\delta P_{ba}} \Bigg|_{\sigma = \bar \sigma, S = {\mathcal{S}}, P = {\mathcal{P}}},
\end{align}
are the meson self-energies, which also depend on ${\mathcal{S}}$ and ${\mathcal{P}}$.
Notice that if we take $V_2 = 0$, in the gap equations in Eq.\eqref{eq:generalGapEquations}, 
the propagators are replaced by the tree-level ones, thus the 2PI effective potential will be reduced to the 1PI effective action at one-loop level.

Since $V_2$ includes infinite 2PI diagrams, we have to make a truncation to close the stationary condition and gap equations in Eq.~\eqref{eq:generalGapEquations}. 
In the present work, we apply the double-bubble diagram truncation, as depicted in Fig.~\ref{fig:HartreeApprox}, 
for which the truncation scheme is equivalent to the so-called Hartree approximation. 
The thus truncated 2PI diagrams 
are generated from the four-meson interactions in $V_0$ given in Eq.~\eqref{eq:LSMTreePotential}, which reads  
\begin{align}
    V_2[S, P] = \mathcal{F}^{abcd} \biggl[ \int_k S_{ab}(k) \int_q S_{cd}(q) + \int_k P_{ab}(k) \int_q P_{cd}(q) \biggl] \, + \, 2 \mathcal{H}^{abcd} \int_k S_{ab}(k) \int_q P_{cd}(q),
    \label{eq:V2UnderHartree}
\end{align}
where
\begin{align}
    \mathcal{F}^{abcd} &\equiv \frac{1}{8}\frac{\partial^4 V_0}{\partial \sigma_a \partial \sigma_b \partial \sigma_c \partial \sigma_d}\Biggl|_{\Phi=0} = \frac{\lambda_1}{8} \Big( d^{abn}d^{ncd} + d^{adn}d^{nbc} + d^{acn}d^{nbd} \Big) + \frac{\lambda_2}{4} \Big( \delta^{ab}\delta^{cd} + \delta^{ad}\delta^{bc} + \delta^{ac}\delta^{bd} \Big), \nonumber\\
    \mathcal{H}^{abcd}&\equiv \frac{1}{8} \frac{\partial^4 V_0}{\partial \sigma_a \partial \sigma_b \partial \pi_c \partial \pi_d}\Biggl|_{\Phi=0} = \frac{\lambda_1}{8} \Big( d^{abn}d^{ncd} + f^{acn}f^{nbd} + f^{bcn}f^{nad} \Big) + \frac{\lambda_2}{4} \delta^{ab}\delta^{cd},
    \label{eq:definitionOfHandF}
\end{align}
and the structure constants are defined as 
\begin{align}
    d^{abc} = 2 \operatorname{tr} \Big[ \{ T^a, T^b \} T^c \Big], \qquad f^{abc} = -2i \operatorname{tr} \Big[ [T^a, T^b] T^c \Big].
\end{align}

\subsection{Gap equations of meson masses}


Under the Hartree approximation as in Eq.~\eqref{eq:V2UnderHartree}, 
the loop correction does not generate the momentum-dependent terms.
As has been done in Eq.~\eqref{eq:parameterizationOfFullPropagatorsMainText} of the main text, 
we may therefore parameterize the full propagators as
\begin{align}
    \Big[ {\mathcal S}^{-1}(k) \Big]^{ab} &= k^2 \delta^{ab} + \Big[ M_S^2 \Big]^{ab}, \nonumber\\
    \Big[ {\mathcal P}^{-1}(k) \Big]^{ab} &= k^2 \delta^{ab} + \Big[ M_P^2 \Big]^{ab},
\label{mom-indep}
\end{align}
where $\Big[ M_S \Big]^{ab}$ and $ \Big[ M_P \Big]^{ab}$ are the full masses which include quantum and thermal corrections.
Then, the gap equations in Eq.\eqref{eq:generalGapEquations} are reduced to
\begin{align}
    \Big[ M_S^2 \Big]^{ab} &= \Big[ m_S^2(\bar\sigma) \Big]^{ab} + 4 \mathcal{F}^{abcd} \int_k {\mathcal S}_{cd}(k) + 4 \mathcal{H}^{abcd} \int_k {\mathcal P}_{cd}(k), \nonumber\\
    \Big[ M_P^2 \Big]^{ab} &= \Big[ m_P^2(\bar\sigma) \Big]^{ab} + 4 \mathcal{F}^{abcd} \int_k {\mathcal P}_{cd}(k) + 4 \mathcal{H}^{abcd} \int_k {\mathcal S}_{cd}(k).
    \label{eq:gapEquationsCJT}
\end{align}

At the classical level, we diagonalize the mass matrices to obtain the mass eigenstates, see Eqs.~\eqref{eq:TreeLevelDiagonalizationScalar} and \eqref{eq:TreeLevelDiagonalizationPsuedoScalar}.
After incorporating the quantum and thermal corrections into the meson propagators, we also introduce the following orthogonal transformations to diagonalize the full masses,  in such a way that 
\begin{align}
    \Big[ \tilde M_{S/P}^2 \Big]_{(i)} \delta^{ij} = \big( O_{S/P}^{\prime,-1} \big)^{i}_{\,\, a} \Big[ M_{S/P}^2 \Big]^{ab} \big( O_{S/P}^\prime \big)_{b}^{\,\, j}.
\label{rot-loop}
\end{align}
The rotation angles are expressed as
\begin{align}
    \theta_{S/P} = \frac{1}{2} \arctan  \left[ \frac{2 \Big[ M_{S/P}^2 \Big]^{08}}{\Big[ M_{S/P}^2 \Big]^{00} - \Big[ M_{S/P}^2 \Big]^{88}} \right].
\label{angle-loop}
\end{align}
In general, the transformation matrices $O_{S/P}^\prime$ are different from those at the tree-level, i.e., Eqs.~\eqref{eq:OrthogonalTransForScalarsTree} and \eqref{eq:OrthogonalTransForPseudoScalarsTree}. 

The full-propagators $\mathcal S$ and $\mathcal P$ can also be diagonalized simultaneously by $O_{S/P}^\prime$ in Eq.~\eqref{rot-loop}, which goes like  
\begin{align}
    \tilde{\mathcal S}_{(i)}(k) \delta_{ij} &= \big( O_{S}^{\prime,-1} \big)_{i}^{\,\, a} {\mathcal S}_{ab}(k) \big( O_{S}^\prime \big)^{b}_{\,\, j}, \nonumber\\
    \tilde{\mathcal P}_{(i)}(k) \delta_{ij} &= \big( O_{P}^{\prime,-1} \big)_{i}^{\,\, a} {\mathcal P}_{ab}(k) \big( O_{P}^\prime \big)^{b}_{\,\, j}.
\end{align}
Then, the diagonalized meson propagators are represented fully in terms of the mass eigenvalues, i.e.,
\begin{align}
    \tilde{\mathcal S}^{-1}_{(i)}(k) = k^2 + \Big[ \tilde M_{S}^2 \Big]_{(i)}, 
    \qquad
    \tilde{\mathcal P}^{-1}_{(i)}(k) = k^2 + \Big[ \tilde M_{P}^2 \Big]_{(i)},
\end{align}
where the loop corrections constructed from the meson propagators only include the form 
\begin{align}
    \int_k \tilde{\mathcal S}_{(i)}(k) \quad \text{and} \quad \int_k \tilde{\mathcal P}_{(i)}(k).
\end{align}

Here, we show the expressions of the diagonalized scalar meson self-energies $\tilde\sigma_{(i)}$ explicitly as
\begin{align}
    \tilde\Sigma_{(0)} =& \int_k \tilde{\mathcal S}_{(0)}(k) \, \cdot \frac{1}{16} \Big[ 27 \lambda_1 + 48 \lambda_2 - \lambda_1 \Big( 4 \cos(2 \theta_S) + 7 \cos(4\theta_S) + 32\sqrt{2} \cos{\theta_S} \sin^3 \theta_S \Big) \Big] \nonumber\\
    +& \int_k \tilde{\mathcal S}_{(8)}(k) \, \cdot \frac{1}{16} \Big[ 9 \lambda_1 + 16 \lambda_2 + \lambda_1 \Big( 7 \cos(4 \theta_S) - 4 \sqrt{2} \sin(4\theta_S) \Big) \Big] \nonumber\\
    +& \int_k \tilde{\mathcal S}_{(1)}(k) \, \cdot \frac{3}{4} \Big[ 3 \lambda_1 + 4 \lambda_2 + \lambda_1 \Big( \cos(2 \theta_S) + 2 \sqrt{2} \sin(2\theta_S) \Big) \Big] \nonumber\\
    +& \int_k \tilde{\mathcal S}_{(4)}(k) \, \cdot \Big[ 3 \lambda_1 + 4 \lambda_2 + \lambda_1 \Big( \cos(2 \theta_S) - \sqrt{2} \sin(2\theta_S) \Big) \Big] \nonumber\\
    +& \int_k \tilde{\mathcal P}_{(0)}(k) \, \cdot \frac{1}{24} \Big[ 9 \lambda_1 + 24 \lambda_2 - \lambda_1 \cos(2\theta_P) \nonumber\\
    &\qquad+ \lambda_1 \sin(\theta_P) \Big( -4\sqrt{2}\cos\theta_P + 4\sqrt{2}\cos(\theta_P+2\theta_S) - 9\sin(\theta_P-2\theta_S) + 7\sin(\theta_P+2\theta_S) \Big) \Big] \nonumber\\
    +& \int_k \tilde{\mathcal P}_{(8)}(k) \, \cdot \frac{1}{24} \Big[ 9 \lambda_1 + 24 \lambda_2 + \lambda_1 \cos(2\theta_P) \nonumber\\
    &\qquad+ \lambda_1 \cos(\theta_P) \Big( - 9\cos(\theta_P-2\theta_S) + 7\cos(\theta_P+2\theta_S) -8\sqrt{2}\cos(\theta_P+\theta_S)\sin\theta_S  \Big) \Big] \nonumber\\
    +& \int_k \tilde{\mathcal P}_{(1)}(k) \, \cdot \frac{1}{4} \Big[ 3 \lambda_1 + 12 \lambda_2 + \lambda_1 \Big( \cos(2 \theta_S) + 2 \sqrt{2} \sin(2\theta_S) \Big) \Big] \nonumber\\
    +& \int_k \tilde{\mathcal P}_{(4)}(k) \, \cdot \frac{1}{3} \Big[ 9 \lambda_1 + 12 \lambda_2 - \lambda_1 \Big( 5 \cos(2 \theta_S) + \sqrt{2} \sin(2\theta_S) \Big) \Big], 
\end{align}
\begin{align}
    \tilde\Sigma_{(8)} =& \int_k \tilde{\mathcal S}_{(0)}(k) \, \cdot \frac{1}{16} \Big[ 9 \lambda_1 + 16 \lambda_2 + \lambda_1 \Big( 7 \cos(4 \theta_S) - 4 \sqrt{2} \sin(4\theta_S) \Big) \Big] \nonumber\\
    +& \int_k \tilde{\mathcal S}_{(8)}(k) \, \cdot \frac{1}{16} \Big[ 27 \lambda_1 + 48 \lambda_2 + \lambda_1 \Big( 4 \cos(2 \theta_S) - 7 \cos(4\theta_S) + 32\sqrt{2} \cos^3{\theta_S} \sin \theta_S \Big) \Big] \nonumber\\
    +& \int_k \tilde{\mathcal S}_{(1)}(k) \, \cdot \frac{3}{4} \Big[ 3 \lambda_1 + 4 \lambda_2 - \lambda_1 \Big( \cos(2 \theta_S) + 2 \sqrt{2} \sin(2\theta_S) \Big) \Big] \nonumber\\
    +& \int_k \tilde{\mathcal S}_{(4)}(k) \, \cdot \Big[ 3 \lambda_1 + 4 \lambda_2 - \lambda_1 \Big( \cos(2 \theta_S) - \sqrt{2} \sin(2\theta_S) \Big) \Big] \nonumber\\
    +& \int_k \tilde{\mathcal P}_{(0)}(k) \, \cdot \frac{1}{24} \Big[ 9 \lambda_1 + 24 \lambda_2 - \lambda_1 \cos(2\theta_P) \nonumber\\
    &\qquad+ 2 \lambda_1 \sin(\theta_P) \Big( \cos(2\theta_S)\sin\theta_P - 4 \cos\theta_S(\sqrt{2}\cos(\theta_S+\theta_P)+4\cos(\theta_P)\sin(\theta_S)) \Big) \Big] \nonumber\\
    +& \int_k \tilde{\mathcal P}_{(8)}(k) \, \cdot \frac{1}{24} \Big[ 9 \lambda_1 + 24 \lambda_2 + \lambda_1 \cos(2\theta_P) \nonumber\\
    &\qquad + \lambda_1 \cos(\theta_P) \Big( 9\cos(\theta_P-2\theta_S) - 7\cos(\theta_P+2\theta_S) + 8\sqrt{2}\cos\theta_S \sin(\theta_P+\theta_S)  \Big) \Big] \nonumber\\
    +& \int_k \tilde{\mathcal P}_{(1)}(k) \, \cdot \frac{1}{4} \Big[ 3 \lambda_1 + 12 \lambda_2 - \lambda_1 \Big( \cos(2 \theta_S) + 2 \sqrt{2} \sin(2\theta_S) \Big) \Big] \nonumber\\
    +& \int_k \tilde{\mathcal P}_{(4)}(k) \, \cdot \frac{1}{3} \Big[ 9 \lambda_1 + 12 \lambda_2 + \lambda_1 \Big( 5 \cos(2 \theta_S) + \sqrt{2} \sin(2\theta_S) \Big) \Big],
\end{align}
\begin{align}
    \tilde\Sigma_{(1)} =& \int_k \tilde{\mathcal S}_{(0)}(k) \, \cdot \frac{1}{4} \Big[ 3 \lambda_1 + 4 \lambda_2 + \lambda_1 \Big( \cos(2 \theta_S) + 2 \sqrt{2} \sin(2\theta_S) \Big) \Big] \nonumber\\
    +& \int_k \tilde{\mathcal S}_{(8)}(k) \, \cdot \frac{1}{4} \Big[ 3 \lambda_1 + 4 \lambda_2 - \lambda_1 \Big( \cos(2 \theta_S) + 2\sqrt{2}\sin(2\theta_S) \Big) \Big] \nonumber\\
    +& \int_k \tilde{\mathcal S}_{(1)}(k) \, \cdot \frac{5}{2} \Big[ \lambda_1 + 2\lambda_2 \Big] \nonumber\\
    +& \int_k \tilde{\mathcal S}_{(4)}(k) \, \cdot 2 \Big[ \lambda_1 + 2 \lambda_2 \Big] \nonumber\\
    +& \int_k \tilde{\mathcal P}_{(0)}(k) \, \cdot \frac{1}{12} \Big[ 3 \lambda_1 + 12 \lambda_2 + \lambda_1 \Big( \cos(2\theta_P) + 2\sqrt{2}\sin(2\theta_P) \Big) \Big] \nonumber\\
    +& \int_k \tilde{\mathcal P}_{(8)}(k) \, \cdot \frac{1}{12} \Big[ 3 \lambda_1 + 12 \lambda_2 - \lambda_1 \Big( \cos(2\theta_P) + 2\sqrt{2}\sin(2\theta_P) \Big) \Big] \nonumber\\
    +& \int_k \tilde{\mathcal P}_{(1)}(k) \, \cdot \frac{1}{2} \Big[ 7 \lambda_1 + 6 \lambda_2 \Big] \nonumber\\
    +& \int_k \tilde{\mathcal P}_{(4)}(k) \, \cdot 2 \Big[ \lambda_1 + 2 \lambda_2 \Big],
\end{align}
\begin{align}
    \tilde\Sigma_{(4)} =& \int_k \tilde{\mathcal S}_{(0)}(k) \, \cdot \frac{1}{4} \Big[ 3 \lambda_1 + 4 \lambda_2 + \lambda_1 \Big( \cos(2 \theta_S) - \sqrt{2} \sin(2\theta_S) \Big) \Big] \nonumber\\
    +& \int_k \tilde{\mathcal S}_{(8)}(k) \, \cdot \frac{1}{4} \Big[ 3 \lambda_1 + 4 \lambda_2 - \lambda_1 \Big( \cos(2 \theta_S) - \sqrt{2}\sin(2\theta_S) \Big) \Big] \nonumber\\
    +& \int_k \tilde{\mathcal S}_{(1)}(k) \, \cdot \frac{3}{2} \Big[ \lambda_1 + 2\lambda_2 \Big] \nonumber\\
    +& \int_k \tilde{\mathcal S}_{(4)}(k) \, \cdot 3 \Big[ \lambda_1 + 2 \lambda_2 \Big] \nonumber\\
    +& \int_k \tilde{\mathcal P}_{(0)}(k) \, \cdot \frac{1}{12} \Big[ 9 \lambda_1 + 12 \lambda_2 - \lambda_1 \Big( 5 \cos(2\theta_P) + \sqrt{2}\sin(2\theta_P) \Big) \Big] \nonumber\\
    +& \int_k \tilde{\mathcal P}_{(8)}(k) \, \cdot \frac{1}{12} \Big[ 9 \lambda_1 + 12 \lambda_2 + \lambda_1 \Big( 5 \cos(2\theta_P) + \sqrt{2}\sin(2\theta_P) \Big) \Big] \nonumber\\
    +& \int_k \tilde{\mathcal P}_{(1)}(k) \, \cdot \frac{3}{2} \Big[ \lambda_1 + 2 \lambda_2 \Big] \nonumber\\
    +& \int_k \tilde{\mathcal P}_{(4)}(k) \, \cdot \Big[ 3 \lambda_1 + 4 \lambda_2 \Big].
\end{align}
For the diagonalized pseudoscalar meson self-energies $\tilde\Pi_{(i)}$, they are the same as $\tilde\Sigma_{(i)}$ with the replacement 
\begin{align}
    \tilde{\mathcal S}_{(0)}(k) \longleftrightarrow \tilde{\mathcal P}_{(0)}(k) \quad \text{and} \quad \theta_S \longleftrightarrow \theta_P.
    \label{eq:replacementOfSelfEnergy}
\end{align}
We do not show them explicitly for readability.

To obtain the value of the rotation angles $\theta_{S/P}$ in Eq.~\eqref{angle-loop}, we look at the $08$-component of the gap equations in Eq.~\eqref{eq:gapEquationsCJT} after performing the orthogonal transformation, which turns out to give the following constraints: 
\begin{align}
    &0 = \Big[ \tilde m_S^2(\bar\sigma) \Big]^{08} + \tilde \Sigma^{08}, \nonumber\\
    &0 = \Big[ \tilde m_P^2(\bar\sigma) \Big]^{08} + \tilde \Pi^{08},
    \label{eq:MixingElementOfGapEq}
\end{align}
where
\begin{align}
    \Big[ \tilde m_{S/P}^2(\bar\sigma) \Big]^{08} = \Big[ m_{S/P}^2(\bar\sigma) \Big]^{08} \cos(2\theta_{S/P}) - \biggl\{ \Big[ m_{S/P}^2(\bar\sigma) \Big]^{00} - \Big[ m_{S/P}^2(\bar\sigma) \Big]^{88} \biggl\} \cos(\theta_{S/P}) \sin(\theta_{S/P}),
\end{align}
and 
\begin{align}
    \tilde\Sigma^{08} 
    =& \int_k \tilde{\mathcal S}_{(0)}(k) \, \cdot \frac{\lambda_1}{8} \Big[ \sin\theta_S \Big( 9\cos\theta_S + 7\cos(3\theta_S) - 4\sqrt{2}\sin(3\theta_S) \Big) \Big] \nonumber\\
    +& \int_k \tilde{\mathcal S}_{(8)}(k) \, \cdot \frac{\lambda_1}{8} \Big[ \cos\theta_S \Big( 9\sin\theta_S - 7\sin(3\theta_S) - 4\sqrt{2}\cos(3\theta_S) \Big) \Big] \nonumber\\
    +& \int_k \tilde{\mathcal S}_{(1)}(k) \, \cdot \frac{3\lambda_1}{4} \Big[ 2\sqrt{2}\cos(2\theta_S) - \sin(2\theta_S) \Big] \nonumber\\
    -& \int_k \tilde{\mathcal S}_{(4)}(k) \, \cdot \lambda_1 \Big[ \sqrt{2}\cos(2\theta_S) + \sin(2\theta_S) \Big] \nonumber\\
    +& \int_k \tilde{\mathcal P}_{(0)}(k) \, \cdot \frac{\lambda_1}{24} \sin\theta_P \Big[ 9\cos(\theta_P-2\theta_S) + 7\cos(\theta_P+2\theta_S) - 4\sqrt{2} \sin(\theta_P+2\theta_S) \Big] \nonumber\\
    -& \int_k \tilde{\mathcal P}_{(8)}(k) \, \cdot \frac{\lambda_1}{24}\cos\theta_P \Big[ 9\sin(\theta_P-2\theta_S) + 7\sin(\theta_P+2\theta_S) + 4\sqrt{2} \cos(\theta_P+2\theta_S) \Big] \nonumber\\
    +& \int_k \tilde{\mathcal P}_{(1)}(k) \, \cdot \frac{\lambda_1}{4} \Big[ 2\sqrt{2}\cos(2\theta_S) - \sin(2\theta_S) \Big] \nonumber\\
    -& \int_k \tilde{\mathcal P}_{(4)}(k) \, \cdot \frac{\lambda_1}{3}\Big[ \sqrt{2}\cos(2\theta_S) - 5\sin(2\theta_S) \Big].    
\end{align}
Again, the pseudoscalar mixing self-energy $\tilde\Pi^{08}$ is obtained directly through $\tilde\Sigma^{08}$ simply by the replacement in Eq.~\eqref{eq:replacementOfSelfEnergy}.

\subsection{The stationary condition for VEVs}

The stationary conditions derived from the 2PI effective potential \eqref{eq:GeneralCJTPotnetial} read
\begin{align}
    0 &= \frac{\partial V(\bar\sigma)}{\partial \bar\sigma_0} - 3 \mathcal G^{0bc} \biggl[ \int_k {\mathcal S}_{cb}(k) - \int_k {\mathcal P}_{cb}(k) \biggl] + 4 \mathcal F^{0bcd} \bar\sigma_b \int_k {\mathcal S}_{dc}(k) + 4 \mathcal H^{0bcd} \bar\sigma_b \int_k {\mathcal P}_{dc}(k), \nonumber\\
    0 &= \frac{\partial V(\bar\sigma)}{\partial \bar\sigma_8} - 3 \mathcal G^{8bc} \biggl[ \int_k {\mathcal S}_{cb}(k) - \int_k {\mathcal P}_{cb}(k) \biggl] + 4 \mathcal F^{8bcd} \bar\sigma_b \int_k {\mathcal S}_{dc}(k) + 4 \mathcal H^{8bcd} \bar\sigma_b \int_k {\mathcal P}_{dc}(k),
    \label{eq:StationaryConditionsCJT}
\end{align}
where $V(\bar\sigma)$ is given in Eq.~\eqref{eq:treePotentialAroundMeanField}, and
\begin{align}
    \mathcal G^{abc} &\equiv -\frac{1}{6} \frac{\partial^3 V_{\rm anom}}{\partial \sigma_a \partial \sigma_b \partial \sigma_c} \Biggl|_{\Phi=0} = \frac{1}{6} \frac{\partial^3 V_{\rm anom}}{\partial \sigma_a \partial \pi_b \partial \pi_c} \Biggl|_{\Phi=0} \nonumber\\
    &= \frac{B}{6} \biggl[ d^{abc} - \frac{3}{2}\Big( \delta^{a0}d^{0bc} + \delta^{b0} d^{a0c} + \delta^{c0}d^{ab0} \Big) + \frac{9}{2} d^{000}\delta^{a0}\delta^{b0}\delta^{c0} \biggl].
    \label{eq:definitionOfG}
\end{align}
%

To regularize the meson loops, we perform the vacuum-subtraction prescription, 
such that
\begin{align}
   & \int_k \tilde{\mathcal S}_{(i)}(k) = \int \frac{\df^3 \bvec{k}}{(2\pi)^3} \frac{1}{\epsilon^{S,(i)}_{\bvec{k}}} \left( \exp \left\{ \frac{\epsilon^{S,(i)}_{\bvec{k}}}{T} \right\} -1 \right)^{-1},\nonumber\\
    &\int_k \tilde{\mathcal P}_{(i)}(k)  = \int \frac{\df^3 \bvec{k}}{(2\pi)^3} \frac{1}{\epsilon^{P,(i)}_{\bvec{k}}} \left( \exp \left\{ \frac{\epsilon^{P,(i)}_{\bvec{k}}}{T} \right\} -1 \right)^{-1},
\end{align}
where
\begin{align}
    \epsilon^{S/P,(i)}_{\bvec{k}} \equiv \sqrt{{\bvec{k}}^2 + \Big[ \tilde M_{S/P}^2 \Big]_{(i)} }.
\end{align}
%

\section{Derivations of WTIs and extension to anomalous WTIs}
\label{app:DerivationsOfWTIS}

\subsection{Chiral-limit WTIs in 1PI formalism and threshold property of pseudo NG masses}

Consider 
a continuous symmetry ($\in \, \,{\rm group}\,\,G$) associated with the infinitesimal variation $(\delta_\epsilon)$ of the field $\phi$ (in the linear symmetry realization) and the action $S[\phi]$ as follows: 
%
\begin{align}
    \delta_\epsilon S[\phi] = \frac{\delta S}{\delta \phi_a} \cdot \delta_\epsilon \phi_a = 0\,, \qquad \delta_\epsilon \phi_a = 
    d_a^{\,\, b} \phi_b 
    \label{eq:symmetricActionUnderdeltaepsilon}
\end{align}
where $d_a^{\,\, b}$ is arbitrary coefficient matrix in the field space, and 
$A \cdot B \equiv \int_x \, A(x) B(x)$. 
The generating functional for the action $S[\phi]$ 
is defined together with the source $J_a$ for $\phi_a$ as
\begin{align}
     Z[J] = \int \big[ \mathcal{D} \phi_a \big] e^{-S[\phi] + J^a \cdot \phi_a} = e^{W[J]},
\end{align}
Imposing this $Z[J]$ to be invariant 
under the symmetry transformation in Eq.~\eqref{eq:symmetricActionUnderdeltaepsilon}, 
we have 
\begin{align}
    \int \big[ \mathcal{D} \phi_a^\prime \big] e^{-S[\phi^\prime] + J^a \cdot \phi_a^\prime} \Bigg|_{\phi_a^\prime = \phi_a + i \epsilon \delta_\epsilon \phi_a}= \int \big[ \mathcal{D} \phi_a \big] e^{-S[\phi] + J^a \cdot \phi_a},
\end{align}
%
Suppose that the path integral measure $\big[ \mathcal{D} \phi_a \big]$ is invariant under this symmetry, i.e., no anomaly there,  
we find the following identity: 
\begin{align}
    &\int \big[ \mathcal{D} \phi_a \big] \left[ e^{-S[\phi^\prime] + J^a \cdot \phi_a^\prime } - e^{-S[\phi] + J^a \cdot \phi_a } \right]_{\phi_a^\prime = \phi_a + i \epsilon \delta_\epsilon \phi_a}\nonumber\\
    &= -i\epsilon \int \big[ \mathcal{D} \phi_a \big] \left[ \frac{\delta S}{\delta \phi_a} \cdot \delta_\epsilon \phi_a - J^a\cdot \delta_\epsilon \phi_a \right] e^{-S[\phi] + J^a \cdot \phi_a } + \mathcal O(\epsilon^2) \nonumber\\
    &= i\epsilon \Big[ J^a\cdot \langle \delta_\epsilon \phi_a \rangle_J \Big] Z[J] + \mathcal O(\epsilon^2) \nonumber\\
    &=0,
    \label{eq:derivationOfWTIin1PI}
\end{align}
where
\begin{align}
    \langle O \rangle_J \equiv \frac{1}{Z[J,\mathcal{M}]} \int \left[ \mathcal{D} \phi_a \right] \, O \, e^{- S[\phi,\mathcal{M}] + J^a \cdot \phi_a}.
\end{align}
Equation~\eqref{eq:derivationOfWTIin1PI} gives the WTI at the generating functional level, 
\begin{align}
    J^a\cdot \langle \delta_\epsilon \phi_a \rangle_J = 0.
    \label{eq:WTIatGeneratingFunctionalLevel}
\end{align}

Defining the 1PI effective action
\begin{align}
    \Gamma[\phi_{\rm cl}]  = \sup_J \Big( J \cdot \phi_{\rm cl} - W[J] \Big),
    \label{eq:definning1PIaction}
\end{align}
we have 
\begin{align}
    \frac{\delta W}{\delta J^a} = \langle \phi_a \rangle_J \equiv (\phi_{\rm cl})_a, \qquad \frac{\delta \Gamma}{\delta (\phi_{\rm cl})_a} = J^a.
\end{align}
Replacing $J_a$ with $\Gamma$ as above, 
the WTI in Eq.\eqref{eq:WTIatGeneratingFunctionalLevel} is cast into the form 
\begin{align}
     \frac{\delta \Gamma}{\delta (\phi_{\rm cl})_a} \cdot \delta_\epsilon (\phi_{\rm cl})_a = 0,
     \label{eq:WTIfor1PIEA}
\end{align}
where $\delta_\epsilon (\phi_{\rm cl})_a =  d_a^{\,\,\,b} (\phi_{\rm cl})_b$.

We apply the WTI arguments above to the present LSM case. 
First of all, consider the chiral limit. 
In the chiral limit, the LSM action is invariant under the chiral $SU(3)_L \times SU(3)_R$ transformation, 
\begin{align}
    \Phi \rightarrow g_L \cdot \Phi \cdot g_R^\dagger,
    \label{eq:FieldTransformation}
\end{align}
with $g_{L/R} = \exp\{ i \left( \theta_{L/R}\right)_\alpha T^\alpha \}$ and $\alpha = 1,\cdots,8$.
Under the $SU(3)$ axial rotation with $\left( \theta_L\right)_\alpha = - \left( \theta_R \right)_\alpha = \left( \theta_A\right)_\alpha$, 
the $\sigma_a$ and $\pi_a$ thus infinitesimally transform like 
\begin{align}
    \delta^\alpha \Big( \sigma_a T^a \Big) &= \{ T^\alpha, \pi_a T^a \}, \nonumber\\
    \delta^\alpha \Big( \pi_a T^a \Big) &= -\{ T^\alpha, \sigma_a T^a \}\,, 
\end{align}
%
%
Or equivalently, 
\begin{align}
    \delta^\alpha \sigma_a = d^{\alpha b}_{\quad a} \pi_b, \qquad \delta^\alpha \pi_a = - d^{\alpha b}_{\quad a} \sigma_b,
    \label{eq:transLawOfFields}
\end{align}
where
\begin{align}
    d^{\alpha b}_{\quad a} \equiv 2 \operatorname{tr} \Big[ \{ T^\alpha, T^b \} T_a \Big], \qquad d^{\alpha b}_{\quad 0} = \sqrt{\frac{2}{3}} \delta^{\alpha b}.
\end{align}
When the general WTI in Eq.~\eqref{eq:WTIfor1PIEA} is applied to the 1PI effective action in the LSM, we have 
\begin{align}
    0 &= \frac{\delta \Gamma[\sigma_{\rm cl},\pi_{\rm cl}]}{\delta (\sigma_{\rm cl})_a} \cdot \delta^\alpha (\sigma_{\rm cl})_a + \frac{\delta \Gamma[\sigma_{\rm cl},\pi_{\rm cl}]}{\delta (\pi_{\rm cl})_a} \cdot \delta^\alpha (\pi_{\rm cl})_a = \frac{\delta \Gamma[\sigma_{\rm cl},\pi_{\rm cl}]}{\delta (\sigma_{\rm cl})_a} \cdot d^{\alpha b}_{\quad a} (\pi_{\rm cl})_b + \frac{\delta \Gamma[\sigma_{\rm cl},\pi_{\rm cl}]}{\delta (\pi_{\rm cl})_a} \cdot d^{\alpha b}_{\quad a} (\sigma_{\rm cl})_b\,.
    \label{eq:WTIofSU3ChiralLimit}
\end{align}
Performing one more functional derivative with respect to the external (background) pseudoscalar field $(\pi_{\rm cl}(y))_{b_1}$ on the WTIs in Eq.\eqref{eq:WTIofSU3ChiralLimit},  
we get 
\begin{align}
    &\frac{\delta}{\delta (\pi_{\rm cl}(y))_{b_1}} \biggl( \frac{\delta \Gamma[\sigma_{\rm cl},\pi_{\rm cl}]}{\delta (\sigma_{\rm cl})_a} \cdot d^{\alpha b}_{\quad a} (\pi_{\rm cl})_b + \frac{\delta \Gamma[\sigma_{\rm cl},\pi_{\rm cl}]}{\delta (\pi_{\rm cl})_a} \cdot d^{\alpha b}_{\quad a} (\sigma_{\rm cl})_b \biggl) \nonumber\\
    &= \frac{\delta \Gamma}{\delta (\sigma_{\rm cl})_{a}} \cdot \delta^{(4)}(x-y) d^{\alpha b_1}_{\quad a} - \frac{\delta^2 \Gamma}{\delta (\pi_{\rm cl}(y))_{b_1} \delta (\pi_{\rm cl}(x))_{a}} \cdot d^{\alpha b}_{\quad a} (\sigma_{\rm cl})_{b}.
\label{WTI:LSM:chi}
\end{align}
Imposing the quantum equations of motion such that $\frac{\delta \Gamma}{\delta (\sigma_{\rm cl})_{a}} = 0$ at the background field values $\bar\Phi$, 
Eq.~\eqref{WTI:LSM:chi} is simplified to 
%
\begin{align}
    - \frac{\delta^2 \Gamma}{\delta (\pi_{\rm cl}(y))_{b_1} \delta (\pi_{\rm cl}(x))_{a}} \cdot d^{\alpha b}_{\quad a} \bar\sigma_{b} = - \int_x \, \Big[ \mathcal P^{-1} (y,x) \Big]^{b_1 a} d^{\alpha b}_{\quad a} \bar\sigma_{b},
\label{2nd-deri}
\end{align}
where $\left( \mathcal P^{-1} (y,x) \right)^{b_1 a}$ is the full inverse propagator of the pseudoscalar mesons satisfying the gap equations in Eq.~\eqref{eq:quantumEOMsinCJT}.
In the present system, the translation invariance holds when we turn off the external sources, thus in Euclidean momentum space, $\left( \mathcal P^{-1} (y,x) \right)^{b_1 a}$ can be Fourier transformed like 
\begin{align}
    \Big[ \mathcal P^{-1} (y,x) \Big]^{b_1 a}= \int \frac{\df^4 k}{(2\pi)^4} \Big[ {\mathcal{P}}^{-1} (k) \Big]^{b_1 a} e^{i (y-x) \cdot k}. 
\label{FT}
\end{align}
In the Hartree approximation, as done in Eq.~\eqref{mom-indep}, 
${\mathcal{P}}^{-1} (k)$ can have momentum dependence only from the canonical kinetic part, which reads 
\begin{align}
    \Big[ {\mathcal{P}}^{-1} (k) \Big]^{b_1 a} = k^2 \delta^{b_1 a} + \Big[ M_P^2 \Big]^{b_1 a}\,. 
    \label{eq:parameterizationInWTIs}
\end{align}
%
Putting Eqs.~\eqref{FT} and \eqref{eq:parameterizationInWTIs} into Eq.~\eqref{2nd-deri}, 
we are able to perform the above $x$-integral in Eq.~\eqref{2nd-deri}, to get 
\begin{align}
    - \int_x \, \Big[ \mathcal P^{-1} (y,x) \Big]^{b_1 a} d^{\alpha b}_{\quad a} \bar\sigma_{b} &= - \int \df^4 x \, \int \frac{\df^4 k}{(2\pi)^4} \Big[ {\mathcal{P}}^{-1} (k) \Big]^{b_1 a} e^{i (y-x) \cdot k} d^{\alpha b}_{\quad a} \bar\sigma_{b} \nonumber\\
    &= - \Big[ {\mathcal{P}}^{-1} (0) \Big]^{b_1 a} d^{\alpha b}_{\quad a} \bar\sigma_{b} \nonumber\\
    &= - \Big[ M_P^2 \Big]^{b_1 a} d^{\alpha b}_{\quad a} \bar\sigma_{b}.
\label{x-int-done}
\end{align}
%

Taking $b_1 = 1$ in Eq.~\eqref{x-int-done}, 
we find the right-hand side to be 
\begin{align}
    - \Big[ M_P^2 \Big]^{1 a} d^{\alpha b}_{\quad a} \big( \bar\sigma_{0} \delta_{0b} + \bar\sigma_{8} \delta_{8b} \big).
\end{align}
Since the $SU(3)$ 
vectorial symmetry is present in the system, 
the $SU(2)$ triplet pions and the $SU(2)$ doublet kaons do not mix each other, 
hence we have $\Big[ M_P^2 \Big]^{1 a} = \Big[ M_P^2 \Big]^{1 1} \delta^{1 a}$.
Noting also that $d^{\alpha 0}_{\quad 1} = \sqrt{2/3} \delta^\alpha_1$ and $d^{\alpha 8}_{\quad 1} = 1/\sqrt{3} \delta^\alpha_1$, we then get 
\begin{align}
    - \Big[ M_P^2 \Big]^{1 1} d^{\alpha b}_{\quad 1} \big( \bar\sigma_{0} \delta_{0b} + \bar\sigma_{8} \delta_{8b} \big) = - M_\pi^2 \Big( \sqrt{\frac{2}{3}}\bar\sigma_{0} + \frac{1}{\sqrt{3}} \bar\sigma_{0} \Big)\delta^\alpha_1 = -2 M_\pi^2 \bar\Phi_1 \delta^\alpha_1.
    \label{eq:pionGMORLHS}
\end{align}
%

By taking $b_1 = 4$ in Eq.~\eqref{x-int-done}, 
the right-hand side is evaluated to be  
\begin{align}
    - \Big[ M_P^2 \Big]^{4 4} d^{\alpha b}_{\quad 4} \big( \bar\sigma_{0} \delta_{0b} + \bar\sigma_{8} \delta_{8b} \big).
\end{align}
Here, we have $d^{\alpha 0}_{\quad 4} = \sqrt{2/3} \delta^\alpha_0$ and $d^{\alpha 8}_{\quad 4} = -1/(2\sqrt{3}) \delta^\alpha_4$, such that
\begin{align}
    - \Big[ M_P^2 \Big]^{4 4} d^{\alpha b}_{\quad 4} \big( \bar\sigma_{0} \delta_{0b} + \bar\sigma_{8} \delta_{8b} \big) = - M_\kappa^2 \Big( \sqrt{\frac{2}{3}}\bar\sigma_{0} - \frac{1}{2 \sqrt{3}} \bar\sigma_{0} \Big)\delta^\alpha_4 = - M_\kappa^2 \Big( \bar \Phi_1 + \bar \Phi_3 \Big) \delta^\alpha_4.
    \label{eq:kaonGMORLHS}
\end{align}
%
Taking $\alpha = 1$ and $\alpha = 4$ individually, 
we thus find the following threshold property for the pion and kaon masses
\begin{align}
    M_\pi^2 \bar\Phi_1 = 0, \qquad  M_\kappa^2 \Big( \bar \Phi_1 + \bar \Phi_3 \Big) = 0,
\end{align}
which precisely follows the massless NG boson when $\bar{\Phi}_1$ or $\bar{\Phi}_8$ 
takes a nonzero value, i.e., the spontaneous breaking of the chiral symmetry takes place. 

\subsection{Anomalous WTIs and GMOR relations}

In the present LSM, actually, the chiral $SU(3)$ symmetry is explicitly broken by the current quark masses encoded in the spurion field $\mathcal M$. 
We therefore extend the formulation described above on the chiral-limit WTIs by including ${\cal M}$.  

As in the main text, the spurion field ${\cal M}$ is introduced so as to compensate for the explicit chiral breaking by quark masses in the action. 
The spurion field ${\cal M}$ therefore transforms under the chiral $SU(3)$ symmetry 
in the same way as $\phi$ does in Eq.~\eqref{eq:FieldTransformation}: 
\begin{align}
    \mathcal{M} \rightarrow g_L \cdot \mathcal{M} \cdot g_R^\dagger.
\label{trans:calM:app}
\end{align}
%
Thus, the chiral variation of the action, as generically given in Eq.~\eqref{eq:symmetricActionUnderdeltaepsilon}, gets shifted as  
\begin{align}
    \delta_\epsilon S[\phi] = \frac{\delta S}{\delta \phi_a} \cdot \delta_\epsilon \phi_a + \frac{\delta S}{\delta \mathcal{M}} \cdot \delta_\epsilon \mathcal{M} + \frac{\delta S}{\delta \mathcal{M}^\dagger} \cdot \delta_\epsilon \mathcal{M}^\dagger = 0.
    \label{eq:symmetricActionUnderdeltaepsilonBroken}
\end{align}
On the level of the generating functional, Eq.~\eqref{eq:derivationOfWTIin1PI} is then modified as
\begin{align}
    &\int \left[ \mathcal{D} \phi_a \right] \, \left[ e^{- S[\phi^\prime,\mathcal{M}] + J^a \cdot \phi_a^\prime} - e^{- S[\phi,\mathcal{M}] + J^a \cdot \phi_a} \right]_{\phi_a^\prime = \phi_a + i \epsilon \delta_\epsilon \phi_a} \nonumber\\
    &= - i \epsilon \int \left[ \mathcal{D} \phi_a \right] \, \left[ \frac{\delta S}{\delta \phi_a} \cdot \delta_\epsilon \phi_a - J^a \cdot  \delta_\epsilon \phi_a \right]e^{- S[\phi,\mathcal{M}] + J^a \cdot \phi_a} + \mathcal{O}(\epsilon^2) \nonumber\\
    &= i\epsilon \int \left[ \mathcal{D} \phi_a \right] \, \left[ \frac{\delta S}{\delta \mathcal{M}} \cdot \delta_\epsilon \mathcal{M} + \frac{\delta S}{\delta \mathcal{M}^\dagger} \cdot \delta_\epsilon \mathcal{M}^\dagger + J^a \cdot  \delta_\epsilon \phi_a \right]e^{- S[\phi,\mathcal{M}] + J^a \cdot \phi_a} + \mathcal{O}(\epsilon^2) \nonumber\\
    &= i\epsilon \biggl[ \biggl\langle \frac{\delta S}{\delta \mathcal{M}} \biggl\rangle_J \cdot \delta_\epsilon \mathcal{M} + \biggl\langle \frac{\delta S}{\delta \mathcal{M}^\dagger} \biggl\rangle_J \cdot \delta_\epsilon \mathcal{M}^\dagger + J^a \cdot \langle \delta_\epsilon \phi_a \rangle_J \biggl] Z[J,\mathcal{M}] + \mathcal{O}(\epsilon^2) \nonumber\\
    &= 0\,. 
    \label{eq:derivationOfWTIAtGeneFuncLevelBroken}
\end{align}
%
We then find the general form of the anomalous WTIs including ${\cal M}$,  
\begin{align}
    \biggl\langle \frac{\delta S}{\delta \mathcal{M}} \biggl\rangle_J \cdot \delta_\epsilon \mathcal{M} + \biggl\langle \frac{\delta S}{\delta \mathcal{M}^\dagger} \biggl\rangle_J \cdot \delta_\epsilon \mathcal{M}^\dagger + J^a \cdot \langle \delta_\epsilon \phi_a \rangle_J = 0.
    \label{eq:WTIatQuantumLevel}
\end{align}
%
Written in terms of the 1PI effective action in Eq.\eqref{eq:definning1PIaction}, the anomalous WTIs take the form 
\begin{align}
    \frac{\delta S}{\delta \mathcal{M}} \Biggl|_{\phi_a = (\phi_{\rm cl})_a + \Delta_{ab} \cdot \frac{\delta}{\delta (\phi_{\rm cl})_b}} \cdot \delta_\epsilon \mathcal{M} + \frac{\delta S}{\delta \mathcal{M}^\dagger} \Biggl|_{\phi_a = (\phi_{\rm cl})_a + \Delta_{ab} \cdot \frac{\delta}{\delta (\phi_{\rm cl})_b}} \cdot \delta_\epsilon \mathcal{M}^\dagger + \frac{\delta \Gamma}{\delta (\phi_{\rm cl})_a} \cdot \delta_\epsilon (\phi_{\rm cl})_a = 0,
     \label{eq:BrokenWTIfor1PIEA}
\end{align}
where 
\begin{align}
    \Delta_{ab}(x,y) = \frac{\delta^2 W}{\delta J(x) \delta J(y)}
\end{align}
denotes the full connected two-point function. 

\begin{figure}[t]
    \centering
    \includegraphics[width=0.25\linewidth]{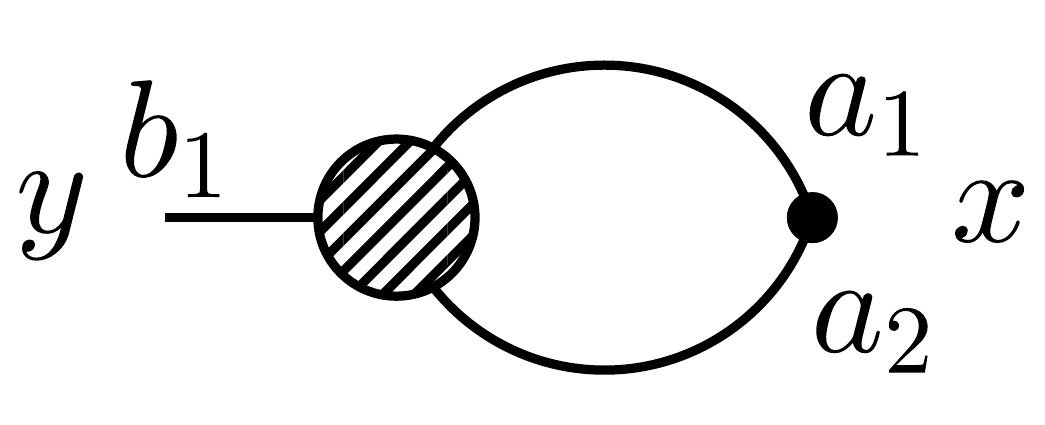}
    \caption{A Feynman diagram of the meson scattering kernel defined in Eq.~\eqref{eq:epionScalarScatteringKernel}. The solid line connected to the point $y$ denotes the pseudoscalar meson propagator, and the lines involving the point $x$ represent the scalar or pseudoscalar meson propagators. The hatched blob stands for the dressed three-point meson vertex in the 1PI formulation. }
    \label{fig:epionScalarScatteringKernel}
\end{figure}

With Eq.~\eqref{eq:BrokenWTIfor1PIEA} at hand, we are ready to derive the GMOR relations at the 1PI level. 
In the last subsection, we have dealt with the first term in Eq.~\eqref{eq:BrokenWTIfor1PIEA}, hence, we focus on the latter two terms.
Following Eq.~\eqref{trans:calM:app}
The infinitesimal transformation of $\mathcal M$ goes like 
\begin{align}
    \delta^\alpha \mathcal M_{mn} = \{ T^\alpha, \mathcal{M} \}_{mn}.
\end{align}
Decomposing ${\cal M}$ into scalar and pseudoscalar parts as $\mathcal M = \mathcal M^s + i \mathcal M^p = \mathcal M^s_a T^a + i \mathcal M^p_a T^a$, the transformation of $\mathcal M^{s/p}_a$ then reads as 
\begin{align}
    \delta^\alpha \mathcal M^s_a = d^{\alpha b}_{\quad a} \mathcal M^p_b, \qquad \delta^\alpha \mathcal M^p_a = - d^{\alpha b}_{\quad a} \mathcal M^s_b.
\end{align}
Consider
\begin{align}
    \frac{\delta S_{\rm LSM}}{\delta \mathcal M_{mn}} &= -c \big( \Phi^\dagger \big)_{nm} - kc \epsilon_{mbc} \epsilon^{nef} \Phi^b_e \Phi^c_f \nonumber\\
    &= -c \big( \sigma_a - i \pi_a \big) T^a_{nm} - kc \big( \sigma_{a_1} + i \pi_{a_1} \big) \big( \sigma_{a_2} + i \pi_{a_2} \big) \epsilon_{mbc} \epsilon^{nef} ( T^{a_1} )^b_e ( T^{a_2} )^c_f.
\end{align}
Replacing $\phi_a$ by $(\phi_{\rm cl})_a + \Delta_{ab} \cdot \frac{\delta}{\delta (\phi_{\rm cl})_b}$, we obtain
\begin{align}
    &\frac{\delta S_{\rm LSM}}{\delta \mathcal M_{mn}}\Biggl|_{\phi_a = (\phi_{\rm cl})_a + \Delta_{ab} \cdot \frac{\delta}{\delta (\phi_{\rm cl})_b}} \cdot \delta^\alpha \mathcal{M}_{mn} \nonumber\\
    &= -c \Big[ (\sigma_{\rm cl})_a - i (\pi_{\rm cl})_a \Big] \cdot \Big[  \mathcal M^p_{a_2} - i \mathcal M^s_{a_2} \Big] T^b_{nm} T^a_{nm} d^{\alpha a_2}_{\quad b} \nonumber\\
    &\qquad- kc \Big[ \mathcal S_{a_1 a_2}(x,x) + i \mathcal P_{a_1 a_2}(x,x) \Big] \cdot \Big[  \mathcal M^p_{a_3} - i \mathcal M^s_{a_3} \Big] \epsilon_{mbc} \epsilon^{nef} ( T^g )^m_n ( T^{a_1} )^b_e ( T^{a_2} )^c_f d^{\alpha a_3}_{\quad g}\nonumber\\
    &\qquad - kc \Big[ (\sigma_{\rm cl})_{a_1} (\sigma_{\rm cl})_{a_2} + i (\sigma_{\rm cl})_{a_1} (\pi_{\rm cl})_{a_2} + i (\pi_{\rm cl})_{a_1} (\sigma_{\rm cl})_{a_2}  - (\pi_{\rm cl})_{a_1} (\pi_{\rm cl})_{a_2} \Big] \cdot \Big[  \mathcal M^p_{a_3} - i \mathcal M^s_{a_3} \Big] \nonumber\\
    &\qquad \quad \times \epsilon_{mbc} \epsilon^{nef} ( T^g )^m_n ( T^{a_1} )^b_e ( T^{a_2} )^c_f d^{\alpha a_3}_{\quad g},
    \label{eq:BreakingTermOfWTIExplicit}
\end{align}
where $\mathcal S_{a_1 a_2}(x,y)$ and $\mathcal P_{a_1 a_2}(x,y)$ are the full-dressed scalar and pseudoscalar propagators.
We further get the following.   
\begin{align}
    &\frac{\delta}{\delta (\pi_{\rm cl}(y))_{b_1}} \eqref{eq:BreakingTermOfWTIExplicit} \nonumber\\
    &= c \delta^{(4)}(x-y) \Big[ \delta^{b_1}_a \Big] \cdot \Big[ \mathcal M^s_{a_2} + i \mathcal M^p_{a_2} \Big] T^b_{mn} T^a_{nm} d^{\alpha a_2}_{\quad b} \nonumber\\
    &\qquad- kc \Big[ \frac{\delta}{\delta (\pi_{\rm cl}(y))_{b_1}} \mathcal S_{a_1 a_2}(x,x) + i \frac{\delta}{\delta (\pi_{\rm cl}(y))_{b_1}} \mathcal P_{a_1 a_2}(x,x) \Big] \cdot \Big[  \mathcal M^p_{a_3} - i \mathcal M^s_{a_3} \Big] \nonumber\\
    &\qquad \quad \times \epsilon_{mbc} \epsilon^{nef} ( T^g )^m_n ( T^{a_1} )^b_e ( T^{a_2} )^c_f d^{\alpha a_3}_{\quad g} \nonumber\\
    &\qquad - kc \delta^{(4)}(x-y) \Big[ \Big( (\sigma_{\rm cl})_{a_1} + i (\pi_{\rm cl})_{a_1} \Big) \delta^{b_1}_{a_2} + \delta^{b_1}_{a_1} \Big( (\sigma_{\rm cl})_{a_2} + i (\pi_{\rm cl})_{a_2} \Big) \Big] \cdot \Big[ \mathcal M^s_{a_3} + i \mathcal M^p_{a_3} \Big] \nonumber\\
    &\qquad \quad \times \epsilon_{mbc} \epsilon^{nef} ( T^g )^m_n ( T^{a_1} )^b_e ( T^{a_2} )^c_f d^{\alpha a_3}_{\quad g} \nonumber\\
    &= \frac{c}{2} \delta^{(4)}(x-y) \Big[ \delta^{b_1}_a \Big] \cdot \Big[ \mathcal M^s_{a_2} + i \mathcal M^p_{a_2} \Big] \delta^{ab} d^{\alpha a_2}_{\quad b} \nonumber\\
    &\qquad- kc \Big[ H^{b_1}_{\quad a_1 a_2}(y,x) \Big] \cdot \Big[  \mathcal M^p_{a_3} - i \mathcal M^s_{a_3} \Big] \mathcal T^{g a_1 a_2} d^{\alpha a_3}_{\quad g}\nonumber\\
    &\qquad - kc \delta^{(4)}(x-y) \Big[ \Big( (\sigma_{\rm cl})_{a_1} + i (\pi_{\rm cl})_{a_1} \Big) \delta^{b_1}_{a_2} + \delta^{b_1}_{a_1} \Big( (\sigma_{\rm cl})_{a_2} + i (\pi_{\rm cl})_{a_2} \Big) \Big] \cdot \Big[ \mathcal M^s_{a_3} + i \mathcal M^p_{a_3} \Big] \mathcal T^{g a_1 a_2} d^{\alpha a_3}_{\quad g},
    \label{eq:DerOfBreakingTermOfWTIExplicit}
\end{align}
where
\begin{align}
    \mathcal T^{g a_1 a_2} \equiv \epsilon_{mbc} \epsilon^{nef} ( T^g )^m_n ( T^{a_1} )^b_e ( T^{a_2} )^c_f. 
\end{align}
In Eq.~\eqref{eq:DerOfBreakingTermOfWTIExplicit}, we have defined the \textit{meson scattering kernel} $H^{b_1}_{\quad a_1 a_2}(y,x)$ as 
\begin{align}
    H^{b_1}_{\quad a_1 a_2}(y,x) \equiv \frac{\delta}{\delta (\pi_{\rm cl}(y))_{b_1}} \mathcal S_{a_1 a_2}(x,x) + i \frac{\delta}{\delta (\pi_{\rm cl}(y))_{b_1}} \mathcal P_{a_1 a_2}(x,x). 
    \label{eq:epionScalarScatteringKernel}
\end{align}
The corresponding diagrammatic interpretation is available in 
Fig.~\ref{fig:epionScalarScatteringKernel}.

The meson scattering kernel $H^{b_1}_{\quad a_1 a_2}(y,x)$ includes the fully dressed $3$-point meson vertex. 
In the present work, we adopt the truncation scheme such that the $3$-point meson vertex is replaced by the tree-level vertex, i.e., 
\begin{align}
    \frac{\delta^3 \Gamma}{\delta \sigma_a(x) \delta \sigma_b(y) \delta \sigma_c(z)} = -\frac{\delta^3 \Gamma}{\delta \sigma_a(x) \delta \pi_b(y) \delta \pi_c(z)} = \frac{\delta^3 S_{\rm LSM}}{\delta \sigma_a(x) \delta \sigma_b(y) \delta \sigma_c(z)} = -6 \mathcal{G}^{a b c} \delta^{(4)}(x-y) \delta^{(4)}(x-z),
\end{align}
where $\mathcal{G}^{a b c}$ is defined in Eq.~\eqref{eq:definitionOfG}.
We find that within the current truncation scheme, the contribution from the meson scattering kernel vanishes in the anomalous WTIs, in Eq.~\eqref{eq:DerOfBreakingTermOfWTIExplicit}.

Taking the background field values $\bar \sigma_a T^a = \operatorname{diag}\{ \bar \Phi_1, \bar \Phi_1, \bar \Phi_3 \}$, $\mathcal M = \operatorname{diag}\{ m_l, m_l, m_s \}$, and $\bar \pi_a = 0$,  we see that for $b_1 = \alpha = 1$,  
\begin{align}
    \eqref{eq:DerOfBreakingTermOfWTIExplicit} = c m_l + 2k c m_l \bar \Phi_3,
    \label{eq:pionGMORRHS}
\end{align}
and for $b_1 = \alpha = 4$,
\begin{align}
    \eqref{eq:DerOfBreakingTermOfWTIExplicit} = \frac{1}{2} \big( c m_l + cm_s \big) + k \big( c m_l + cm_s \big) \bar \Phi_1,
    \label{eq:kaonGMORRHS}
\end{align}
where we have integrated spacetime over $x$. 
The hermitian conjugate term $\propto \mathcal M^\dagger$ gives the same result as above.

Combining Eqs.~\eqref{eq:pionGMORLHS}, \eqref{eq:kaonGMORLHS}, \eqref{eq:pionGMORRHS},  and \eqref{eq:kaonGMORRHS}, we get the GMOR relations expressed in terms of the LSM parameters, as in Eq.~\eqref{eq:GMORrelations1PIMainText}, 
\begin{align}
    M_\pi^2 \bar\Phi_1 &= c m_l \Big( 1 + 2k \bar \Phi_3 \Big) \,, \nonumber\\
    M_\kappa^2 \Big( \bar \Phi_1 + \bar \Phi_3 \Big) &= \big( c m_l + cm_s \big) \Big( 1 + 2 k \bar \Phi_1 \Big)\,. 
    \label{eq:GMORrelations1PI}
\end{align}
These two relations coincide with those obtained from the tree-level action, as we expected, meaning that the contributions to the explicit chiral breaking terms generated from the quantum fluctuations are dropped in the current truncation scheme. 
In the 1PI formulation, the above GMOR relations are actually equivalent to the stationary conditions for $\bar\Phi_1$ and $\bar\Phi_3$, which substantially deviate from each other in the 2PI formalism, as noted in the main text and will be futher clarified in the next section.

\subsection{(Anomalous) WTIs in 2PI formalism}

Considering the generating functional with a bilocal source $K^{ab}(x,y)$, 
\begin{align}
    Z[J,K] = \int \big[ \mathcal{D} \phi_a \big] e^{-S[\phi] + J^a \cdot \phi_a + \frac{1}{2} \phi_a \cdot K^{ab} \cdot \phi_b} = e^{W[J,K]},
    \label{eq:bilocalGeneratingFunctional}
\end{align}
we have
\begin{align}
    \frac{\delta W[J,K]}{\delta J^a} = \langle \phi_a \rangle_{J,K} \equiv ( \phi_{\rm cl} )_a, \qquad \frac{\delta W}{\delta J^a \delta J^b} = \langle \phi_a \phi_b \rangle_{J,K} - \langle \phi_a \rangle_{J,K} \langle \phi_b \rangle_{J,K} \equiv \Delta_{ab},
    \label{eq:FuncDerToWJK}
\end{align}
and
\begin{align}
     \frac{\delta W[J,K]}{\delta K^{ab}} = \frac{1}{2} \Big( \Delta_{ab} + ( \phi_{\rm cl} )_a ( \phi_{\rm cl} )_b \Big).
\end{align}
The 2PI effective action is obtained by performing the following double-Legendre transformation: 
\begin{align}
    \Gamma[\phi_{\rm cl}, \Delta] &= \sup_{K} \biggl[ K^{ab} \cdot \frac{\delta \Gamma[\phi_{\rm cl},K]}{\delta K^{ab}}  - \sup_J \Big( J^a \cdot ( \phi_{\rm cl} )_a - W[J,K] \Big) \biggl] \nonumber\\
    &= \sup_{K} \inf_J \biggl[ W[J,K] - J^a \cdot ( \phi_{\rm cl} )_a - \frac{1}{2} K^{ab} \cdot \Big( \Delta_{ab} + ( \phi_{\rm cl} )_a ( \phi_{\rm cl} )_b \Big) \biggl],
\end{align}
where we have used 
\begin{align}
    \frac{\delta \Gamma[\phi_{\rm cl},K]}{\delta K^{ab}} = -\frac{\delta W[J,K]}{\delta K^{ab}}.
\end{align}
We then have
\begin{align}
    \frac{\delta \Gamma[\phi_{\rm cl}, \Delta]}{\delta ( \phi_{\rm cl} )_a} = - J^a - K^{ab} \cdot ( \phi_{\rm cl} )_b, \qquad \frac{\delta \Gamma[\phi_{\rm cl}, \Delta]}{\delta \Delta_{ab}} = -\frac{1}{2} K^{ab}.
    \label{eq:FuncDerToGammaJK}
\end{align}

In a manner similar to what has been done in Eq.~\eqref{eq:derivationOfWTIin1PI}, 
the symmetry variation as in Eq.~\eqref{eq:symmetricActionUnderdeltaepsilon} 
and its invariantce of the generating functional $Z[J,K]$ in Eq.~\eqref{eq:bilocalGeneratingFunctional} 
%
leads to the WTIs in terms of $Z[J,K]$, 
\begin{align}
    &\int \big[ \mathcal{D} \phi_a \big] \left[ e^{-S[\phi^\prime] + J^a \cdot \phi_a^\prime + \frac{1}{2} \phi_a^\prime \cdot K^{ab} \cdot \phi_b^\prime} - e^{-S[\phi] + J^a \cdot \phi_a + \frac{1}{2} \phi_a \cdot K^{ab} \cdot \phi_b} \right] \nonumber\\
    &= -i\epsilon \int \big[ \mathcal{D} \phi_a \big] \left[ \frac{\delta S}{\delta \phi_a} \cdot \delta_\epsilon \phi_a - J^a \cdot \delta_\epsilon \phi_a - \frac{1}{2} \phi_a \cdot K^{ab} \cdot \delta_\epsilon \phi_b - \frac{1}{2} \delta_\epsilon \phi_a \cdot K^{ab} \cdot \phi_b \right] e^{-S[\phi] + J^a \cdot \phi_a + \frac{1}{2} \phi_a \cdot K^{ab} \cdot \phi_b} \nonumber\\
    &\qquad + \mathcal{O}(\epsilon^2) \nonumber\\
    &=i \epsilon \biggl[ J^a \cdot \langle \delta_\epsilon \phi_a \rangle + \frac{1}{2} K^{ab} \cdot \langle \phi_a \cdot \delta_\epsilon \phi_b \rangle + \frac{1}{2} K^{ab} \cdot \langle \delta_\epsilon\phi_a \cdot \phi_b \rangle \biggl] Z[J,K] + \mathcal{O}(\epsilon^2) \nonumber\\
    &=0\,. 
    \label{K-WTI}
\end{align}
Taking into account Eqs.\eqref{eq:FuncDerToWJK} and \eqref{eq:FuncDerToGammaJK},  
we rewrite Eq.~\eqref{K-WTI} into the form 
\begin{align}
    \frac{\delta \Gamma[\phi_{\rm cl}, \Delta]}{\delta (\phi_{\rm cl})_a} \cdot d_a^{\,\, b} ( \phi_{\rm cl} )_b + \frac{\delta \Gamma[\phi_{\rm cl}, \Delta]}{\delta \Delta_{ab}} \cdot \Big( d_b^{\,\, c} \Delta_{ac} + d_a^{\,\, c} \Delta_{cb} \Big) = 0. 
    \label{eq:WTIat2PI}
\end{align}
This is the exact form of WTIs in the 2PI formalism. 
Taking the functional derivative of both sides in Eq.~\eqref{eq:WTIat2PI}
with respect to $( \phi_{\rm cl} )_c$, 
we reach the form of Eq.~\eqref{eq:2PIWTIderivativeMainText} in the main text: 
\begin{align}
    \frac{\delta^2 \Gamma[\phi_{\rm cl}, \Delta]}{\delta (\phi_{\rm cl})_c \delta (\phi_{\rm cl})_a} \cdot d_a^{\,\, b} ( \phi_{\rm cl} )_b + \frac{\delta \Gamma[\phi_{\rm cl}, \Delta]}{\delta (\phi_{\rm cl})_a} \cdot d_a^{\,\, c} + \frac{\delta^2 \Gamma[\phi_{\rm cl}, \Delta]}{\delta (\phi_{\rm cl})_c \delta \Delta_{ab}} \cdot \Big( d_b^{\,\, c} \Delta_{ac} + d_a^{\,\, c} \Delta_{cb} \Big) = 0.
    \label{eq:2PIWTIderivative}
\end{align}
%

\begin{figure}[t]
    \begin{minipage}[b]{1\textwidth}
        \includegraphics[width=0.32\linewidth]{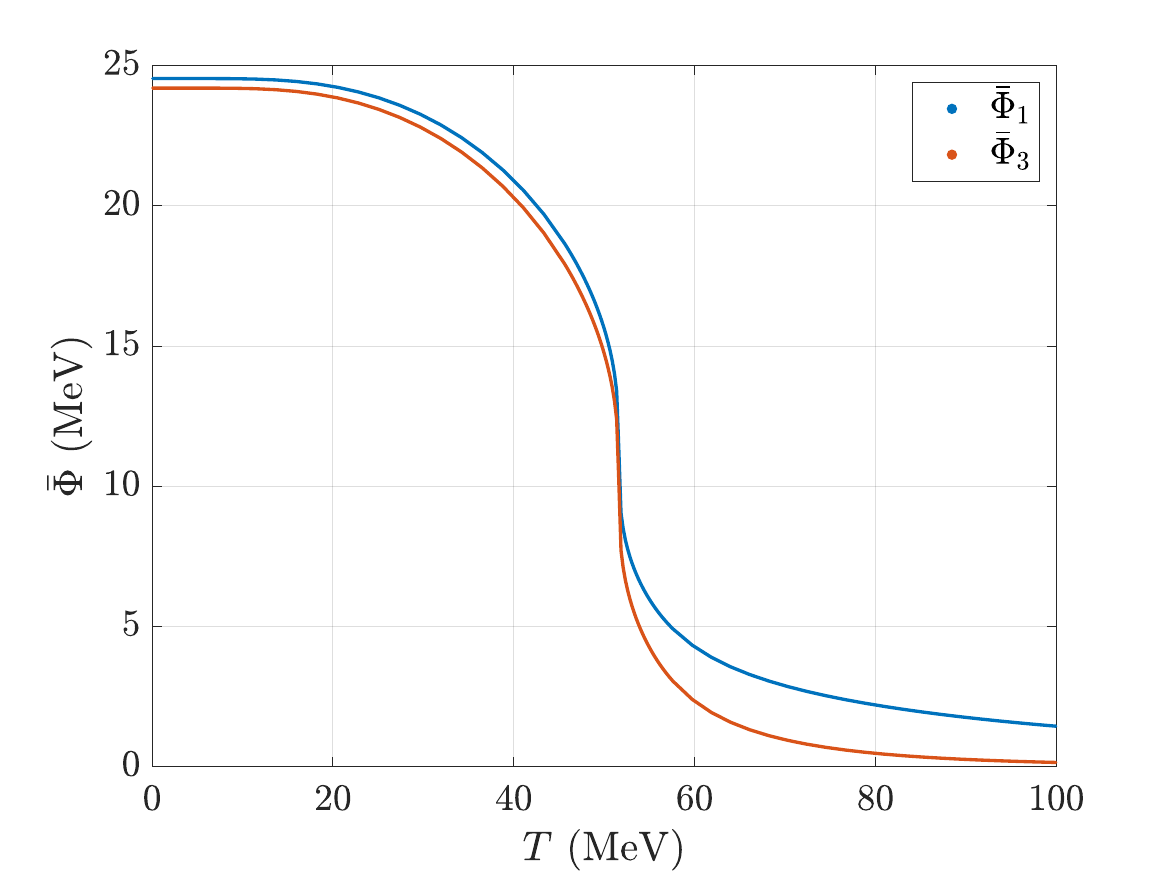}
        \includegraphics[width=0.32\linewidth]{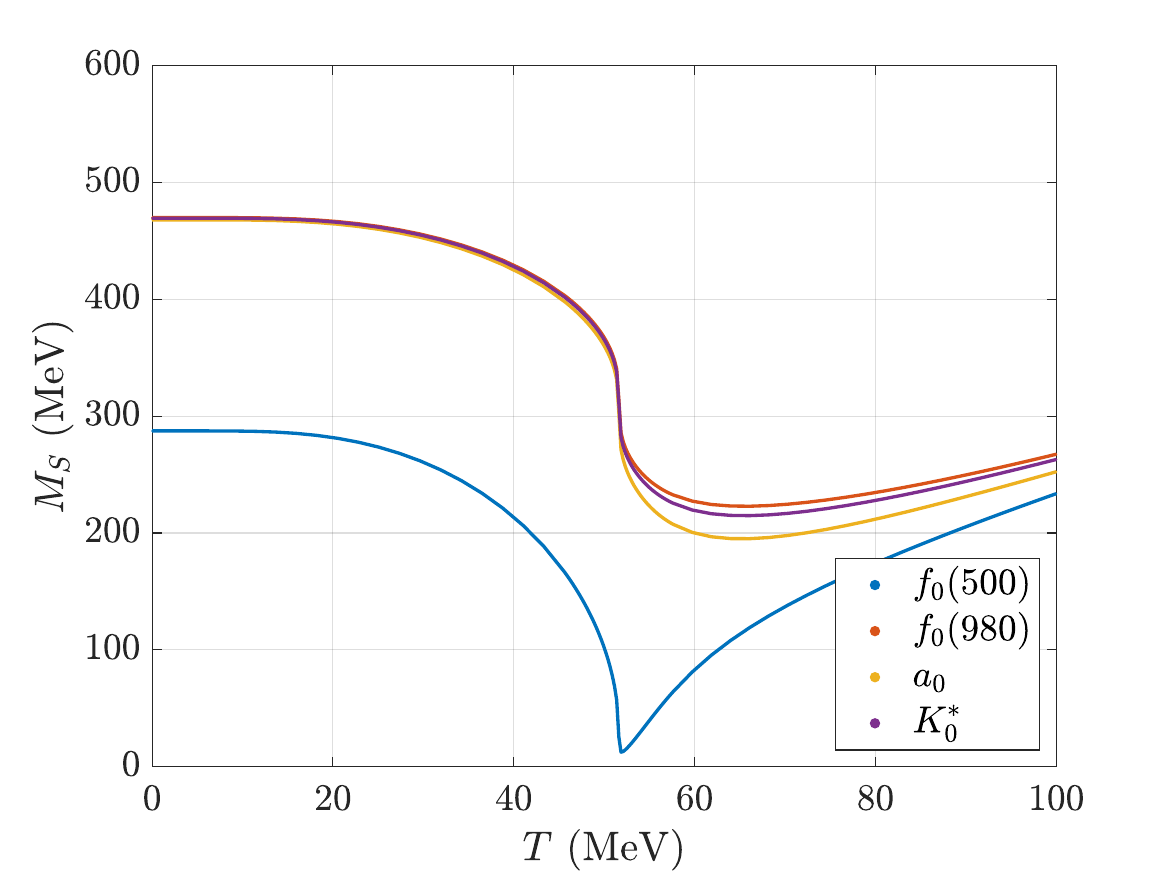}
        \includegraphics[width=0.32\linewidth]{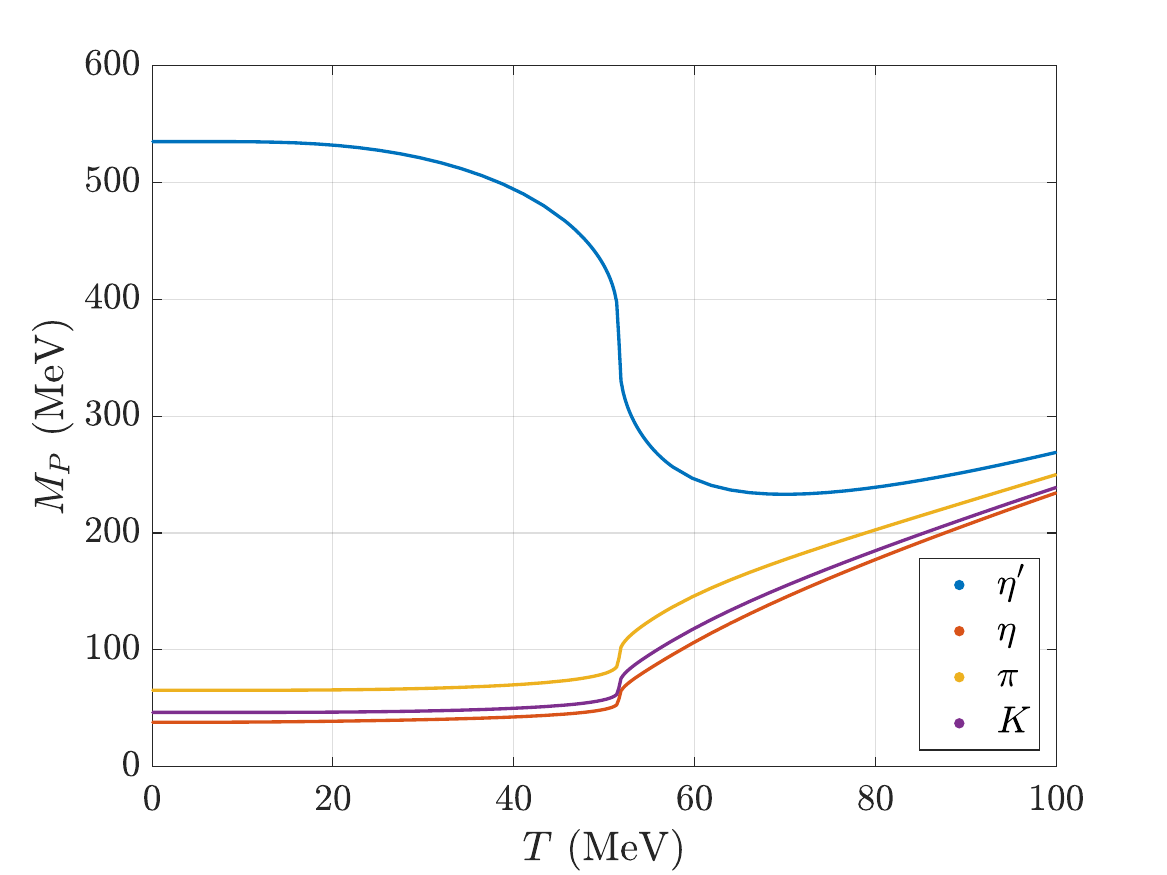}
        \\ \vspace{0.5em}
        \small (a)
    \end{minipage}
    \begin{minipage}[b]{1\textwidth}
        \includegraphics[width=0.32\linewidth]{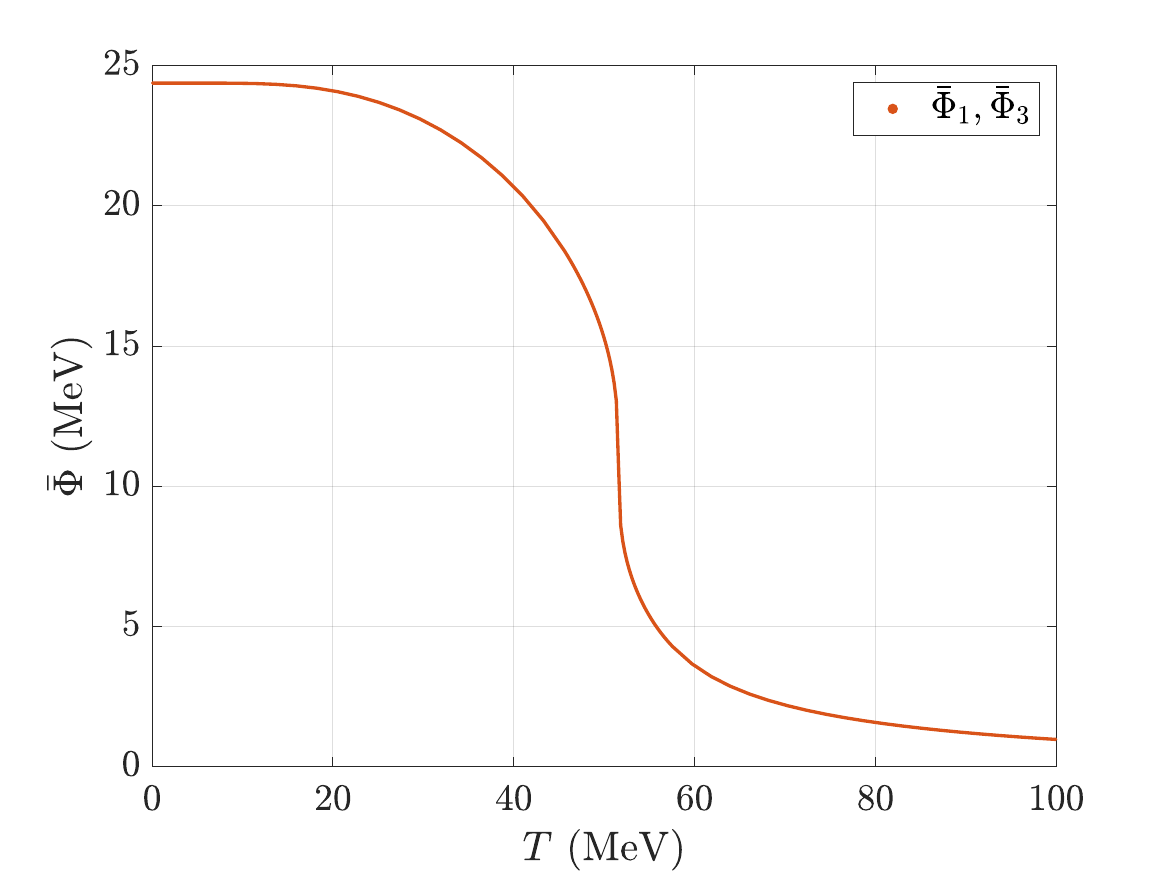}
        \includegraphics[width=0.32\linewidth]{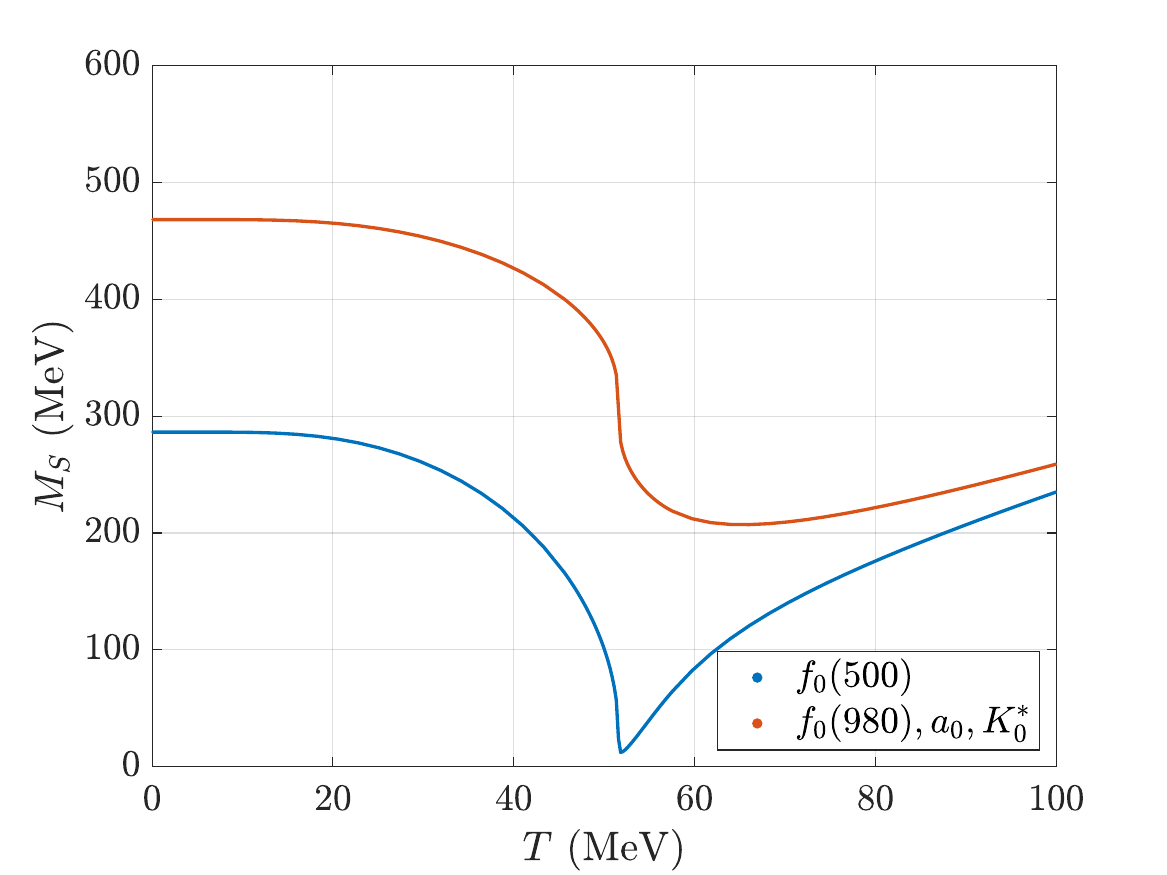}
        \includegraphics[width=0.32\linewidth]{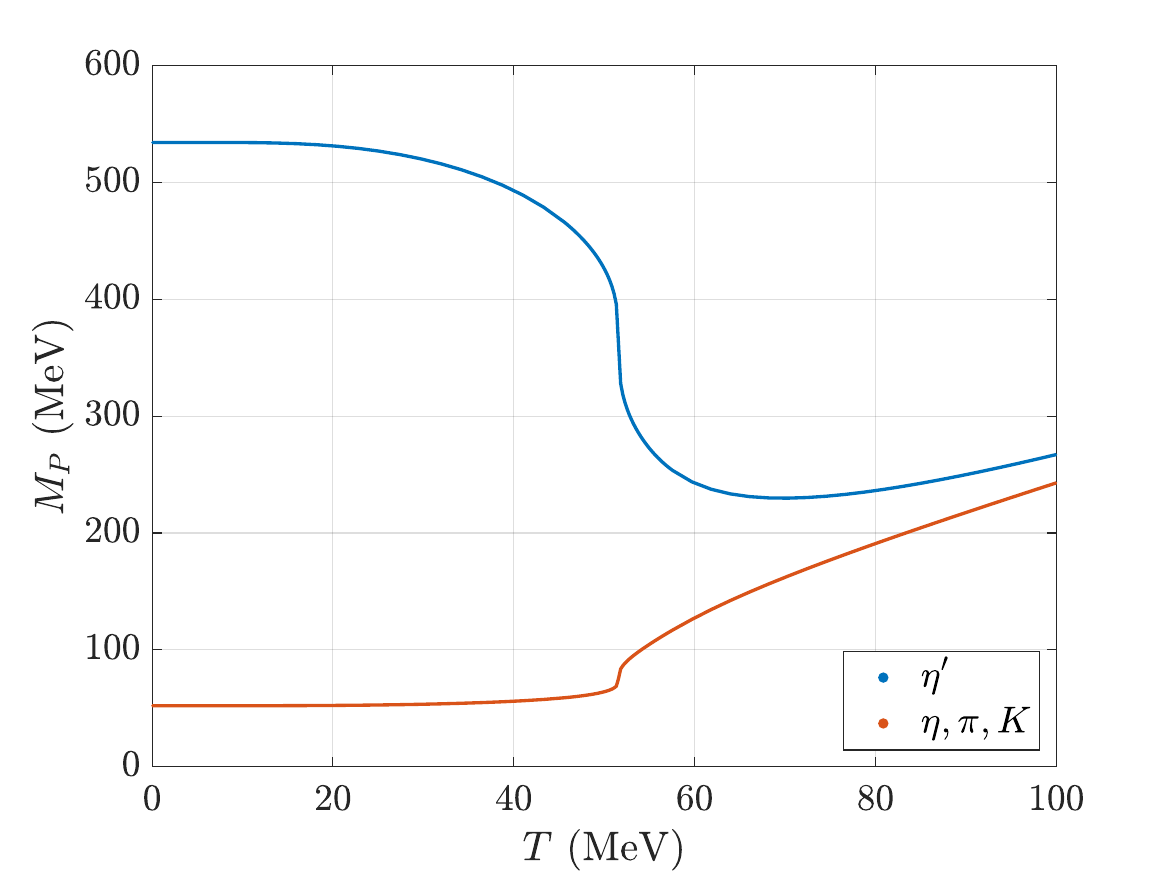}
        \\ \vspace{0.5em}
        \small (b)
    \end{minipage}
    \begin{minipage}[b]{1\textwidth}
        \includegraphics[width=0.32\linewidth]{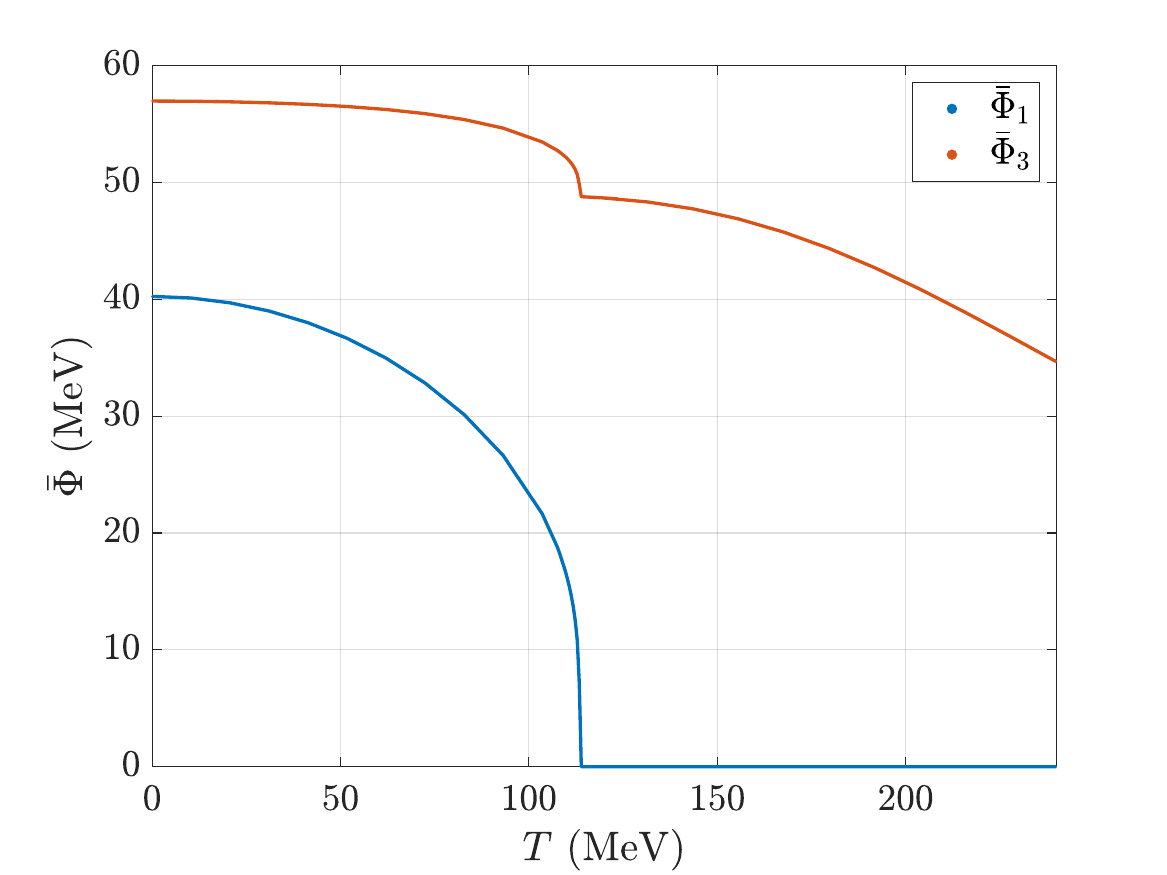}
        \includegraphics[width=0.32\linewidth]{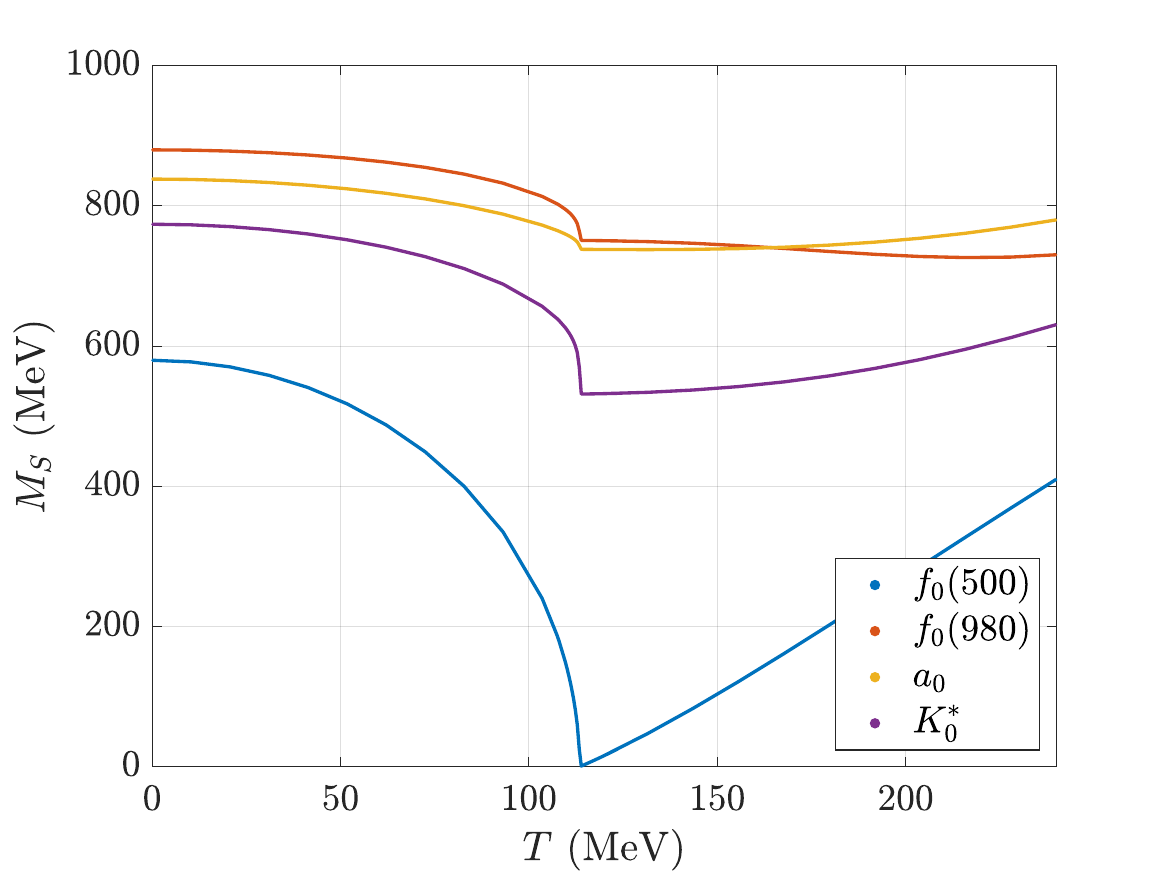}
        \includegraphics[width=0.32\linewidth]{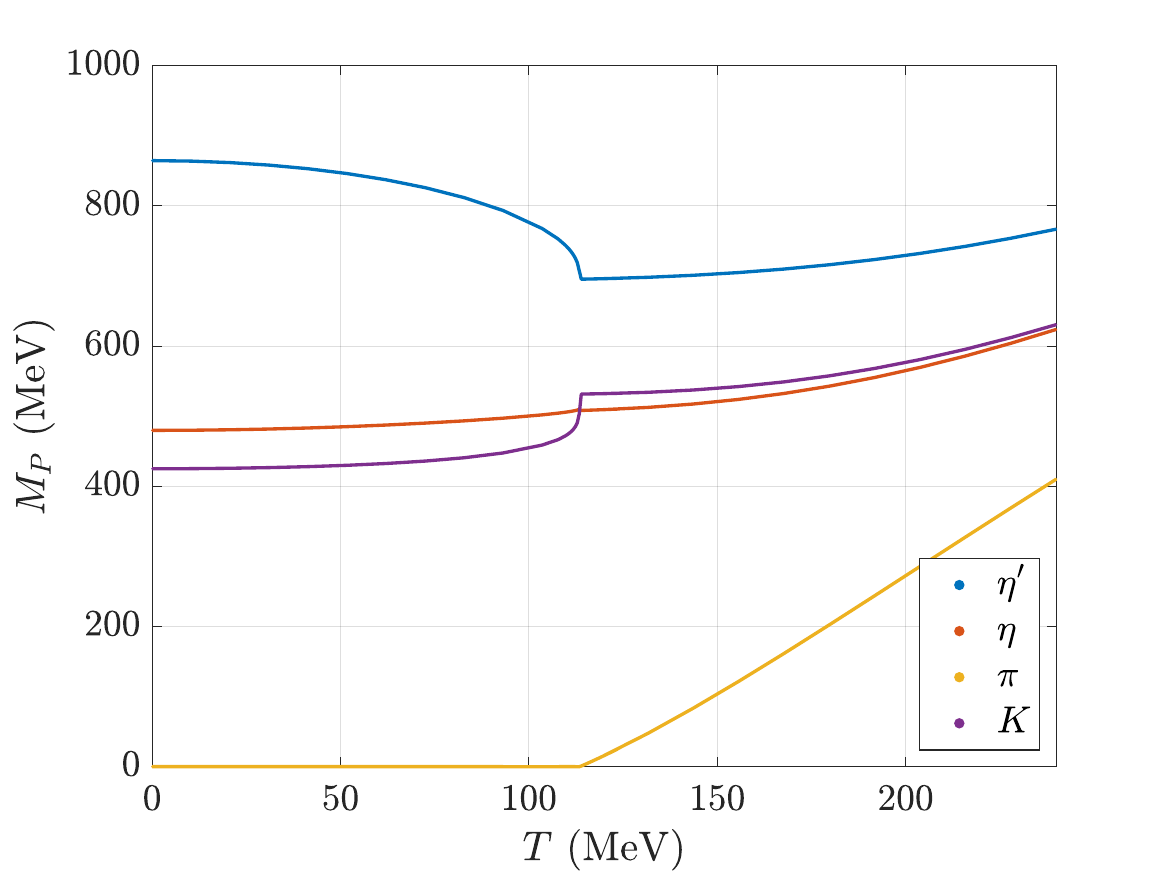}
        \\ \vspace{0.5em}
        \small (c)
    \end{minipage}
    \caption{
    Benchmarked chiral phase transitions at critical points (a), (b), and (c) defined in the text. 
    }
    \label{fig:BenchmarksSICJT}
\end{figure}

\section{Thermal evolutions of chiral order parameters and meson masses at critical points}
\label{app:PhaseTransitionsAtSomeTypicalPoints}

In this appendix, we present the results based on the SICJT formalism on 
the $T$-dependences of the chiral order parameters 
$\bar{\Phi}_{1,3}$ and scalar and pseudoscalar meson masses at several critical points 
in the Columbia plot, as seen in Fig.~\ref{fig:ColumbiaPlotsSigma}. 
We take the parameter set I in Sec.~\ref{Para-I} to fix the $f_0(500)$ mass: $m[f_0(500)] = 672.4 \ {\rm MeV}$. 

The chosen critical points are: 
\begin{itemize} 
\item[(a)] 
$(m_l, m_s)/m_l^{\rm phys.} \simeq (0.1474, 0)$: 
the second-order critical line on the $m_l$-axis \,, 

\item[(b)] 
$(m_l, m_s)/m_l^{\rm phys.} \simeq (0.0931, 0.0939)$: 
the second-order critical line on the three-flavor symmetric curve
\,,  

\item[(c)]  
$(m_l, m_s)/m_s^{\rm phys.} \simeq (0, 0.696)$:  
the second-order critical line on the $m_s$-axis, i.e., vicinity of the tricritical point.  

\end{itemize}
In Fig.~\ref{fig:BenchmarksSICJT}, we plot 
the $T$-dependences of the chiral order parameters 
$\bar{\Phi}_{1,3}$ and scalar and pseudoscalar meson masses 
at those critical points.

\section{Systematics on evaluation of the critical exponents}
\label{app:EvaluationOfTheCriticalExponents}

\begin{figure}[t]
    \centering
    \includegraphics[width=0.43\linewidth]{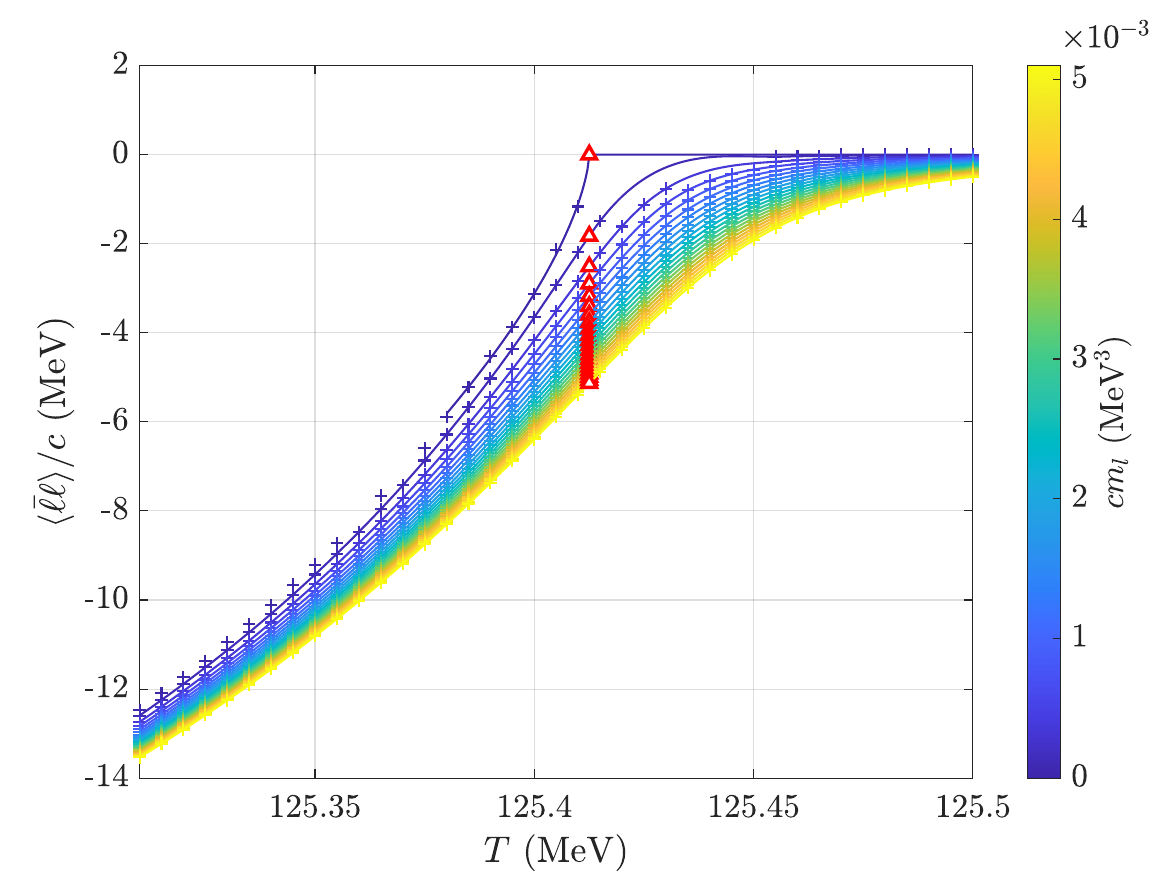}
    \includegraphics[width=0.43\linewidth]{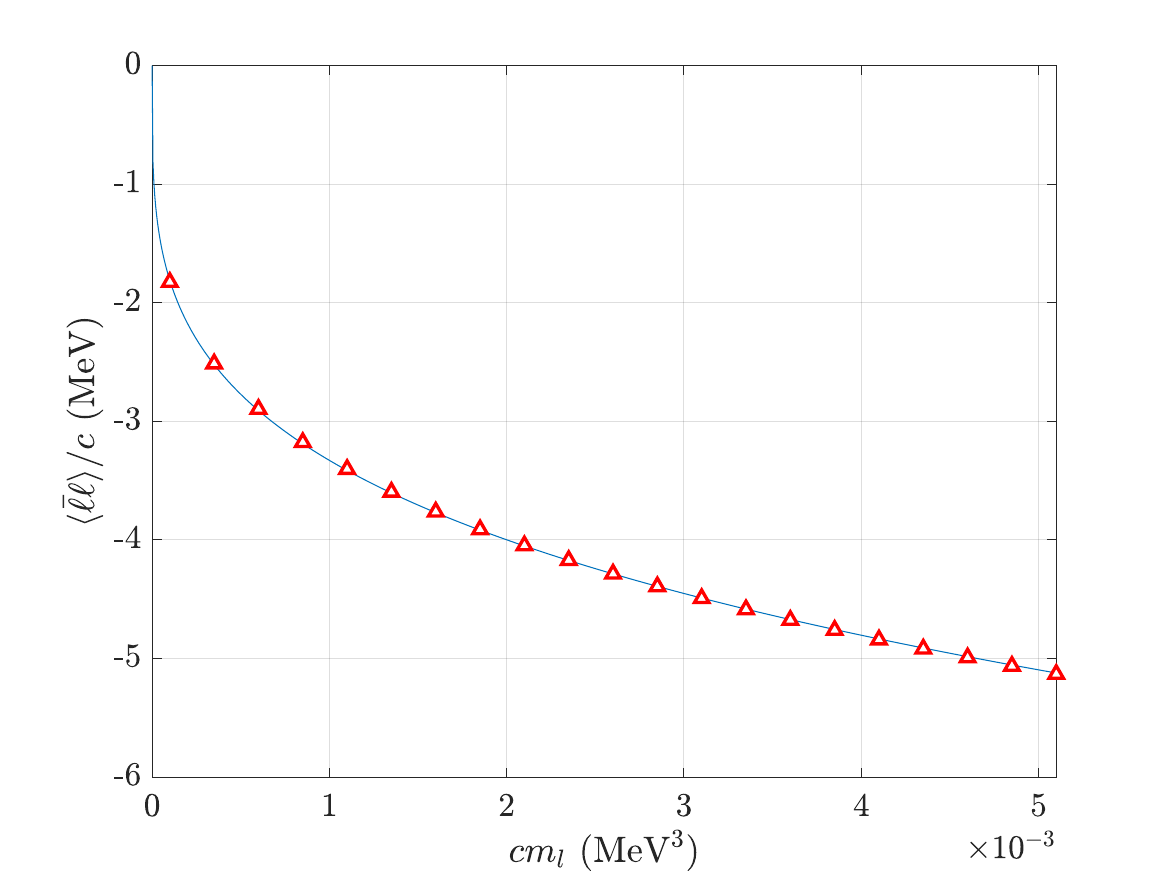}
    \caption{
    Plots of the light quark condensate $\langle \bar \ell \ell \rangle$ around the critical points where $m_l^c = 0$ and $m_s^c = m_s^{\rm phys.}$ as a function of the temperature $T$ (left panel) and the light current quark mass $c m_l$ (right panel).
    The plus markers in the left panel denote the benchmark points evaluated from the numerical approach, the solid line in the top denotes the fit result of the scaling function, and the rest of the solid lines represent the spline-interpolation functions.
    The correspondence between the benchmarks and the light current quark mass is shown in the colorbar.
    The red-triangle markers denote the points obtained by the spline interpolation at the critical temperature $T_c$.
    In the right panel, the red-triangle markers denote the same points obtained from the left panel, and the blue-solid line shows the scaling fit-function.
    }
    \label{fig:scalingBenchmarks}
\end{figure}

We read off the critical exponents of the light quark condensate $\langle \bar \ell \ell \rangle$ in the vicinity of the critical points based on the formulae in Eq.~\eqref{eq:scalingAnsatz}. 
To this end, we first try to observe the order of the phase transition by varying the temperature for selected points in the Columbia plot on the $(c m_l,c m_s)$ plane. 
Once the critical point is found out, we next shrink the range of the external parameters $(T, cm_l)$ with the step sizes $\Delta T = 0.3 \ {\rm MeV}$ and $\Delta (c m_l) = 5 \times 10^{-3} \ ({\rm MeV}^3)$.
For the current strange quark mass $c m_s$, we fix it to the critical values listed in Table~\ref{tab:CriticalExponents}, as also reflected in Fig.~\ref{fig:BenchmarksSICJT}. 
We then read off the $c m_l$-dependence of the light quark condensate to fit the scaling functions with the spline interpolation. 

For the case of $c m_l^{c} = 0$, we use the following three-parameter ansatz to fit the phase transition curve across the critical point with the variation of $T$: 
\begin{align}
    \langle \bar \ell \ell \rangle (T,0) = \theta(T_c - T) \ A \ \left( \frac{T - T_c}{T_c} \right)^\beta,
\end{align}
where $A$, $B$, and $T_c$ are parameters to be determined. 
When going along the $c m_l$-direction, we apply the following scaling function: 
\begin{align}
    \langle \bar \ell \ell \rangle (T_c,c m_l) = C \ (c m_l)^{\frac{1}{\delta}}
    \,, 
\end{align}
to extract the exponent $\delta$, where $C$ and $\delta$ are the paraameters to be determined.

For the case of $c m_l^{c} \neq 0$, to fit the scaling condensate with $T$, we introduce the following fit ansatz:  
\begin{align}
    \langle \bar \ell \ell \rangle (T,c m_l^c) = A \ \left( \frac{T - T_c}{T_c} \right)^\beta + B\,,
\end{align}
with a constant $B$ part. In moving in the $c m_l$-direction, we follow the following scaling function: 
\begin{align}
    \langle \bar \ell \ell \rangle (T_c,c m_l) = C \ \left(\frac{c m_l - c m_l^{c}}{c m_l^{c}}\right)^{\frac{1}{\delta}} + B
\,. 
\end{align}
When the $B$ part becomes relevant, the size of the scaling steps for ($T, c m_l$) would be crucial to determine the critical exponent. 
In that case, we have taken the step sizes as $\Delta T = 1 \times 10^{-3} \ {\rm MeV}$ and $\Delta(c m_l)/ c m_l^{c} = 1 \times 10^{-5} $. 

We use the least-squares fitting to determine the parameters in the scaling functions versus our numerical data.
One example at the critical point is shown in Fig.~\ref{fig:scalingBenchmarks}, 
where we have taken $m_l^c = 0$ and $m_s^c = m_s^{\rm phys.}$ together with the parameter set I in Sec.~\ref{Para-I}.

\bibliographystyle{JHEP} 
\bibliography{refs}

\end{document}